\documentclass[useAMS, fleqn, usenatbib]{mnras}
\usepackage{amsmath}
\usepackage{graphicx}
\usepackage[T1]{fontenc}
\usepackage{ae,aecompl}
\usepackage{amsmath}
\usepackage{amsfonts}
\usepackage{amssymb}
\usepackage{txfonts}
\usepackage{mathtools}
\usepackage{enumerate}
\usepackage{enumitem}
\usepackage{braket}
\usepackage{layouts}
\usepackage{float}
\usepackage{multirow}
\usepackage{tabularx}
\usepackage{booktabs}
\usepackage{array}
\usepackage[caption = false]{subfig}
\usepackage{amssymb}

\def\aap{{ A\&A}}

\def\apj{{ApJ}}
\def\apjl{{ApJL}}

\def\mnras{{MNRAS}}

\def\apjs{{ApJS}}

\makeatletter
\DeclareRobustCommand{\EmphMeas}[1]{\ifdim\fontdimen\@ne\font>\z@{\upshape#1}\else{\itshape#1}\fi}
\makeatother

\newcommand{\tr}{\textrm{Tr}}
\newcommand{\Mgal}{M_{\textrm{gal}}}

\newcommand{\Msun}{M_{\odot}}
\newcommand{\ts}[1]{\textsuperscript{#1}}

\DeclarePairedDelimiter\abs{\lvert}{\rvert}%
\DeclarePairedDelimiter\norm{\lVert}{\rVert}%

\makeatletter
\let\oldabs\abs
\def\abs{\@ifstar{\oldabs}{\oldabs*}}
\let\oldnorm\norm
\def\norm{\@ifstar{\oldnorm}{\oldnorm*}}
\makeatother

\newcolumntype{C}{>{\centering\arraybackslash}X}%
\newcolumntype{R}{>{\raggedleft\arraybackslash}X}%
\newcolumntype{L}{>{\raggedright\arraybackslash}X}%

\makeatletter
\newsavebox\myboxA
\newsavebox\myboxB
\newlength\mylenA

\newcommand*\xoverline[2][0.75]{%
    \sbox{\myboxA}{$\m@th#2$}%
    \setbox\myboxB\null% Phantom box
    \ht\myboxB=\ht\myboxA%
    \dp\myboxB=\dp\myboxA%
    \wd\myboxB=#1\wd\myboxA% Scale phantom
    \sbox\myboxB{$\m@th\overline{\copy\myboxB}$}%  Overlined phantom
    \setlength\mylenA{\the\wd\myboxA}%   calc width diff
    \addtolength\mylenA{-\the\wd\myboxB}%
    \ifdim\wd\myboxB<\wd\myboxA%
       \rlap{\hskip 0.5\mylenA\usebox\myboxB}{\usebox\myboxA}%
    \else
        \hskip -0.5\mylenA\rlap{\usebox\myboxA}{\hskip 0.5\mylenA\usebox\myboxB}%
    \fi}
\makeatother
\title[Probabilistic trained cosmic web classification]
{Probabilistic cosmic web classification using fast-generated training data}
\author[B. Buncher \& M. Carrasco Kind]{Brandon Buncher$^1$\thanks{buncher2@illinois.edu} and Matias Carrasco Kind$^{2,3}$\\
$^1$Department of Physics, University of Illinois, Champaign, IL 61820 USA\\
$^2$Department of Astronomy, University of Illinois, Urbana, IL 61801 USA\\
$^3$National Center for Supercomputing Applications, Urbana, IL 61801 USA}

\begin{document}

\pagerange{\pageref{firstpage}--\pageref{lastpage}} \pubyear{0000}

\maketitle

\label{firstpage}

\begin{abstract}
We present a novel method of robust probabilistic cosmic web particle classification in three dimensions using a supervised machine learning algorithm.  Training data was generated using a simplified $\Lambda$CDM toy model with pre-determined algorithms for generating halos, filaments, and voids.  While this framework is not constrained by physical modeling, it can be generated substantially more quickly than an N-body simulation without loss in classification accuracy.  For each particle in this dataset, measurements were taken of the local density field magnitude and directionality.  These measurements were used to train a random forest algorithm, which was used to assign class probabilities to each particle in a $\Lambda$CDM, dark matter-only N-body simulation with $256^3$ particles, as well as on another toy model data set.  By comparing the trends in the ROC curves and other statistical metrics of the classes assigned to particles in each dataset using different feature sets, we demonstrate that the combination of measurements of the local density field magnitude and directionality enables accurate and consistent classification of halo, filament, and void particles in varied environments.  We also show that this combination of training features ensures that the construction of our toy model does not affect classification.  The use of a fully supervised algorithm allows greater control over the information deemed important for classification, preventing issues arising from hyperparameters and mode collapse in deep learning models.  Due to the speed of training data generation, our method is highly scalable, making it particularly suited for classifying large datasets, including observed data.
\end{abstract}


\section{Introduction} \label{Introduction}

Large-scale structure (LSS) describes the largest scale inhomogeneities in the universe.  LSS is comprised of clusters, filaments, and voids.  Clusters are small, compact groups of tens to tens of thousands of particles with radii on the scale of Mpc or tens of Mpc.  Typically, galaxies form around dark matter halos, large dark matter overdensities with masses on the scale of $10^{11}$ $-$ $10^{15}$ $M_\odot$ \citep{WhiteStructure}.  Galaxy filaments are long, strand-like visible/dark matter overdensities that connect clusters, with lengths between $50$ and $80$ Mpc $h^{-1}$ \citep{FilLength}.  While filament properties are not well understood, recent data presented by \citet{FilDens} and \citet{FilDens2} has provided constraints on a variety of properties, including the density profile and total mass.  Voids are visible/dark matter underdensities that fill the space between filaments, with a typical density of around $2\times10^{-2}$ Mpc$^{-3}$ $h^3$ \citep{VoidDens}.  Voids are roughly triaxial ellipsoidal, with volumes on the order of $10^4$ - $10^5$ Mpc$^{-3}$ $h^3$ \citep{VoidShape}.

The halo (filament) mass fraction refers to the ratio of the mass of halo (filament) particles relative to the total mass of particles.  For the purposes of this paper, the equivalent metric for void particles, the fraction of particles in underdense regions, will hereafter be referred to as the void mass fraction.  The relative mass fractions of halos, filaments, and voids have been studied extensively; however, there exists substantial disagreements on their values depending on the methodology used.  Using $\Lambda$CDM N-body simulations, the cluster mass fraction has been estimated to be between 9 - 41$\%$ \citep{TWeb,CMFHigh}; the filament mass fraction has been found to be between 18 - 50$\%$ \citep{CMFHigh,LSSStats}; and the void mass fraction has been estimated to be between 13 - 27$\%$ \citep{VWeb,VMFHigh}.  In this work, all particles have the same mass, so the mass fraction is equivalent to the fraction of particles that are a member of a given class.

For the purposes of this work, we model the underlying dark matter distribution for halos, filaments, and voids, which provides an excellent representation of galaxies; hence, particles here refer to dark matter particles.

The formation and evolution of galaxies is controlled by a variety of local properties, including the local density of dark and visible matter \citep{LSSBook}.  The beginning of galaxy formation is primarily guided by the mass and density profile of the proto-galaxy's dark matter halo \citep{GForm}.  The collapse of an overdense region of dark and baryonic matter leaves a dark matter halo, a triaxial ellipsoidal dark matter overdensity.  The baryonic matter that remains in the gravitational well of the dark matter halo cools, it begins to collapse into star-forming regions.  While star formation depends weakly on the local density \citep{LSSBook}, the dynamics of the halo's gravitational collapse, which dictates the proto-galaxy's size and density profile, is governed primarily by this property.  The local density also governs the abundance of dark matter halos and, hence, interactions between local galaxies, such as tidal stripping of matter \citep{LSSBook,GForm}.  As a result, it is crucial to understand the local dark matter density to understand how galaxies form and evolve.

A galaxy's LSS class provides substantial information about the local density.  Thus, it is important to create an efficient, reliable method for determining an individual galaxy's morphological LSS class to understand problems such as these.  However, a universal, deterministic algorithm is too complex to construct explicitly \citep{LSSBook}.  Various classification algorithms, some of which are non-deterministic and/or non-universal, have been created; several of these are summarized in Table \ref{tab:MLCat} (replicated from \citealp{TracCWeb}), which we discuss in greater detail below.

Classifiers that do not use machine learning typically exploit physical or geometric properties of the structures they attempt to classify.  These may be further divided into those that classify individual particles and those that determine the location and extent of individual structures.  To classify particles, cluster finding algorithms frequently utilize connectedness among particles \citep{FOF,MST} and/or rely on local geometric information such as density \citep{CLASSIC}.  Filament finding algorithms, however, must include information on the local and global density field, as well as some additional information that differentiates them from halos.  Filament finding methods are typically geometrical \citep{Bisous,MMF2} or topological in nature \citep{SpineWeb,DisPersE,TRex}, though some graph-based methods exist \citep{MST}.  Geometric algorithms \citep{NEXUS,CLASSIC} have been used to classify both halos and filaments.  However, these algorithms do not assign morphological class values, let alone probabilities, to individual particles.  This substantially hinders the effectiveness of classification: due to difficulties in predicting fundamental properties of LSS analytically, the extent of these structures is highly dependent on arbitrary parameters.  For example, many algorithms differentiate between structures using arbitrary density/scale cutoffs, either implicitly or explicitly \citep{TracCWeb,LSSSup,ORIGAMI}  Differences between these cutoff values lead to substantial inconsistencies between the classes assigned these algorithms, such as structural mass/volume fractions and the halo mass function (HMF); a discussion of these differences can be found at \citet{NEXUS}.

Lagrangian field classifiers have been used for particle-based classification of halos, filaments, sheets, and voids \citep{ORIGAMI,DIVA,LICH}.  One major drawback to Lagrangian field  classifiers is that they require information about the location and velocity fields (or, equivalently, the initial and final positions of particles).  However, these challenges may be circumvented using Bayesian inference methods such as BORG \citep{BORG,HADES}, a Bayesian inference method that can reconstruct the probabilistic history of the mass and velocity fields of a sample of galaxies.  However, differences in the phase-space definitions for each of the structures leads to dramatic difference in class assignments, particularly for voids \citep{LICH}.

Some machine learning-based methods use deep convolutional neural networks; an example of this method may be found in \citep{DCNN}.  Alternative techniques utilize supervised learning from a variety of time snapshots over the evolution of an N-body simulation \citep{Snapshot,ImprintsHMT}.  While ML-based methods are generally more efficient than statistical classifiers, they do have substantial drawbacks.  Due to the inability to trace the internal classification methodology of deep learning algorithms, these algorithms provide little information on the hallmark features of structural classes.  In addition, deep learning methods are generally highly sensitive to initial hyperparameters, further inhibiting our understanding of why a particular class was chosen for a particular region/particle and potentially introducing arbitrarily-selected biases.  This sensitivity may inhibit the generation of a widely-applicable algorithm with reproducible results, as small changes in the test data set may require alteration of these hyperparameters.  Methods to avoid these biases have been implemented in other scenarios, such as in the fast generation of cosmic web images \citep{FastSim}; however, these methods require knowledge of the expected output, for which there is little consensus in the context of LSS classification.  While supervised techniques allow greater understanding of the features utilized to determine a particle or region's class, known algorithms require training data extracted from multiple N-body simulation snapshots, which are computationally expensive.  In addition, these methods are primarily designed to understand the time-evolution of halo mass distributions, which is not the goal of this project.  While it may be possible to adapt these methods to classify halo particles, it would still require multiple time snapshots.

A recent topological cosmic web classifier was presented by \citet{LSSSup}, in which the authors classified LSS particles by treating the cosmic web as a complex network.  Halos were found using a friends-of-friends (FOF) algorithm and used as nodes when constructing the network, and various metrics based on particle position and velocity were used in classification.  Unfortunately, classification was not successful, as demonstrated by an average confusion matrix score of $70\%$.  A major contribution to the poor performance of this method stems from the difficulty in classifying halo particles found in large voids as this method, along with many of the topological models described in Table \ref{tab:MLCat} perform classification using a relatively small range of length scales.  For example, in \citet{LSSSup}, the linking lengths used when constructing the network ranged from $1.6$ - $2.4$ Mpc $h^{-1}$, which fails to cover the radii of even medium-sized voids \citep{VoidShape}.  As is noted in \citep{TracCWeb}, local density is highly scale-dependent, indicating that density magnitude alone is ineffective when classifying halo particles \citep{LSSSup,TracCWeb}.  As such, a robust cosmic web classifier must take into account information beyond the local density, and must also ensure that a strict density magnitude cutoff is implicitly used when distinguishing between structures.

To simplify these classification routines, in this paper, we present an efficient ML-based classification routine that does not fit any of the categories summarized in Table \ref{tab:MLCat}.  Our algorithm requires substantially less information than others; in particular, we demonstrate that training using only information derived from particle positions in a single toy model generates enough information to classify a particles in a substantially larger N-body simulation.  We generate training data using a toy model constructed from pre-determined structural creation algorithms which are distributed pseudo-randomly throughout a particle field.  After performing measurements of the local density magnitude and density field directionality for each particle (each of which retains a ``true'' class inherited from its creation algorithm), we train a random forest ML algorithm to classify particles in an N-body simulation.  The classes assigned are then statistically verified through measurements of known features, such as the HMF.  As these measurements require only information based on the particles' current positions, we avoid requiring calculations of particle velocity, reducing the potential for error propagation due to multiple sources.  Though the toy model lacks many of the physical characteristics of N-body simulation, it requires substantially less time to generate: a 256$^3$-particle N-body simulation required thousands of node-hours on a multi-node cluster, while generating a toy model of the same size requires only a small fraction of this amount of time, and substantially less computational resources.  We aim to show that the robustness of ML will ``fill in'' the information missing from the simulation, allowing accurate classification at a fraction of the cost.

To avoid the issues presented by \citet{TracCWeb} and \citet{LSSSup}, we propose the inclusion of local density field directionality measurements in addition to measurements of the local density magnitude.  These measurements take the place of connectivity measurements \citep{ConnOrig} used in many topological and network-based classifiers.  The connectivity of the filamentary skeleton of a network defines a natural filament length scale, a property that allows the prediction of local filament properties \citep{CosPeak}.  As such, connectivity is especially effective for classification of filament particles \citep{ConnOrig,NetConn}.  The local density field directionality provides an alternative to connectivity and other classification methods based on a particle's proximity to a ridge in a network skeleton, with the benefit that it takes into account only the properties of particles local to a given filament particle.  In addition, as the directionality value is inherently normalized by the local density magnitude, it depends only on the local directionality and density field contrast.  This naturally removes any implicit density magnitude cutoff for classification purposes, making it a robust method for filament classification that takes into account local environmental variations.

An additional benefit provided by supervised models is the ability to assign probabilities to class values.  The classes of particles on the border of a given structure are ambiguous, with the class assignment often being due to arbitrary density cutoffs.  Probabilistic classification, which is most easily achieved using a supervised model, enables us to quantify this ambiguity, providing additional information that may be correlated with other particle properties to more deeply understand how structure class is tied with other properties.

A final major benefit of this classifier is that it could be applied to observed data.  Different regions of a dataset may exhibit differing field depths.  To account for this, a toy model simulation could be created for each region that matches the parameters of that region, preventing the bias that would result from deep learning-based algorithms, which cannot distinguish between local field depth and density.  Also, the only information required for our classifier is particle position (in particular, particle velocities are not required), making it easier to apply to observed data.

The remainder of this paper is structured as follows: in Section \ref{Methods}, we discuss how we generate the toy model and create training data.  In Section \ref{Prediction}, we present our class assignments for particles in an N-body simulation and a toy model.  In Section \ref{Analysis}, we demonstrate the robustness our classifier through an analysis of our results.  Finally, Section \ref{Conclusion} provides a summary of our conclusions.

Supplementary figures and demonstrations of the methodology can be found on GitHub\footnote{https://github.com/bmbuncher/Prob-CWeb}.

\begin{table*}
    \begin{tabular}{@{}lcccl}
        Method & Web types & Input & Type & Main References \\ \hline
        \ \\
        MST & filaments & halos & Graph \& Percolation & \citet{MST} \\
        \ \\
        Bisous; FINE  & filaments & halos & Stochastic & \citet{Bisous,FINE} \\
        \ \\
        T-web; V-web; CLASSIC & all & particles & Hessian & \citet{TWeb,VWeb} \\
        &&&&\citet{CLASSIC} \\
        \ \\
        NEXUS+ & all & particles & Scale-Space, Hessian & \citet{NEXUS} \\
        MMF-2 & all except halos & particles & Scale-Space, Hessian & \citet{MMF2,Kinematic} \\
        \ \\
        Spineweb; DisPersE & all except halos & particles & Topological & \citet{SpineWeb,DisPersE} \\
        T-Rex & filaments & density field & Topological & \citet{TRex} \\
        \ \\
        ORIGAMI; MSWA; LICH & all & particles & Phase-Space & \citet{ORIGAMI,SSVoids} \\
        &&&&\citet{MultiStream,LICH} \\
        DIVA & voids & particles & Phase-Space & \citet{DIVA} \\
        \ \\
        BORG\ts{\textasteriskcentered} & all & particles & Phase-Space & \citet{BORG} \\
        HADES\ts{\textasteriskcentered} & voids & particles & Hamiltonian Monte Carlo & \citet{HADES} \\
        \ \\
        This work & all except sheets & particles & Stochastic Geometric & {} \\
        \ \\
        \hline
    \end{tabular}
    \caption{An expanded overview of the methods compared in \citet{TracCWeb}; ``all'' indicates that the algorithm classifies particles as members of halos, filaments, voids, or sheets/walls.  Algorithms marked with an asterisk indicate that these methodologies do not perform classification independently; rather, they provide a probabilistic reconstruction of the density and velocity field which, when paired with a classifier, may be used to assign class probabilities.}
    \label{tab:MLCat}
\end{table*}

\section{Methods} \label{Methods}

Unless otherwise stated, all parameters found in this section are listed in Table \ref{tab:RosettaStone}.

\begin{table*}
    \begin{tabularx}{\textwidth}{XX}
        \hline \\
        \multicolumn{2}{c}{\bf{Datasets and Measurements}} \\
        \rule{0pt}{2ex}\texttt{TSIM} & Toy model simulation, test dataset \\
        \rule{0pt}{2ex}\texttt{SIM} & N-body simulation, test dataset \\
        \rule{0pt}{2ex}\EmphMeas{VOR} & Voronoi cell volumes (density magnitude) \\
        \rule{0pt}{2ex}\EmphMeas{CMD} & Distance between a particle and the center of mass of particles within a radius $R_{\textrm{CME}}$ (density magnitude) \\
        \rule{0pt}{2ex}\EmphMeas{MI} & Moment of inertia of particles within a radius of $R_{\textrm{CME}}$ (density magnitude) \\
        \rule{0pt}{2ex}\EmphMeas{ENC} & Number of particles within a radius of $R_{\textrm{CME}}$ (density magnitude) \\
        \multicolumn{2}{l}{\hspace{1 mm}--\hspace{1 mm}$R_{\textrm{CME}} \in \{ 0.2,\:0.5,\:0.8,\:1.0,\:1.5,\:2.0,\:3.0,\:5.0,\:7.5,\:10.0,\:12.0\}$ Mpc $h^{-1}$} \\
        \rule{0pt}{2ex}\EmphMeas{KNN} & \multirow{2}{\linewidth}{Distance to the $k^{\textrm{th}}$-nearest neighbor (density magnitude)} \\
        \hspace{1 mm}--\hspace{1 mm}$k \in \{1,\:2,\:3,\:5,\:8,\:10,\:15,\:20,\:25,\:30,\:35,\:40\}$ & \\
        \rule{0pt}{2ex}\EmphMeas{PCA} & \multirow{4}{\linewidth}{Difference between the maximum and minimum explained variance ratio from a \EmphMeas{PCA} decomposition of particles within a radius $R_{\textrm{PCA}}$ (density field directionality)} \\
        \hspace{1 mm}--\hspace{1 mm}$R_{\textrm{PCA}} \in \{1.5,\:1.6,\:1.7,\:1.8,\:1.9,\:2.0\}$ Mpc $h^{-1}$ & \\
        \hspace{1 mm}--\hspace{1 mm}$\sigma_{\textrm{PCA}} = 0.45$ (resampling parameter, see Section \ref{PCA}) & \\
        \hspace{1 mm}--\hspace{1 mm}$\delta n_{\textrm{PCA}} = 5.5$ (resampling parameter, see Section \ref{PCA}) & \\\\
        \hline \\
        \multicolumn{2}{c}{\bf{General}} \\
        \rule{0pt}{2ex}$M_p = 7.55\times10^{10}$ $\Msun$ & Particle mass in toy model and \texttt{SIM} \\
        \rule{0pt}{2ex}$n_{\textrm{tot}} = 1.0$ Mpc$^{-3}$ & Particle number density in toy model and \texttt{SIM} \\
        \rule{0pt}{2ex}$N_{\textrm{tot, \texttt{SIM}}} = 256^3$ & Number of particles in \texttt{SIM} \\
        \rule{0pt}{2ex}$N_{\textrm{tot, \texttt{TSIM}}} = 85^3$ & Number of particles in \texttt{TSIM} \\
        \rule{0pt}{2ex}$L_{\textrm{\texttt{SIM}}} = 256$ Mpc $h^{-1}$ & Side length of \texttt{SIM} particle field \\
        \rule{0pt}{2ex}$L_{\textrm{Toy}} = 85$ Mpc $h^{-1}$ & Side length of training and \texttt{TSIM} particle fields \\
        \rule{0pt}{2ex}$\delta M_{\textrm{H}} = 0.42$ & Halo mass fraction in toy model \\
        \rule{0pt}{2ex}$\delta M_{\textrm{F}} = 0.38$ & Filament mass fraction in toy model \\
        \rule{0pt}{2ex}$\delta M_{\textrm{V}} = 0.42$ & Void mass fraction in toy model \\\\
        \hline \\
        \multicolumn{2}{c}{\bf{Section \ref{HaloGen} (Halo Generation)}} \\
        \rule{0pt}{2ex}$N_{\textrm{min}} = 8$ & Minimum number of particles in toy model halos \\
        \rule{0pt}{2ex}$N_{\textrm{max}} = 13245$ & Maximum number of particles in toy model halos (coorresponds with a mass of $10^{11}$ $\Msun$) \\
        \rule{0pt}{2ex}$R_0 = 0.12$ Mpc $h^{-1}$ & Parameter in Eqn. \eqref{eq:RHalo} \\
        \rule{0pt}{2ex}$\alpha = 0.38$ & Parameter in Eqn. \eqref{eq:RHalo} \\
        \rule{0pt}{2ex}$\sigma_0 = 0.12$ Mpc $h^{-1}$ & Parameter in Eqn. \eqref{eq:SigHalo} \\
        \rule{0pt}{2ex}$\beta = 0.16$ & Parameter in Eqn. \eqref{eq:SigHalo} \\
        \hline \\
        \multicolumn{2}{c}{\bf{Section \ref{FilGen} (Filament Generation)}} \\
        \rule{0pt}{2ex}$R_{\textrm{F,\hspace{0.15em}min}} = 0.3$ Mpc $h^{-1}$ & Minimimum filament radius \\
        \rule{0pt}{2ex}$R_{\textrm{F, max}} = 0.6$ Mpc $h^{-1}$ & Maximum filament radius \\
        \rule{0pt}{2ex}$B_{\textrm{min}} = 0.65$ Mpc$^{-4}$ $h^4$ & Parameter in Eqn. \eqref{eq:nFil} \\
        \rule{0pt}{2ex}$B_{\textrm{max}} = 1.15$ Mpc$^{-4}$ $h^4$ & Parameter in Eqn. \eqref{eq:nFil} \\
        \rule{0pt}{2ex}$\lambda_0 = 2.85$ Mpc$^{-1}$ $h$ & Parameter in Eqn. \eqref{eq:nFil}; represents the filament number density for filaments with radius $R_{\textrm{F}} = R_{\textrm{F, max}}$ \\
        \rule{0pt}{2ex}$\lambda_{\textrm{F, min}} \pm \delta \lambda_{\textrm{F, min}} = 3.75 \pm 0.25$ Mpc$^{-1}$ $h$ & The range of filament number densities for filaments with radius $R_{\textrm{F}} = R_{\textrm{F, min}}$ \\
        \ \\
        \hline
    \end{tabularx}
    \caption{A glossary of acronyms, measurement parameters, and numerical values used throughout this paper}
    \label{tab:RosettaStone}
\end{table*}

We aim to classify individual LSS particles using a random forest \citep{RandomForest} ML algorithm trained using a fast generated data.  We developed a toy model that simulated a particle field comprised of halos, filaments, and voids.  Measurements of the local, global, and isotropic densities and direction fields were taken for each particle and used to train the ML algorithm.  The trained algorithm was used to assign LSS class values for each particle in an N-body simulation we ran, hereafter referred to as ``\texttt{SIM}''.

\texttt{SIM} is a $\Lambda$CDM model simulation consisting of $N_{\textrm{tot,\,\texttt{SIM}}}$ collisionless dark matter particles with particle number density $n_{\textrm{tot}}$, each of which has a mass of $M_p$.  The parameters for this simulation were taken from the WMAP+BAO+$H_0$ results found in \citep{SIMParams}.  The cosmological parameters used were $\Omega_{m,0} = 0.272$, $\Omega_{\Lambda,0} = 0.728$, and $h = 0.704$, where the Hubble parameter $H_0 = 100$ $h$ km s$^{-1}$ Mpc$^{-1}$.  Initial conditions were generated using second-order Lagrangian perturbation theory (2LPT) instead of the standard Zeldovic approximation (see \citet{2LTP} and \citet{Zeldovic} for an explanation of this code).  The primordial linear power spectrum was generated using CAMB.  For this cosmology, the power spectrum was normalized using $\sigma_0 = 0.810$ and spectral index $n_s = 0.967$.  As the simulation included only dark matter particles, we evolved them using the parallel tree N-body/smoothed particle hydrodynamics (SPH) code GADGET-2 \citep{Gadget}.  Only the tree code was used for this simulation.  The simulation started at redshift $z = 50$ (corresponding with scale factor $a = 0.0196$) and evolved until the scale factor $a$ reached 1. For the purposes of this work, we used a snapshot of \texttt{SIM} at $z = 0$.

\subsection{Toy Model Simulation Generation} \label{ToyModelGen}

\begin{figure*}
    \centering
    \begin{minipage}{0.97\textwidth}
        \centering
        \includegraphics[width = 0.97\textwidth,keepaspectratio]{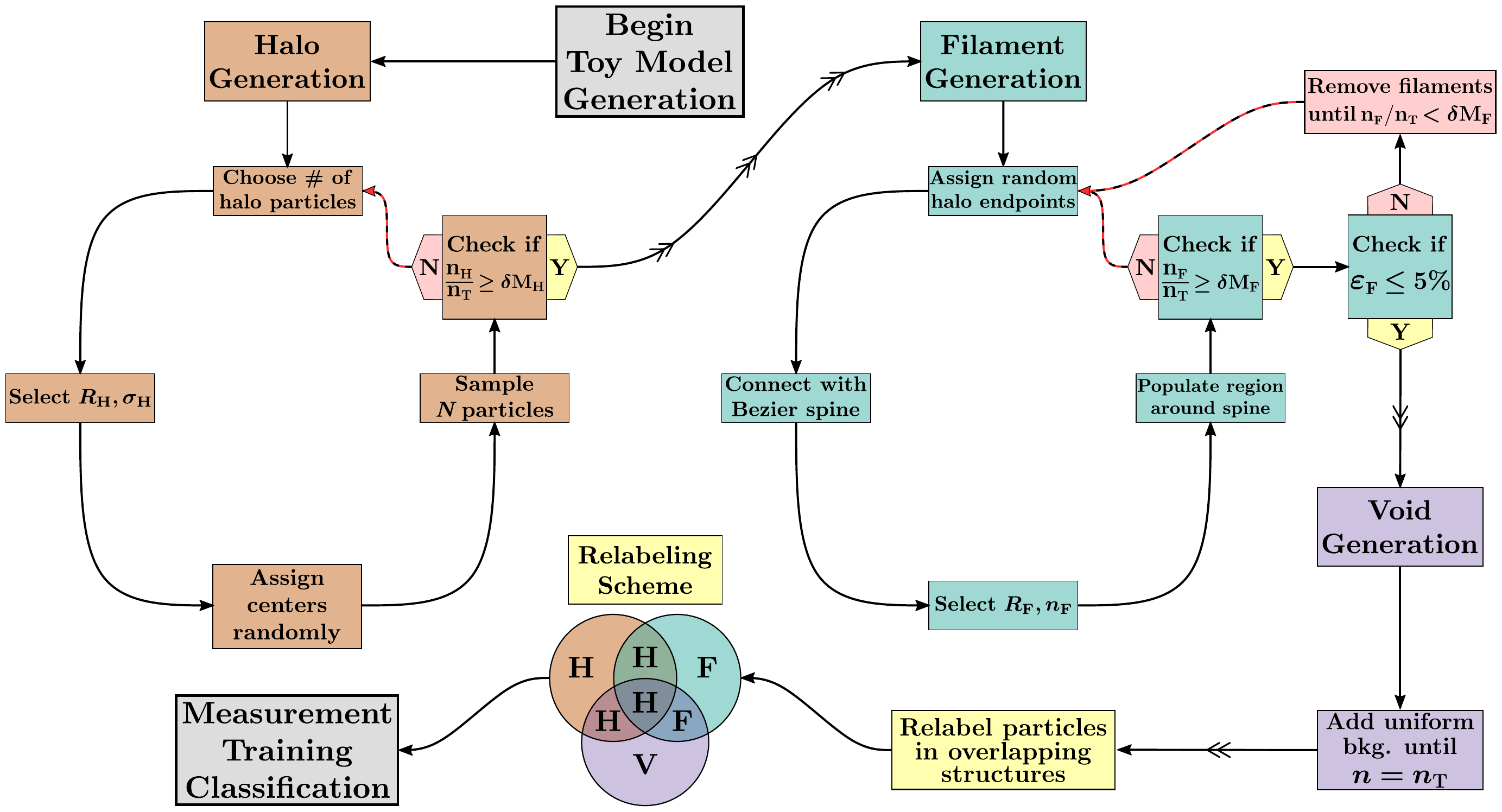}
    \end{minipage}
    \caption{A diagram showing the steps used to generate a toy model dataset.  Values for the listed parameters can be found in Table \ref{tab:RosettaStone}.}
    \label{fig:ToyDiagram}
\end{figure*}

The method for generating a toy model dataset consisted of creation algorithms for halos, filaments, and voids.  Note that the toy model only reproduces general structural features: rather than simulating the time evolution of matter due to gravity from the beginning of the universe, each structure is produced without regard to a physical creation process.  While this toy model lacks the physical processes seen in N-body simulations, the generation process is substantially faster and more computationally efficient.

In the toy model, each particle's mass is defined as $M_{\textrm{p}}$, and the density of the universe $\xoverline{\rho} = M_{\textrm{p}}$ Mpc$^{-3}$ $h^3$, corresponding to a particle number density $n_{\textrm{p}} = 1$ Mpc$^{-3}$ $h^3$.

A diagram of the toy model creation process can be found in Figure \ref{fig:ToyDiagram}

\subsubsection{Halo Generation} \label{HaloGen}

The toy model simulation algorithm begins with halo generation.  Halo masses were sampled from a halo mass function \citep{WarrenHMF} with minimum (maximum) halo sizes $N_{min}$ and $N_{max}$ particles; this HMF model was also used to generate the test data.  To select varied, visually realistic halo densities, we found an empirical probability density function (defined in Eqns. \eqref{eq:RHalo}, \eqref{eq:SigHalo}) that related a halo's radius to its mass.  It was assumed that, for halos of a constant density $\rho_{\textrm{H}} = 200 \xoverline{\rho}$, $R_{200} \sim M_{\textrm{H}}^{1/3}$.  \citep{RtoMRat} demonstrate through analysis of observation and simulation data that $M_{\textrm{H}} \sim N_{200}^\alpha$, where $\alpha$ is close to unity, and thus that $R_{200} \sim N_{200}^{1/3}$.  We performed a similar fit on \texttt{SIM} using the halo mass $M$ and radii $R(M)$ calculated by a friend-of-friend cluster finding algorithm in \citet{DBSCAN} and \citet{yt}, determining that

\begin{equation} \label{eq:RHalo}
    \braket{R(M)}_{\textrm{H}} = R_0 \left(\frac{M}{\Mgal}\right)^\alpha,
\end{equation}

where $\braket{R(M)}_{\textrm{H}}$ is the expected halo radius for a given mass $M$; the empirically-found value for $\alpha$ ($\alpha = 0.38$) agrees well with these prior results.

Based on empirical calculations, we assumed that the probability density function for the radii $R(M)$ for halos of a given mass $M$ followed a log-normal distribution; this assumption was based on observation of the radius histogram for halos of a given mass.  For each $M = M_0$ with at least 100 particles, we fit $R(M = M_0)$ to a log-normal histogram where the mean $\mu = \braket{R(M = M_0)}$.  We then empirically found that the standard deviation $\sigma(M)$ also exhibited a power-law dependence on the mass:

\begin{equation}\label{eq:SigHalo}
    \sigma(M) = \sigma_0 \left(\frac{M}{\Mgal}\right)^\beta
\end{equation}

To create a halo, the halo mass was sampled from the HMF described by \citeauthor{WarrenHMF} using algorithms from \citet{yt} and \citet{HMF}, and the halo radius was sampled from the corresponding log-normal distribution.  Once a halo's mass and radius were determined, particles were generated by sampling their radial distance from the halo's center from a truncated spherical normal spatial distribution with standard deviation $\sigma_{\textrm{H}}(R_{\textrm{H}}) = \frac{R_{\textrm{H}}}{5}$.

Halo masses were sampled from the HMF until the halo mass fraction reached the desired value, i.e. $\frac{M_{\textrm{h, tot}}}{M_{\textrm{tot}}} \geq \delta_{\textrm{h}}$.  Halo centers were pseudo-randomly placed throughout the particle field of side length $L_{\textrm{Toy}}$, then populated via the process described above.

\subsubsection{Filament Generation} \label{FilGen}

Filaments were constructed by first creating a spine, then populating the surrounding volume with particles.  The spine was created by selecting two halo centers as endpoints, then creating a Bezier curve between them of degree 2 \citep{Bezier}.  The Bezier nodes were perturbed from the axis connecting the endpoints by $\Delta r$, were $0 \leq \Delta r \leq L_{\textrm{F}}$, where $L_{\textrm{F}}$ is the distance between the two endpoints.  Particles were populated within a cylinder of radius $R_{\textrm{F, min}} \leq R_{\textrm{F}} \leq R_{\textrm{F, max}}$ around the spine.  The radial number density was calculated by sampling from a uniform distribution with maximum and minimum values

\begin{align} \label{eq:nFil}
    & \lambda_{\textrm{min}}(R_{\textrm{F}}) = B_{\textrm{min}}\left(\frac{R_{\textrm{F, max}} - R_{\textrm{F}}}{R_{\textrm{F, max}} - R_{\textrm{F, min}}}\right)^3 + \lambda_0 \\
    & \lambda_{\textrm{max}}(R_{\textrm{F}}) = B_{\textrm{max}}\left(\frac{R_{\textrm{F, max}} - R_{\textrm{F}}}{R_{\textrm{F, max}} - R_{\textrm{F, min}}}\right)^3 + \lambda_0 \nonumber \\
    & \textrm{for }R_{\textrm{F, min}} \leq R_{\textrm{F}} \leq R_{\textrm{F, max}} \nonumber
\end{align}

This corresponded with fixing the filament radial number density $n_{\textrm{F}}\left(R_{\textrm{F, max}}\right) = n_0$ and allowing the density for the filaments with the minimum radius to vary as $n_{\textrm{F}}\left(R_{\textrm{F, min}}\right) = n_{\textrm{F, min}} \pm \delta n_{\textrm{F, min}}$.  These values were chosen to visually match filaments seen in \texttt{SIM}: we approximated the minimum and maximum filament radii and number densities using several prominent filaments in \texttt{SIM}, then applied a bridging function (Eqn. \ref{eq:nFil}) to ensure that smaller radii correlated with higher densities.  We will demonstrate that these selections do not strongly affect class label assignment, so this process may be easily replicated for another test dataset.

To create a filament, 1) two endpoint halos were selected (excluding halo pairs that already have a filament generated between one another); 2) a density was selected; 3) the spine was generated; and 4) the cylindrical region around the spine was populated.  These particles were placed pseudorandomly along the spine, then perturbed orthogonally from the spine using a truncated normal distribution with standard deviation $\sigma_{\textrm{F}}(R_{\textrm{F}}) = \frac{5}{4}R_{\textrm{F}}$.  This value was chosen so that the density at the edge of the filament was roughly $\frac{3}{4}$ the density of the center, ensuring that the filament boundary corresponded closely with the edge of the particle overdensity.

Filaments were created until the filament number density $n_\textrm{F}$ exceeded the desired density $n_{\textrm{tot}}M_{\textrm{F}}$.  To ensure that the filament number density was close to the desired density, filaments were iteratively destroyed and recreated until

\begin{align}
    \frac{\abs{n_\textrm{F} - n_{\textrm{tot}} \, \delta M_{\textrm{F}}}}{n_{\textrm{tot}} \, \delta M_{\textrm{F}}} \leq 0.05
\end{align}

Once this condition was satisfied, the filament mass fraction's deviation from $\delta M_{\textrm{F}}$ (the desired filament mass fraction) was deemed small enough to begin background generation.

\subsubsection{Background generation} \label{BkgGen}

Background (void) particles were sampled from a uniform distribution so that the number of void particles $N_{\textrm{V, tot}} = N_{\textrm{tot}} - N_{\textrm{H, tot}} - N_{\textrm{F, tot}} \approx N_{\textrm{tot}}\delta M_{\textrm{V}}$.  Note that, due to the fact that a truncated normal distribution was used to populate both halos and filaments, a sharp cutoff exists at the boundary of each halo and filament.  This was a done to simplify the simulation and provide more control over its parameters; we will show that it did not affect our results.

\subsubsection{LSS Labels} \label{Labels}

Each particle inherited an LSS class label (halo, filament, or void) from its creation algorithm; however, to prevent contamination of measurement results from these segments, particles were re-labelled according to a hierarchy.  Particles within the boundaries of a halo were relabelled as halo particles; any remaining particles within the boundaries of a filament were relabelled as filament particles; and the rest remained void particles.

\subsection{Measurements} \label{Measurements}

Next, measurements of the local, global, and isotropic density and direction field were taken to use as training data.  We used five separate measurements of the density magnitude and one measurement of directionality.  While each of them may measure similar properties, each carries different information, so a combination can improve classification accuracy and robustness.  Throughout the remainder of this paper, we discuss which measurements proved most effective.  All measurements were normalized such that all values lay between 0 and 1, and are described below.

\subsubsection{Voronoi Cell Volume \EmphMeas{(VOR)}} \label{VOR}

A Voronoi diagram is a method of partitioning of some multidimensional space.  For each particle, there is a corresponding Voronoi cell, a region bounded by a convex polytope representing the set of all points that are closer to that point (using a Euclidean distance metric) than to any other point.  We created a 3D Voronoi diagram and recorded the Voronoi cell's volume for each particle \citep{SciPy,Qhull,Shapely}.  As a Voronoi cell's volume is closely related to the number of nearby particles, we expect that the volume of a particle's corresponding Voronoi cell will act as an effective measure of local density; in particular, we expect it to effectively classify halo particles.

\subsubsection{Number of Particles Enclosed \EmphMeas{(ENC)}, Center of Mass Distance \EmphMeas{(CMD)}, and Moment of Inertia \EmphMeas{(MI)}} \label{CMDMI}

Using a KD tree, the coordinates for particles within a radius $R_{\textrm{CME}}$ of each particle were found; these values were used as a metric we will refer to as \EmphMeas{ENC}.  We found the center of mass for particles in this region, then used the distance between the center of mass and the particle of interest as a training feature.  Using the same sets of particles, the moment of inertia was calculated.

For small $R_{\textrm{CME}}$, these algorithms measure the local density, while for large $R_{\textrm{CME}}$, they measure the global density.  We expect that the information from \EmphMeas{ENC}, \EmphMeas{CMD}, and \EmphMeas{MI} will be most valuable for halo classification

\subsubsection{Distance to the $k$-Nearest Neighbor \EmphMeas{(KNN)}} \label{KNN}

A ball tree was used to find the distance to the $k$-nearest neighbors for each particle, where the $k$-values used can be found in Table \ref{tab:RosettaStone}.  For small $k$, this algorithms measures the local density, while for large $k$, this algorithm measures the global density.  We expect this algorithm to primarily influence halo classification due to its close similarity to FOF algorithms \citep{FOF}.

\EmphMeas{ENC}, \EmphMeas{CMD}, and \EmphMeas{MI} take into account the properties of all particles within a fixed radius.  As a result, they may fail to account for the spatial extent of the structure a particle is a member of.  On the other hand, \EmphMeas{KNN} measures only the properties of the environment of the closest particles.  By training with very small $k$-values, we can obtain information about not only the density near a particular particle, but also of the natural length scale of the structure that particle is a member of, as the $k^{\textrm{th}}$ nearest neighbor for small $k$-values will likely contain only particles that are a member of that structure.  As a result, we expect \EmphMeas{KNN} measurements to provide information that cannot be obtained with the other density magnitude measurements.

\subsubsection{Principal Component Analysis of Local Particles \EmphMeas{(PCA)}} \label{PCA}

Principal component analysis (PCA) provides a method for determining the principal component axes, an uncorrelated orthogonal basis set such that the first component takes on the highest possible variance.  Using the explained variances, this provided a method for determining the directionality of the data for use in differentiating between filaments and halos.

Prior to performing PCA analysis, the particle field was resampled to ensure an adequate number of particles were contained within each sphere of radius $R_{\textrm{PCA}}$ surrounding a given particle.  First, a Gaussian filter with standard deviation $\sigma_{\textrm{PCA}}$ was applied to the particles within a given sphere to smooth the density distribution.  After binning the coordinates within a given sphere, the density distribution was resampled and particles placed such that the total number density increased by a factor of $\delta n_{\textrm{PCA}}$.  An additional uniform background was added with density $1.0$ Mpc$^{-3}$ $h^3$ to prevent the effects of background particles from being washed out.

After resampling, a PCA decomposition \citep{sklearn,PCA} was performed on all particles within a radius $R_{\textrm{PCA}}$ of each particle, and the explained variance ratio for each axis were found.  The variance of particles within this region may be described by a covariance matrix $c$, with total variance $\sum_{i,j}C_{i,j} = \sigma_0^2$.  After performing the PCA decomposition, $C$ undergoes the transformation $C\,\to\,C'$ such that $C'$ is diagonal and $\tr(C') = \sigma_0^2$.  The explained variance of principal component axis $i$ is $C'_{ii}$, and the explained variance ratio is $\overline{\sigma}_i = \frac{C'_{ii}}{\sigma_0^2}$.  After PCA decomposition is performed, a data set that may initially be correlated is transformed to a data set that exhibits no cross correlation.  The explained variance ratio describes the proportion of the total variance $\sigma_0^2$ that may be attributed to the variance of particles with respect to a given axis.  Intuitively, as the principal component axes correspond to the principal axes when calculating the moment of inertia, this provides information about the spread of particles about an axis such that the mass distribution around that axis is uniform.

After calculating the explained variance ratio $\overline{\sigma}_i$ for each axis for each particle, the directionality value $V$ (the value used for the \EmphMeas{PCA} training metric) was found, where

\begin{align} \label{eq:PCA}
    &V = -\ln\left(\Delta\sigma^2\right), \\
    &\Delta\overline{\sigma}^2 = \overline{\sigma}^2_{\textrm{max}} -\overline{\sigma}^2_{\textrm{min}}. \nonumber
\end{align}

$\Delta\sigma^2$ describes the difference between the minimum and maximum explained variance ratio.  The natural logarithm of this difference was taken to accentuate the differences between the filaments and halos so that, by using $V$ as a training metric, fewer data points would be required to perform accurate classification.

For a filament, it would be expected that the variance about the spine axis would be much smaller than the variance around the other axes due to the density field being preferentially aligned along this axis, producing a small value for $V_{\textrm{F}}$.  In contrast, as the halo density field tends to exhibit very little directionality preference, it would be expected that the explained variances should vary little between the different axes, meaning that $V_{\textrm{H}}$ would be large.  It may be expected that the explained variance ratios for void particles would exhibit similar properties to those of halos, and hence have a large $V$; however, due to the low density of voids, nearby structures would heavily influence the directionality values of void particles.  As a result, it is expected that void particles should exhibit a small value for $V_{\textrm{V}}$, though a larger spread that of filaments.

\subsection{Training and Class Assignment} \label{TrainPred}

A random forest algorithm \citep{RandomForest} is a supervised learning method constructed from several decision trees.  Each tree classifies a given particle using a randomized subset of features, and the class assigned to that particle is the class selected by the plurality of trees.  As we aim to simplify the classifier as much as possible by minimizing the number of features required for classification, a random forest algorithm ensures that our classification is affected less by statistical fluctuations resulting from a small number of features.  In addition, we may assign each particle a class probability based on the number of independent trees that assigned that particle a given class.

A random forest algorithm \citep{sklearn,RandomForest} was trained using the measurements from one simulation.  While it is typical to use multiple datasets to train a classifier, we found that additional training datasets did not influence the results substantially (likely due to the statistical robustness provided by a random forest algorithm), so only one was used.  Using 200 trees, class probabilities were generated for each particle in \texttt{SIM}.  Each particle was then assigned the class selected by the plurality of trees; in cases where multiple classes were assigned a plurality, halos were prioritized over filaments, which were prioritized over voids.  An FOF clustering algorithm \citep{FOF} was applied to halo particles to differentiate halos from one another, which were used to create an HMF.

\section{Classification} \label{Prediction}

\begin{figure}
    \centering\captionsetup[subfloat]{labelfont=bf}
    \begin{tabularx}{\columnwidth}{C}
        \textbf{Training Dataset} \\
        \hline
        \subfloat[]{
            \includegraphics[width = 0.95\columnwidth]{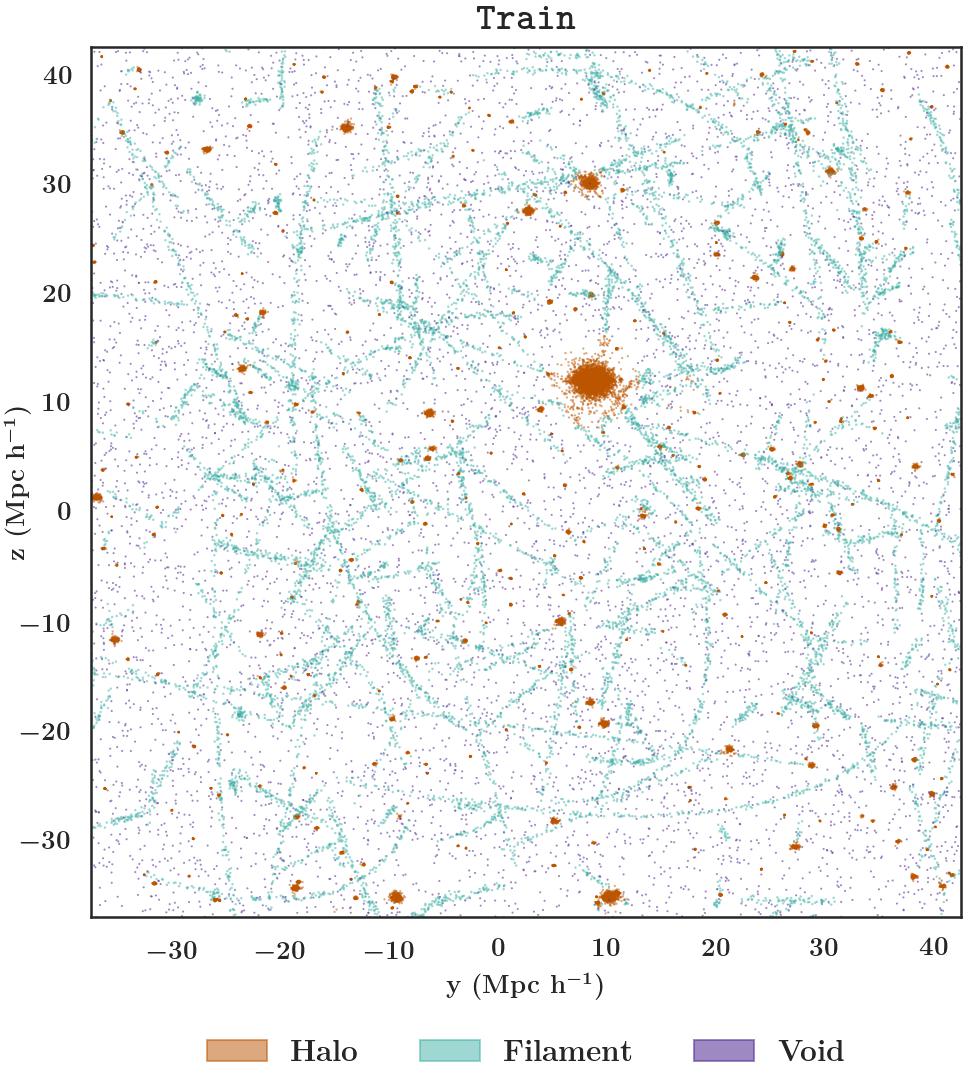}
        } \\
        \subfloat[]{
            \includegraphics[width = 0.95\columnwidth]{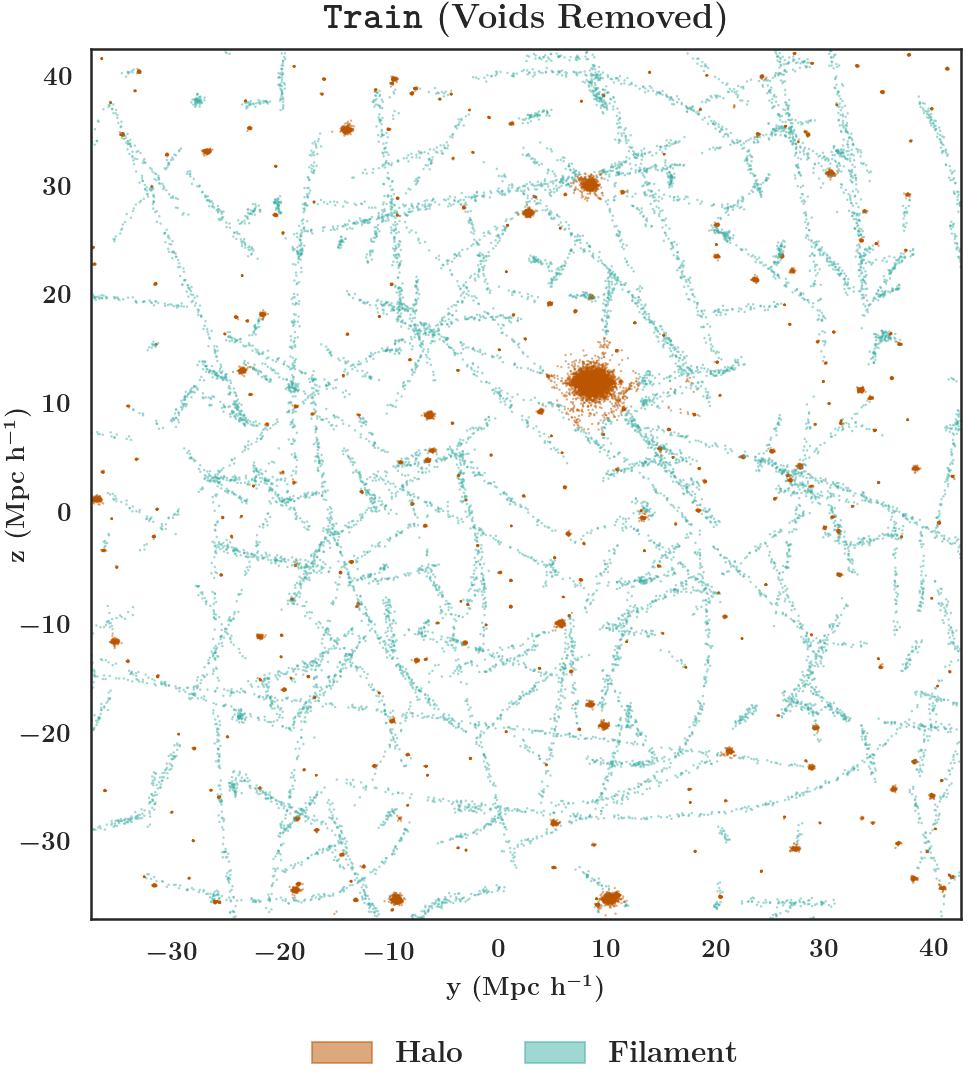}
        }
    \end{tabularx}
    \caption{\textbf{(a)} A 6 Mpc thick slice of the 3D training dataset used throughout this work.  This figure includes void particles (blue).  \textbf{(b)} The same training data set, but with void particles removed.  Removing the void particles provides a better view of the filaments.  Both of these figures include the largest halo.}
    \label{fig:Train}
\end{figure}

It is expected that the primary way to differentiate between structures would be through local density magnitude calculations; however, to account for the arbitrary densities used for filaments, density field directionality measurements were included to differentiate between filaments and halos/voids.  Of the methods used, only \EmphMeas{PCA} provided directionality measurements.

\subsection{Toy-to-\texttt{SIM}} \label{TS}

First, classes were assigned to particles in \texttt{SIM} using a 660,000-particle toy model.  A 6 Mpc thick slice of the toy model is shown in Figure \ref{fig:Train}.

\subsubsection{Measurement Histograms} \label{MeasHist}

\begin{figure*}
    \centering\captionsetup[subfloat]{labelfont=bf}
    \begin{minipage}{1.0\textwidth}
        \centering
        \begin{tabularx}{1.0\textwidth}{CC}
            \multicolumn{2}{c}{\bf{Measurement Histograms from Training Data (Toy Model)}} \\
            \hline
            \subfloat[]{
                \includegraphics[width = 0.5\textwidth,keepaspectratio]{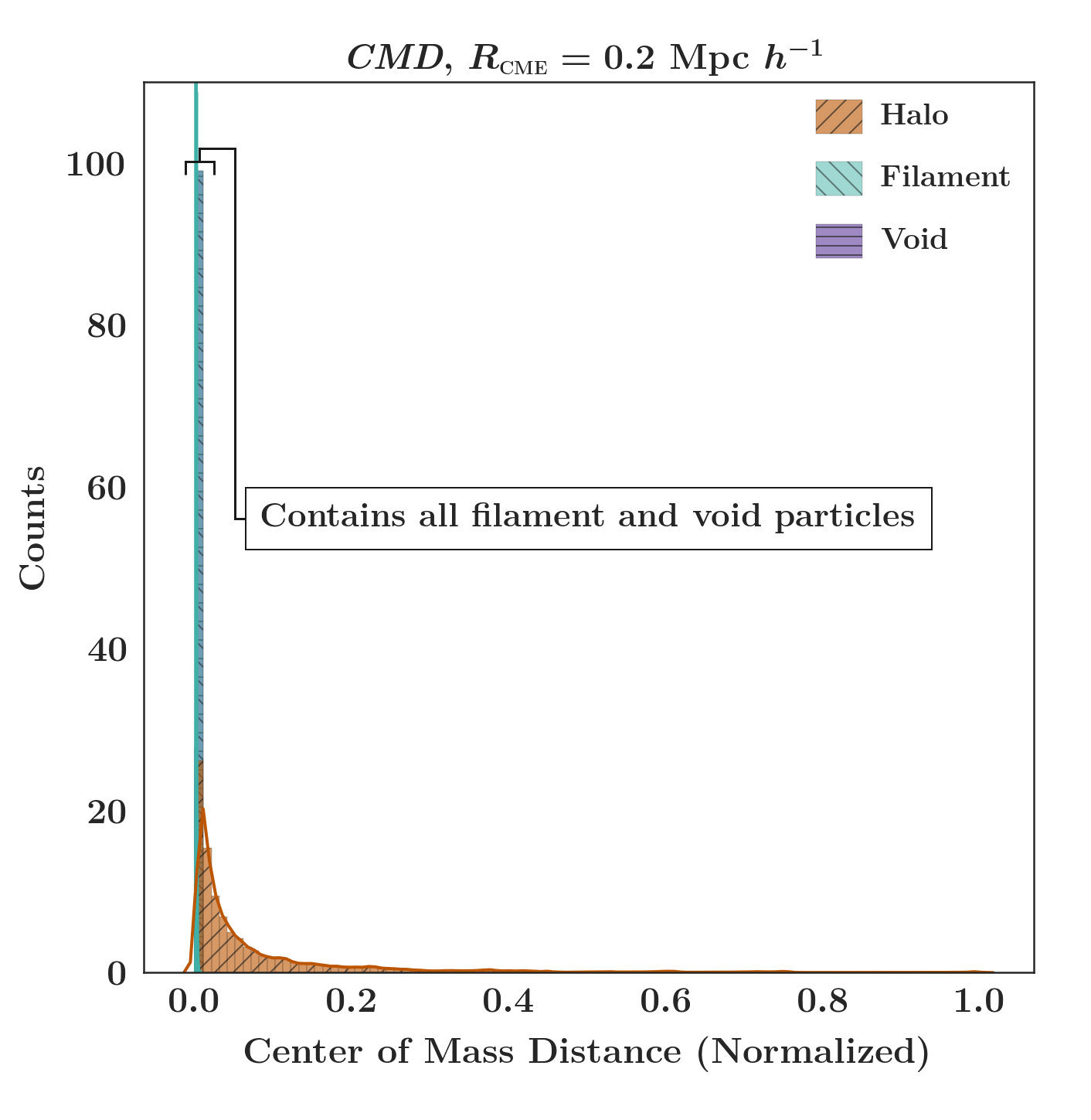}\label{fig:MeasHista}
            } &
            \subfloat[]{
                \includegraphics[width = 0.5\textwidth,keepaspectratio]{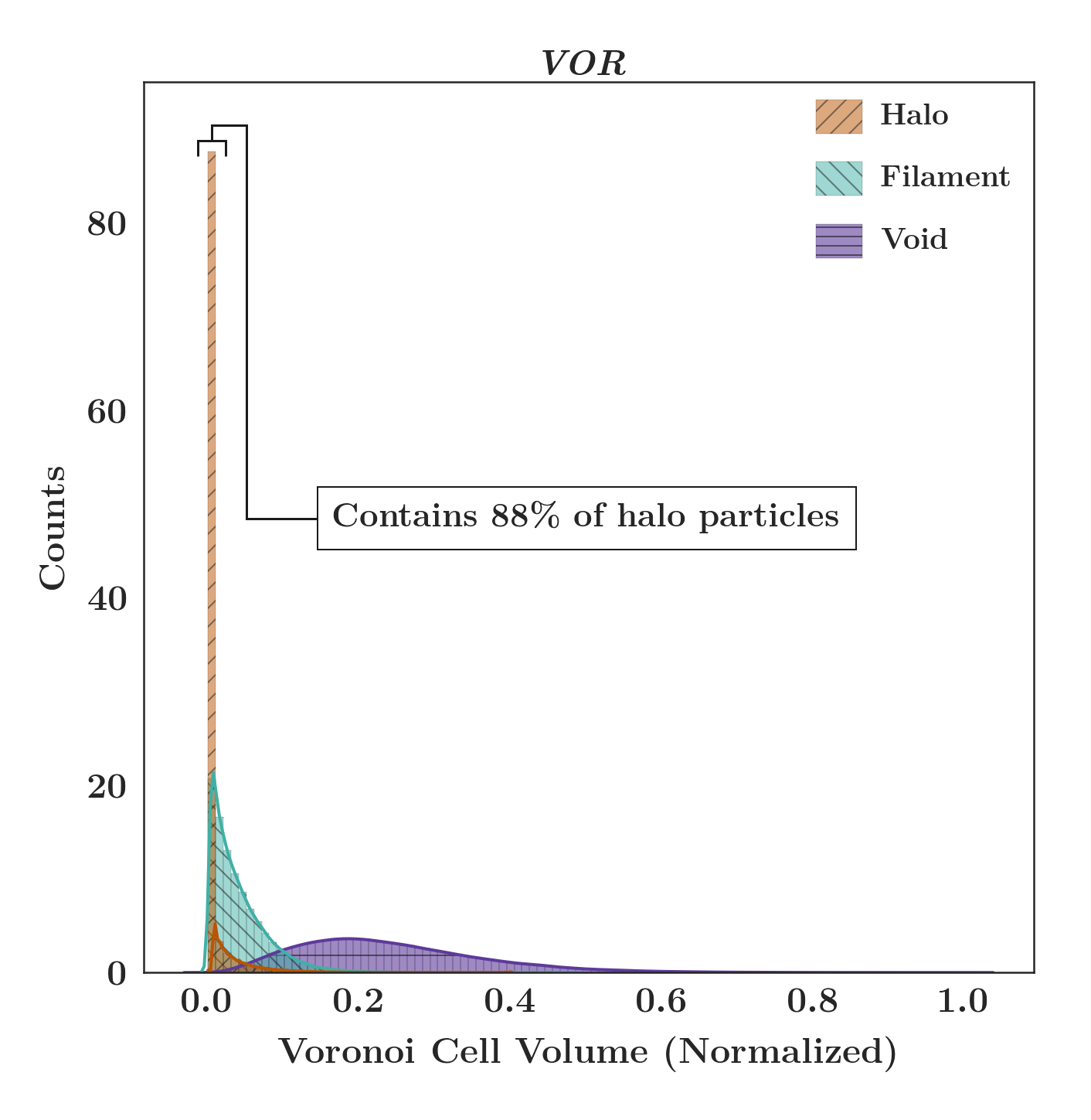}\label{fig:MeasHistb}
            } \\
            \subfloat[]{
                \includegraphics[width =  0.5\textwidth,keepaspectratio]{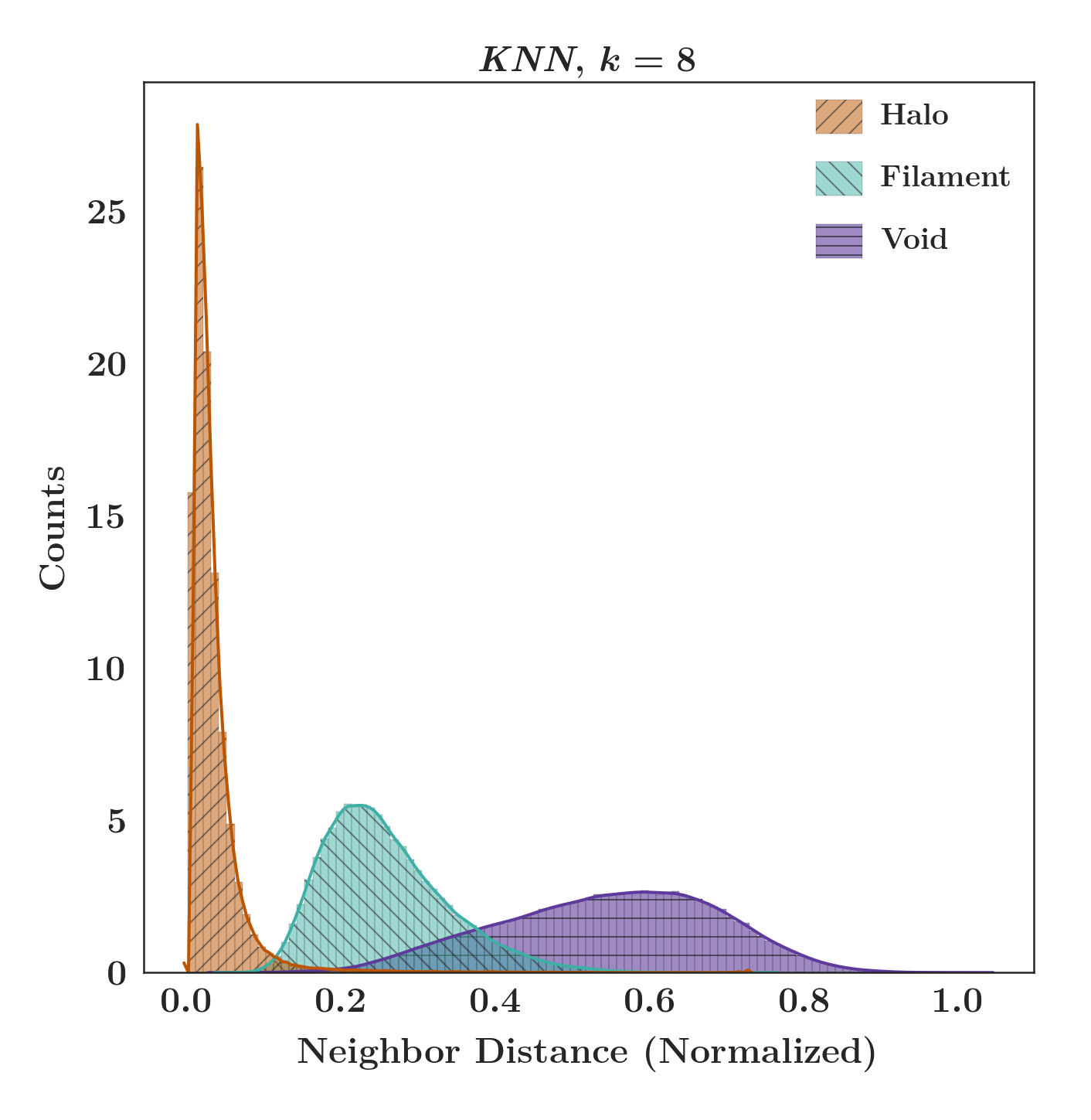}\label{fig:MeasHistc}
            } &
            \subfloat[]{
                \includegraphics[width = 0.5\textwidth,keepaspectratio]{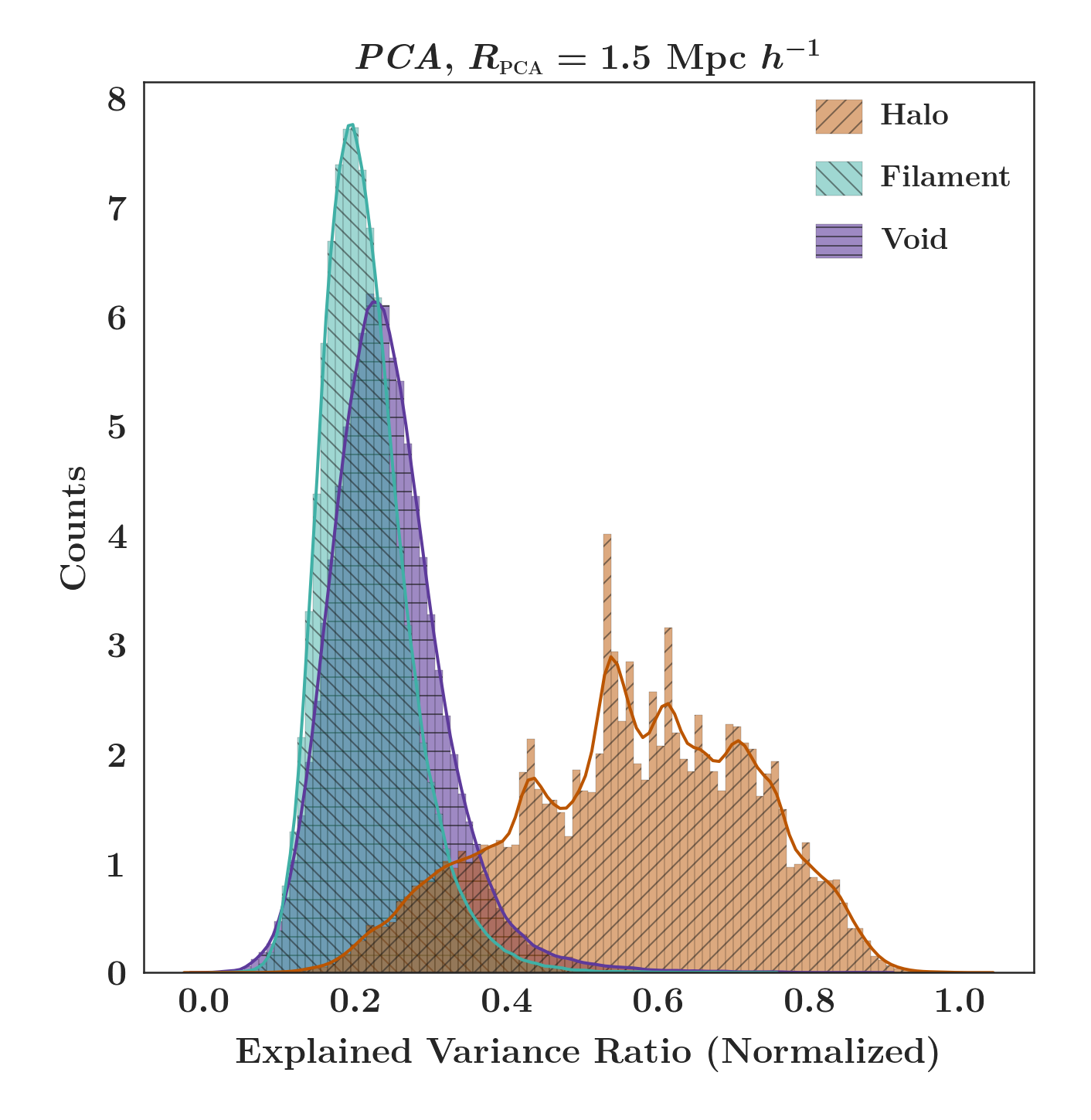}\label{fig:MeasHistd}
            }
        \end{tabularx}
    \end{minipage}
    \caption{Measurement histograms from the true values in the toy model for \textbf{(a)} \EmphMeas{CMD}, \textbf{(b)} \EmphMeas{VOR}, \textbf{(c)} \EmphMeas{KNN}, and \textbf{(d)} \EmphMeas{PCA}.  \textbf{(a)} and \textbf{(b)} show features with less distinguishable distributions, while \textbf{(c)} and \textbf{(d)} are more distinguishable.}
    \label{fig:MeasHist}
\end{figure*}

To provide an initial guess as to which density magnitude methods would provide the most information, histograms were created using measurements of the training dataset.  After performing all measurements on the training data set, each measurement was normalized so that all values fell between 0 and 1, ensuring that all measurement methods would be treated equally when training.  Measurements that provide the most information for use in classification should show little overlap between the measurements on each structure and exhibit large peaks distinct from one another.  Examples of some measurement histograms are displayed in Figure \ref{fig:MeasHist}.

Figure \ref{fig:MeasHist} provides clues as to which measurements will be most effective for classification.  The histograms for \ref{fig:MeasHista} \EmphMeas{CMD} and \ref{fig:MeasHistb} \EmphMeas{VOR} both exhibit a large spike at the origin, indicating that, for each class, the distance from nearly every particle to the center of mass was very small, save for several very large outliers.  Due to the lack of differentiation, it appears that \EmphMeas{CMD} will provide little information that may be used to differentiate between LSS classes.  Similarly, we expect \EmphMeas{VOR} to be ineffective at differentiating halos from filaments.  Though not shown, the measurement histograms for \EmphMeas{MI} and \EmphMeas{ENC} appear similar to those of \EmphMeas{CMD}.

However, the measurements for \ref{fig:MeasHistc} \EmphMeas{KNN} exhibited much more differentiation between each of the different structure classes.  As expected based on local density, halos exhibited the lowest distance to the 8${^{\textrm{th}}}$-nearest neighbor, followed by filaments, and finally voids.  The larger spread on filaments and voids reflects a large chance of contamination by nearby structures due to their low density.  As the classes are strongly differentiated from one another, we expect \EmphMeas{KNN} to be an effective proxy for local density, and, hence, an effective metric for differentiating between all structures.

While not shown, larger $k$-values led to less differentiation between the classes due to each class exhibiting a greater spread.

The measurement histograms for \ref{fig:MeasHistd} \EmphMeas{PCA} show strong peaks for filaments and voids, yet a multi-peaked halo distribution with large spread.  The strong peak for filaments, especially compared to the poorly-defined halo distribution, indicates that these calculations are effectively measuring the density field directionality, as the roughly spherical halos are not expected to exhibit substantial directionality.  The strong peak for voids is likely due to contamination by nearby structures: as voids have very low density, any particles from adjacent halos or filaments would lead to a strong directionality.  These results bode well for the use of \EmphMeas{PCA} in tandem with \EmphMeas{KNN} to create a robust classifier, as \EmphMeas{PCA} provides a natural way to differentiate between halos and filaments independent of the local density magnitude.  While it may seem that these calculations may cause difficulty when distinguishing between filaments and voids, the strong differentiation between these structures from \EmphMeas{KNN} calculations is expected to prevent this issue.

\subsubsection{\texttt{SIM} Class Assignment} \label{TSPred}

\begin{figure}
    \centering\captionsetup[subfloat]{labelfont=bf}
    \begin{tabularx}{\columnwidth}{C}
        \bf{Toy-to-\texttt{SIM} Probability Contrasts for \EmphMeas{CMD} and \EmphMeas{VOR}} \\
        \hline
        \subfloat[]{
            \includegraphics[width = \columnwidth]{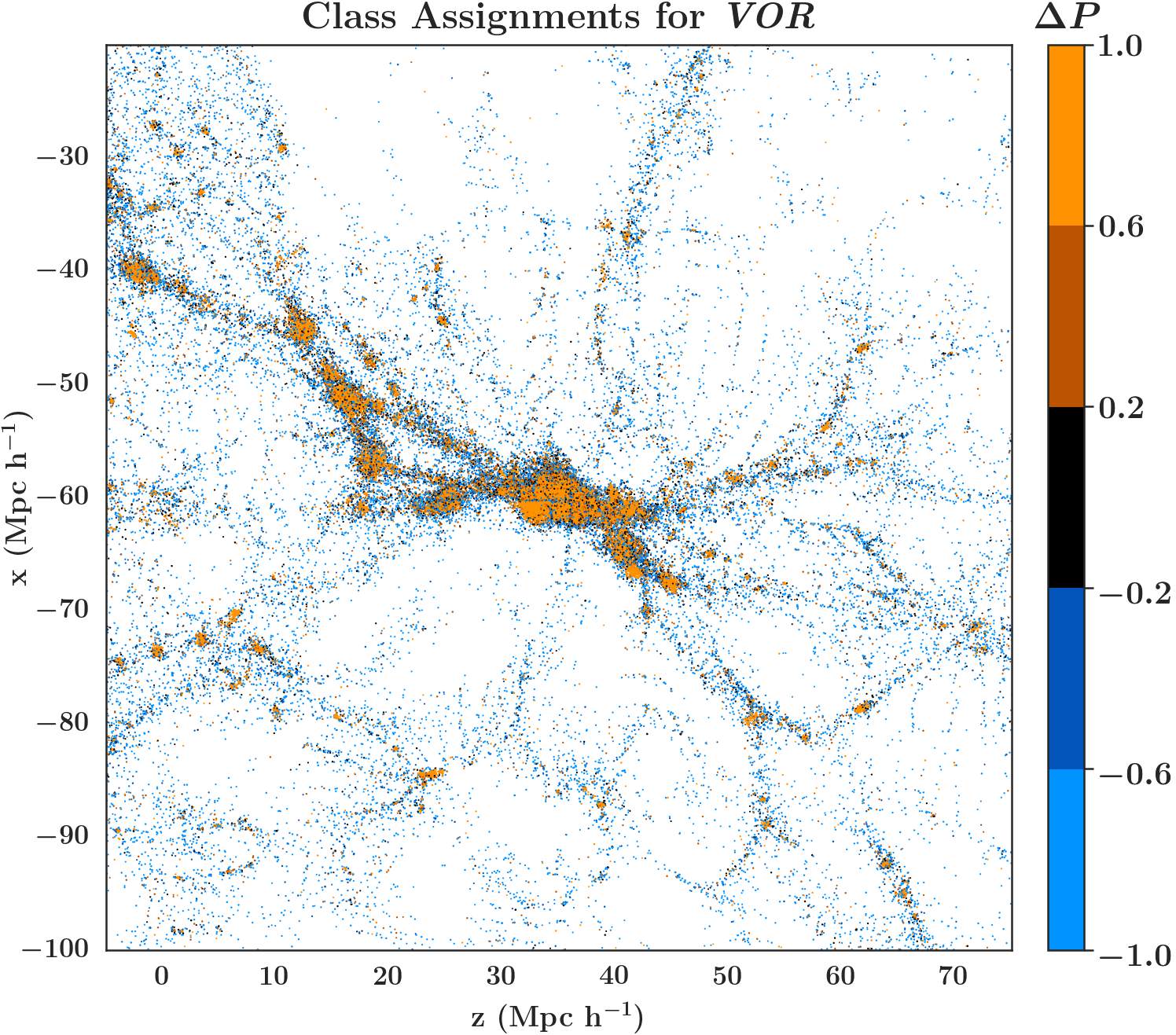}\label{fig:CRadCompa}
        } \\
        \subfloat[]{
            \includegraphics[width = \columnwidth]{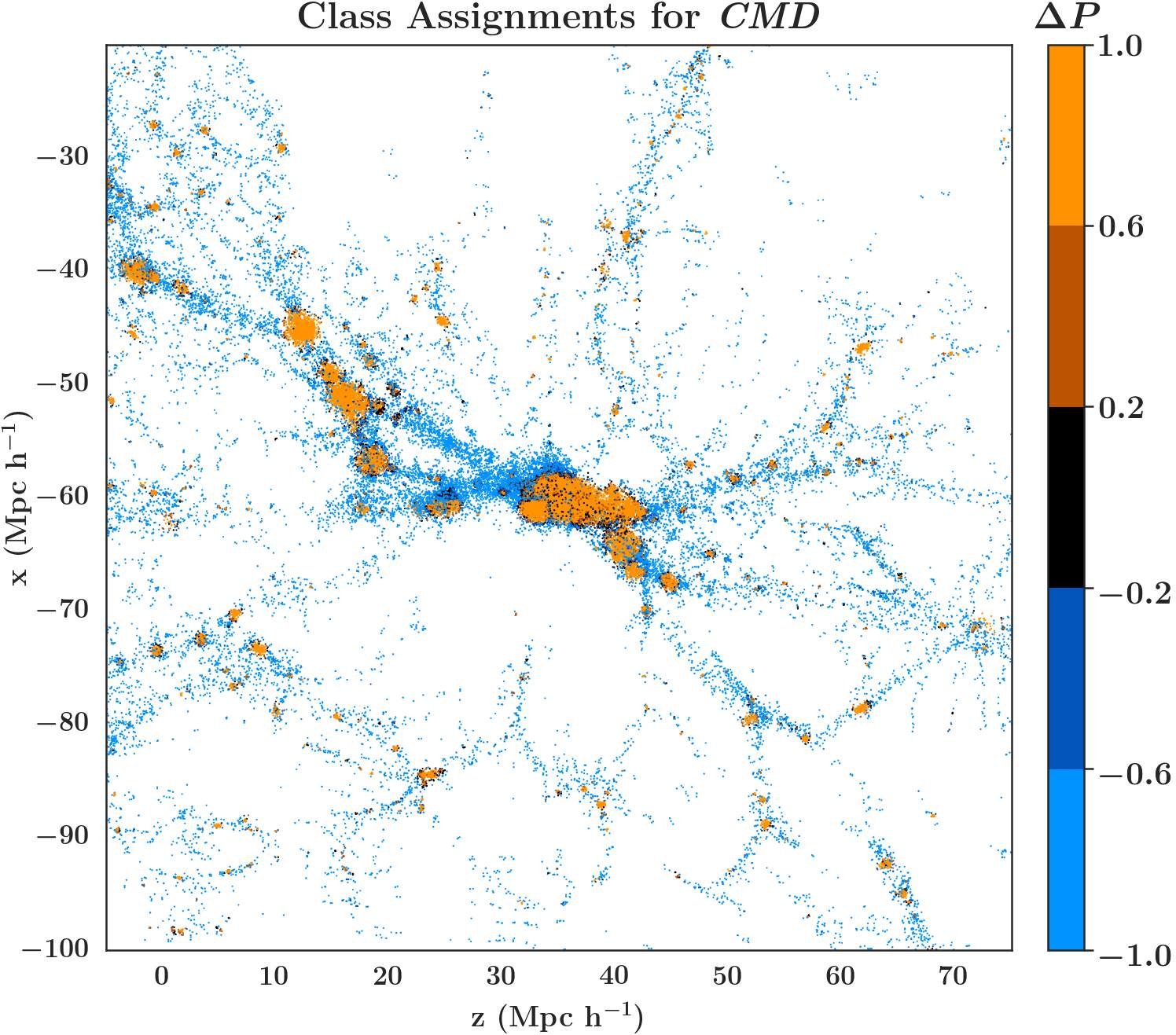}\label{fig:CRadCompb}
        }
    \end{tabularx}
    \caption{Classification results for \texttt{SIM}; for comparison with \texttt{TSIM}, the full particle field, which spans a box with corners at $(x,\,y,\,z) = (\pm128,\,\pm128,\,\pm128)$ Mpc $h^{-1}$ has been restricted to a region centered on the largest halo with side lengths $80$ Mpc $h^{-1}$ and a depth of $7$ Mpc $h^{-1}$.  When trained using all features, this halo has $N = 38798$ particles and is centered at $(x,\,y,\,z) = (-60.6,\,88.4,\,35.0)$.  \textbf{(a)} and \textbf{(b)} show the halo and filament class assignments made using \EmphMeas{VOR} and \EmphMeas{CMD} measurements, respectively, colored based on each particle's probability contrast.}
    \label{fig:CRadComp}
\end{figure}

\begin{figure*}
    \centering\captionsetup[subfloat]{labelfont=bf}
    \begin{minipage}{\textwidth}
        \centering
        \begin{tabularx}{\textwidth}{CC}
        \multicolumn{2}{c}{\bf{Toy-to-\texttt{SIM} Probability Contrasts for \EmphMeas{KNN} (+ \EmphMeas{PCA})}} \\
        \hline
        \subfloat[]{
            \includegraphics[width = 0.5\textwidth]{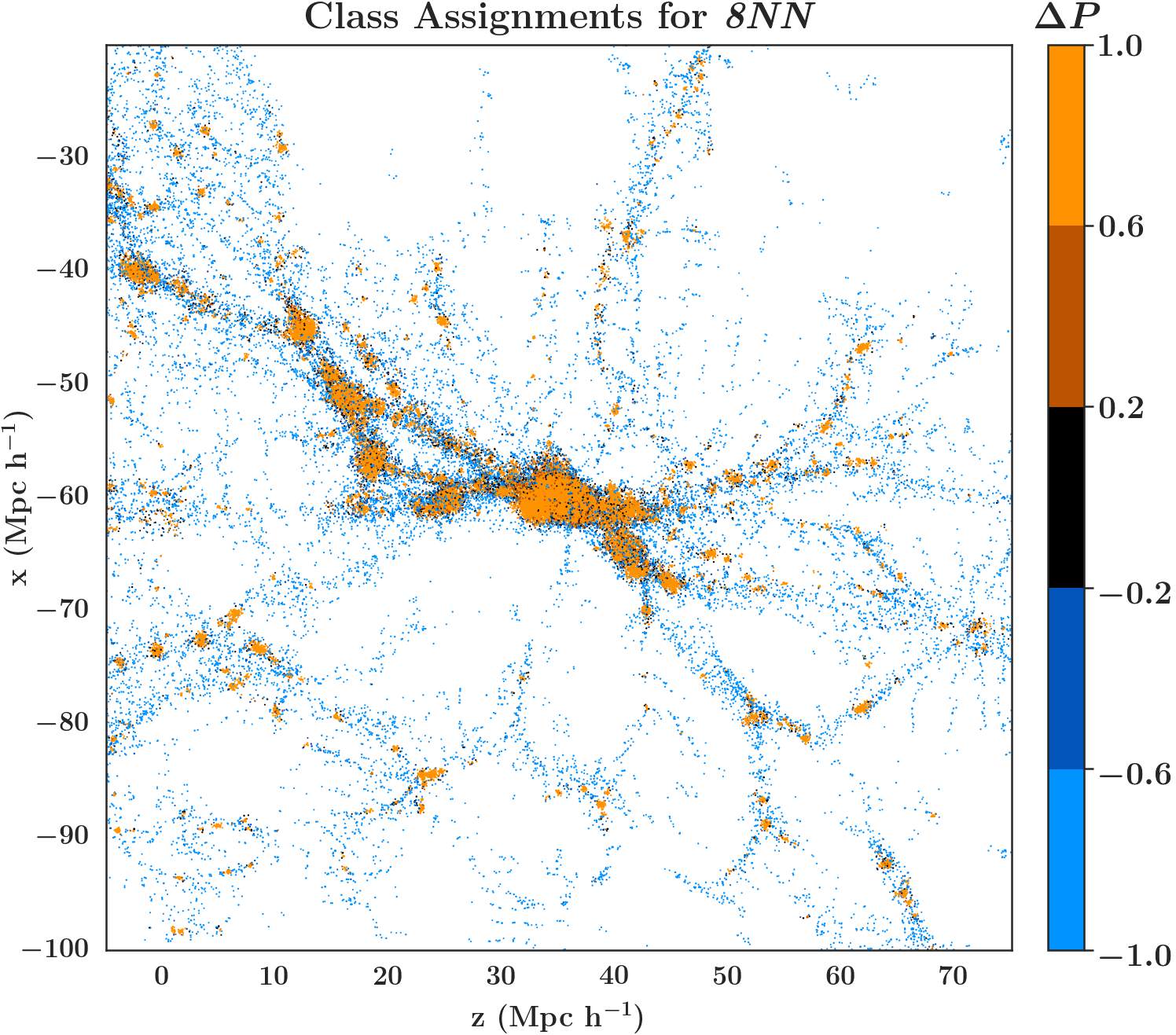}\label{fig:KNNCompa}
        } &
        \subfloat[]{
            \includegraphics[width = 0.5\textwidth]{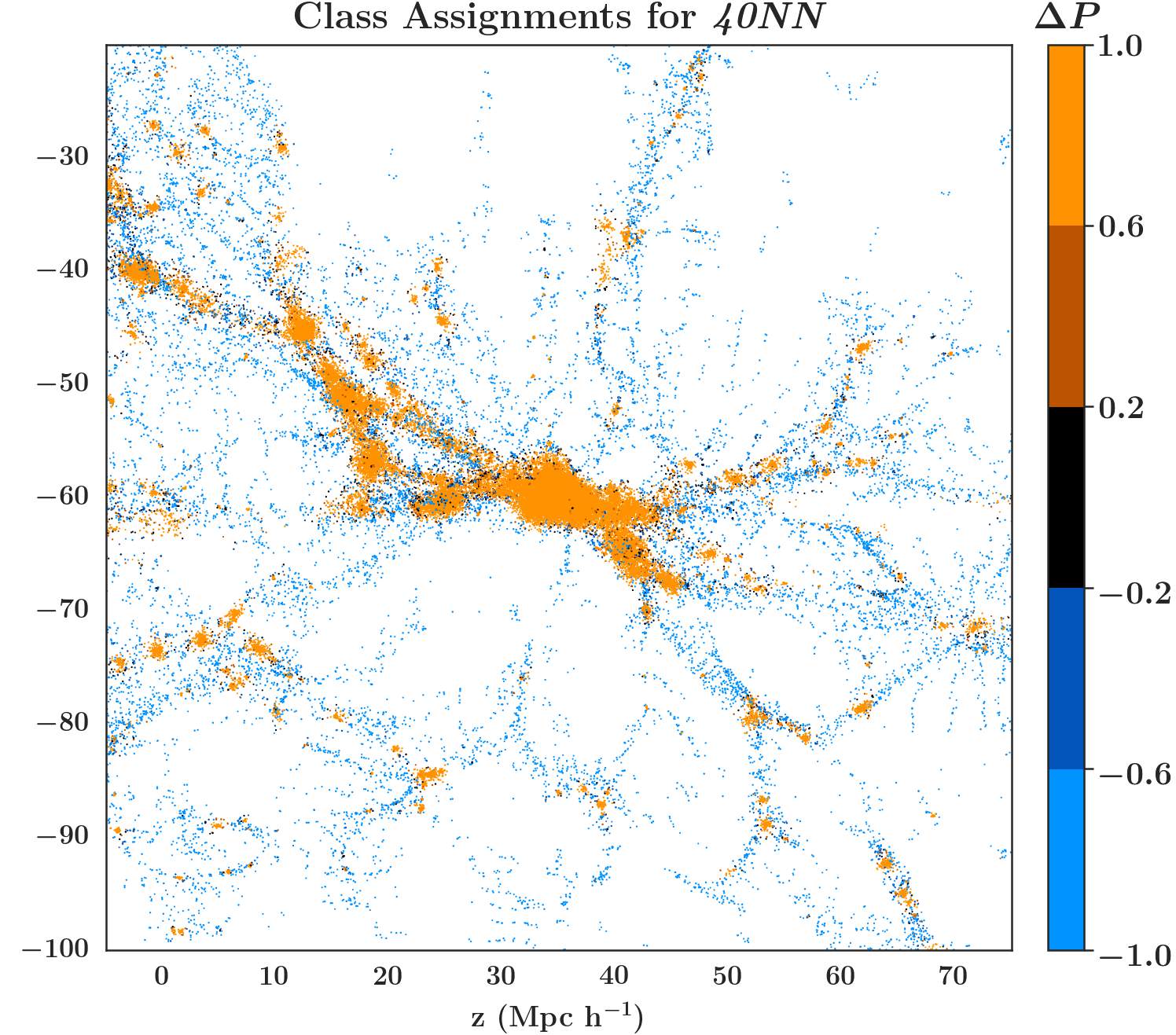}\label{fig:KNNCompb}
        } \\
        \subfloat[]{
            \includegraphics[width = 0.5\textwidth]{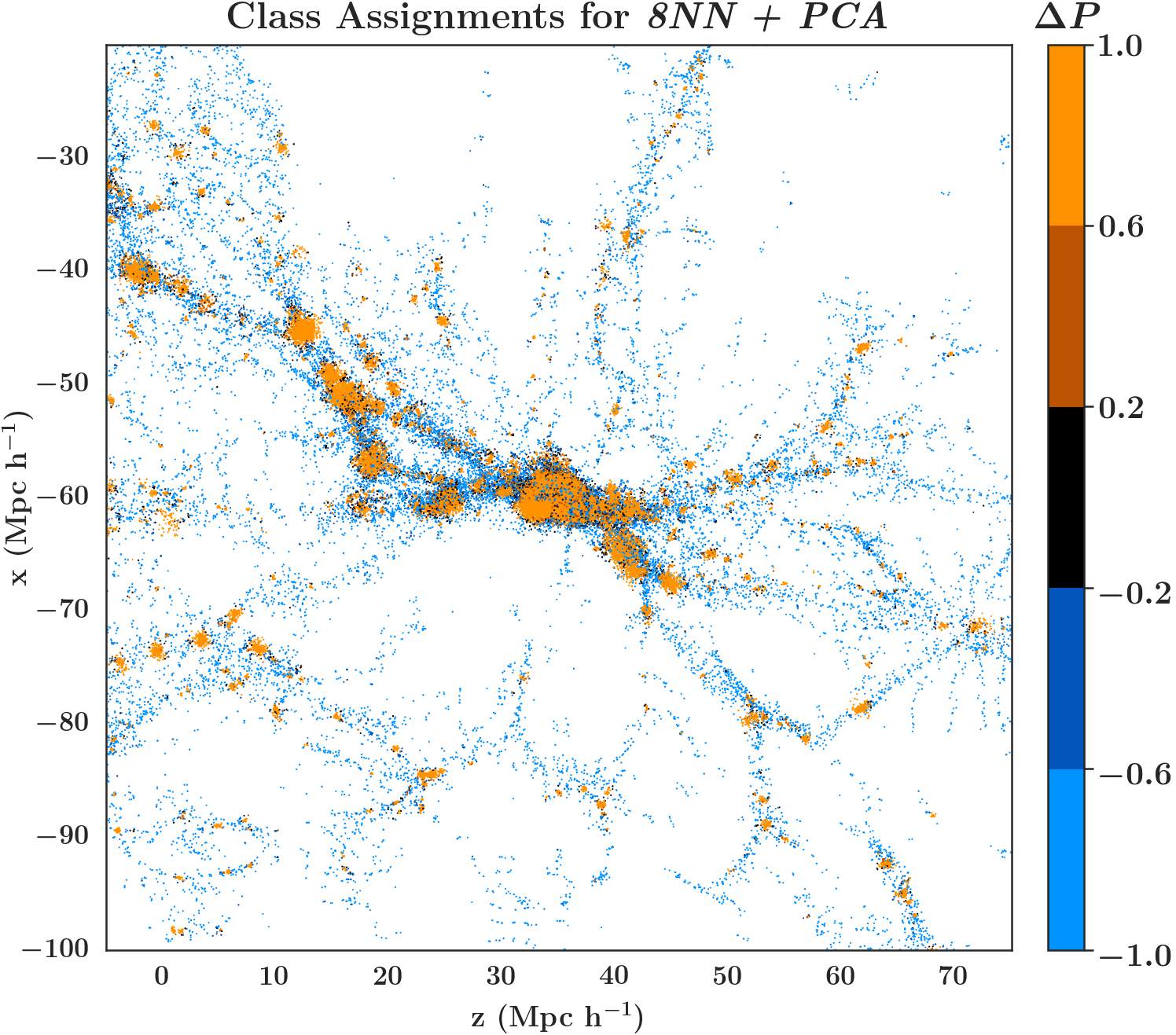}\label{fig:KNNCompc}
        } &
        \subfloat[]{
            \includegraphics[width = 0.5\textwidth]{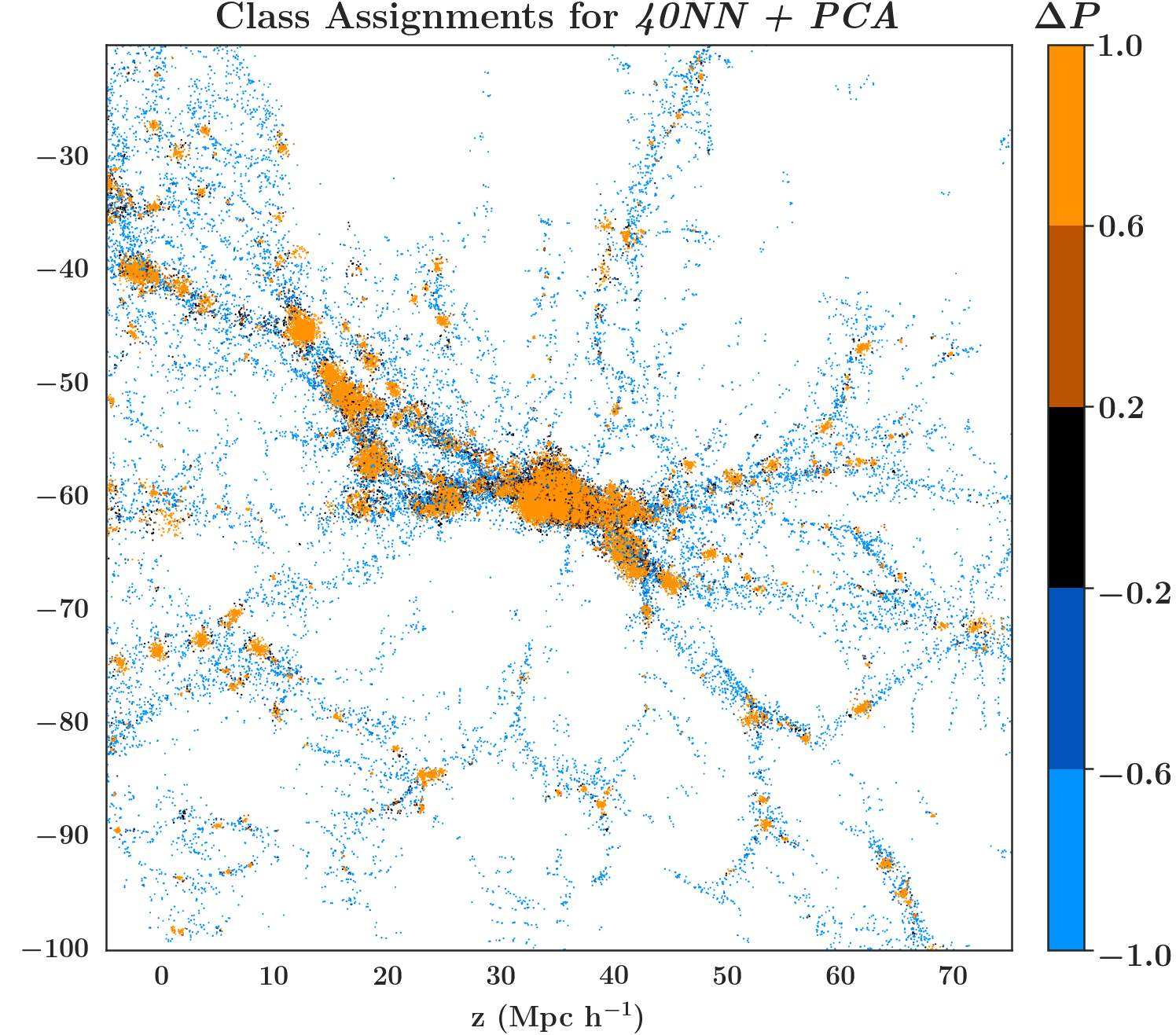}\label{fig:KNNCompd}
        }
        \end{tabularx}
    \end{minipage}
    \caption{Class assignment results for \texttt{SIM}.  The halo-filament particles, colored based on their probability contrast (see Eqn. \eqref{eq:ProbCont}) are shown in \textbf{(a)} (\EmphMeas{KNN}, $k \leq 8$), \textbf{(b)} (\EmphMeas{KNN}, $k \leq 40$), \textbf{(c)} (\EmphMeas{KNN} + \EmphMeas{PCA}, $k \leq 8$), and \textbf{(d)} (\EmphMeas{KNN} + \EmphMeas{PCA}, $k = 40)$.  The displayed particles are from the same 7 Mpc thick slice of the 3D N-body simulation \texttt{SIM} centered on the halo with the greatest mass as in Figure \ref{fig:CRadComp}.}
    \label{fig:KNNComp}
\end{figure*}

\begin{figure}
    \centering\captionsetup[subfloat]{labelfont=bf}
    \begin{tabularx}{\columnwidth}{C}
        \bf{Halo Mass Functions for \texttt{SIM}} \\
        \hline
        \subfloat[]{
            \includegraphics[width = \columnwidth]{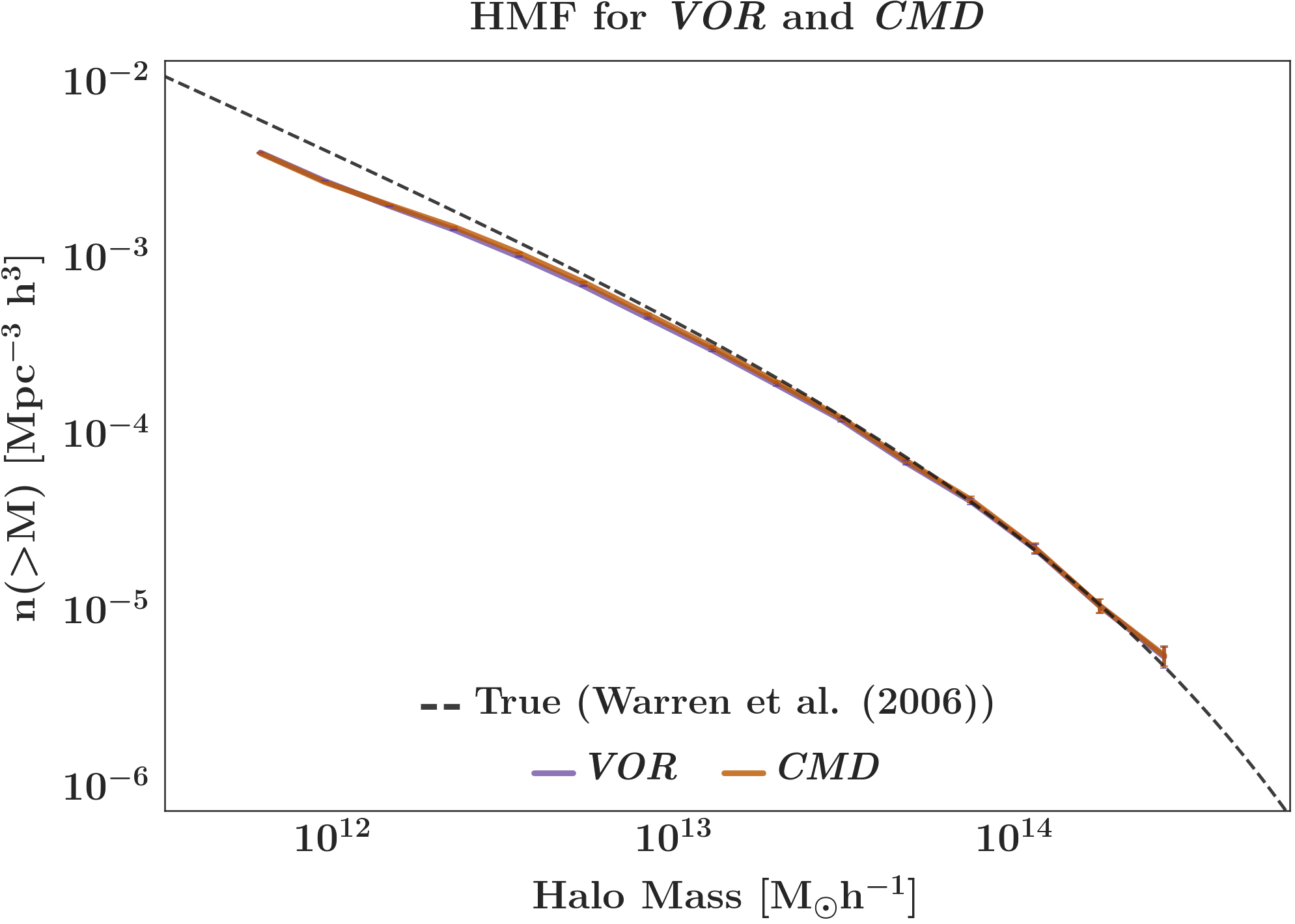}
        } \\
        \subfloat[]{
            \includegraphics[width = \columnwidth]{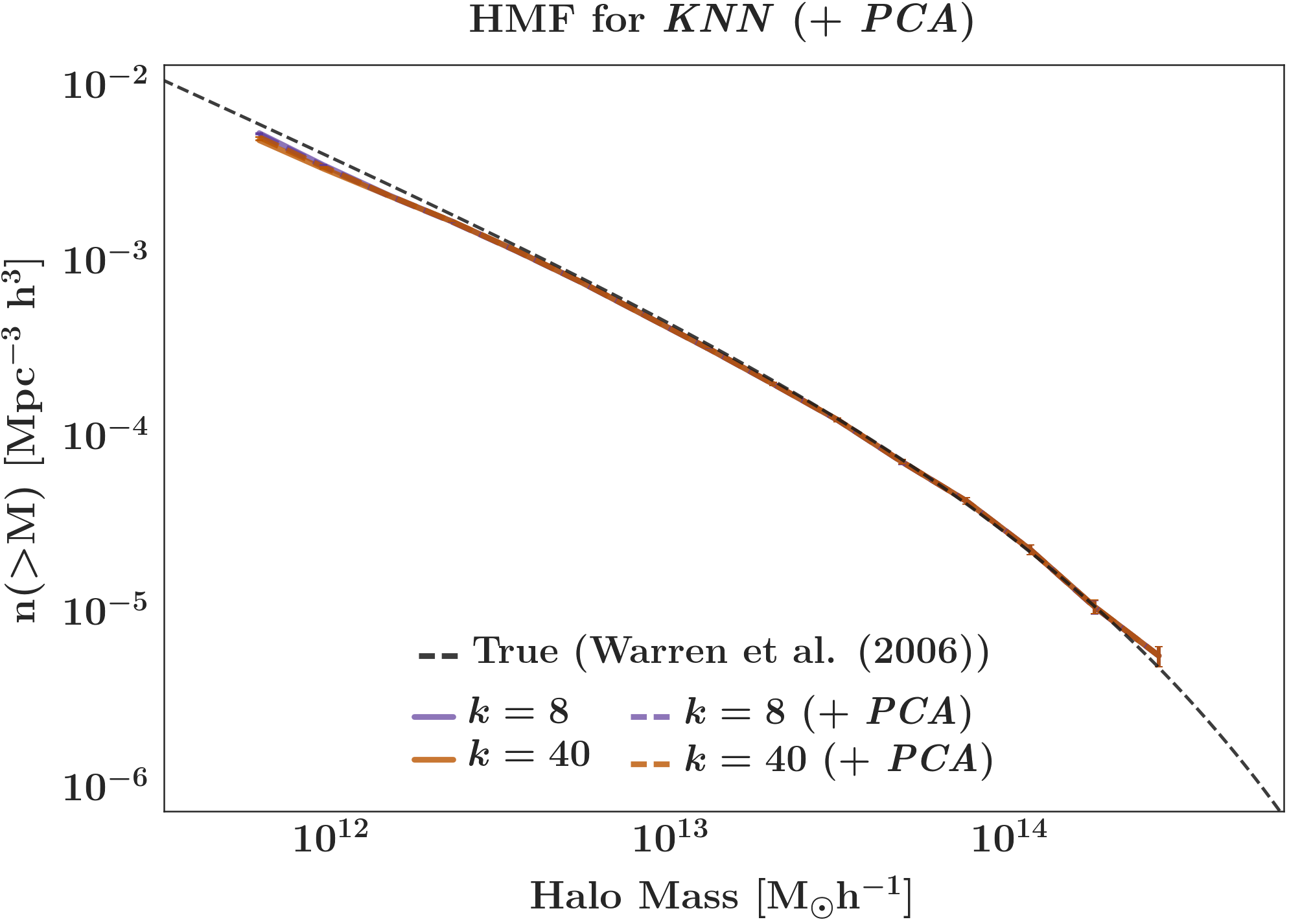}
        }
        \end{tabularx}
    \caption{Halo mass functions for \texttt{SIM} created using classifiers trained using \textbf{(a)} \EmphMeas{VOR} and \EmphMeas{CMD} and \textbf{(b)} \EmphMeas{KNN} (+ \EmphMeas{PCA}) for $k \leq 8,\,40$.  Note that not all lines are easily visible; this is because the HMFs in each figure were deviated little from one another.  The poor correspondence between the \citet{WarrenHMF} HMF and that of \EmphMeas{VOR} and \EmphMeas{CMD}, particularly at small $M$, indicates that \EmphMeas{VOR} and \EmphMeas{CMD} are ineffective proxies for local density magnitude.  In contrast, the HMFs \textbf{(b)} deviated little from the \citet{WarrenHMF} HMF.  This may be due to the similarities between \EmphMeas{KNN} measurements and FOF algorithms.}
    \label{fig:HMF}
\end{figure}

Next, class labels were assigned to particles in \texttt{SIM} (described in Section \ref{Methods}).  In order to achieve probabilistic classification, each particle was classified using 200 trees, each of which independently assigned that particle a class.  The classification probability for a given class was the fraction of estimators that assigned a particular particle that class over all estimators.  To compare the class assignments for particles labeled as halos or filaments, define the probability contrast $\overline{\Delta P}_i$ for a halo or filament particle as

\begin{align} \label{eq:ProbCont}
    \overline{\Delta P}_i = \frac{P_i(H) - P_i(F)}{P_i(H) + P_i(F)},
\end{align}

where $P_i(H)$ ($P_i(F)$) is the probability that particle $i$ is a halo (filament).

Plots of the halo and filament particle distribution, colored by $\overline{\Delta P}$, can be found in Figures \ref{fig:CRadComp} and \ref{fig:KNNComp}.  A probability contrast of 1 indicates that every estimator classified that particle as a member of a halo, while a probability contrast of -1 indicates that every estimator classified that particle as a member of a filament.  A particle colored black indicates that the estimator was unable to precisely differentiate between the particle's class, indicating that there is ambiguity as to whether it is a halo or filament member.  It would be expected that, near the high-density center of halos, the probability contrast would be close to 1, while near the edges, especially where the halo connected to a filament, the probability contrast would be closer to 0.  Note that void particles are not displayed in these plots as void probability assignments were generally close to unity.

All particles in \texttt{SIM} were classified by each metric set as detailed below.  Figures \ref{fig:CRadComp} and \ref{fig:KNNComp} show the halo and filament class probability contrasts assigned to particles within a 7 Mpc thick slice of this 3D simulation.  While the classes assigned by each feature set shown are visually realistic, we discuss their differences below.

Figures \ref{fig:CRadCompa} and \ref{fig:CRadCompb} show the class probabilities assigned by a classifier trained using only \EmphMeas{VOR} and \EmphMeas{CMD}, respectively.  Relative to the assignments by \EmphMeas{CMD}, \EmphMeas{VOR} overestimated the number of halo and filament particles, indicating that it was not sensitive to the low density void regions.  In addition, the class probabilities assigned were generally lower, indicating that \EmphMeas{VOR} alone did not provide enough information to distinguish between the classes easily.  On the other hand, \EmphMeas{CMD} generally assigned very high class probabilities to each particle except for several small, concentrated regions on the border between a halo and filament (appearing as black clumps in Figure \ref{fig:CRadComp}).  These regions had a density magnitude between that of the high-probability halo regions ($P \approx 1$) and the high-probability filament regions ($P \approx -1$).  The small width of these regions implies that using only \EmphMeas{CMD} calculations imposed a strict density field magnitude cutoff when determining halo and filament membership.

While not shown, the class probability contrast plots for \EmphMeas{MI} and \EmphMeas{ENC} were both very similar to \EmphMeas{CMD}.

Figures \ref{fig:KNNCompa} and \ref{fig:KNNCompb} show the class probabilities assigned by a classifier trained using \EmphMeas{KNN} for $k \leq 8,\,40$, respectively.  The assignments made by these classifiers are generally similar; however, classification using a classifier $k \leq 40$ generally produced larger halos, and particles assigned to those halos had a larger $P(H)$.  In addition, both classifiers, particularly the classifier with $k \leq 40$, produced elongated halos, demonstrating difficulty in distinguishing between dense filaments and low-density halos.

The inclusion of \EmphMeas{PCA} calculations (Figure \ref{fig:KNNCompc} and \ref{fig:KNNCompd}) helped eliminate this issue by providing information emphasizing the directionality component of filaments.  Including larger $k$ values for \EmphMeas{KNN} calculations led to many filament particles being classified as halo particles, likely due to the fact that measurements using large $k$ would often include information from a variety of structure classes, blurring their distinction.  However, even though the classifier in Figure \ref{fig:KNNCompd} used $k$-values much larger than those in \ref{fig:KNNCompc}, the class probabilities are very similar, especially when compared to Figures \ref{fig:KNNCompa} and \ref{fig:KNNCompb}.  By including information about the local density field directionality, class assignments were less affected by contamination from distant structures.

Figure \ref{fig:HMF} shows the HMFs for the halos identified in \texttt{SIM}.  For all classifiers, it is clear that classification was least accurate for small halos; elsewhere, the HMFs for \EmphMeas{KNN} (+ \EmphMeas{PCA}) corresponded very closely with \citet{WarrenHMF}; however, the HMFs for \EmphMeas{VOR} and \EmphMeas{CMD} were not as accurate.  This provides evidence that \EmphMeas{KNN} (+ \EmphMeas{PCA}) provides important information for differentiating halos from filaments and voids, further supporting the conclusions drawn from Figure \ref{fig:MeasHist}.  Though not shown, the HMFs for \EmphMeas{MI} and \EmphMeas{ENC} were nearly identical to those of \EmphMeas{VOR} and \EmphMeas{CMD}.  The inaccuracy for small halos had the same source as for FOF calculations \citep{FOFProblems}.

\subsection{Toy-to-\texttt{TSIM}} \label{TT}

\begin{figure}
    \centering\captionsetup[subfloat]{labelfont=bf}
    \begin{tabularx}{\columnwidth}{C}
        \bf{Toy-to-\texttt{TSIM} Probability Contrasts for \EmphMeas{CMD} and \EmphMeas{VOR}} \\
        \hline
        \subfloat[]{
            \includegraphics[width = \columnwidth]{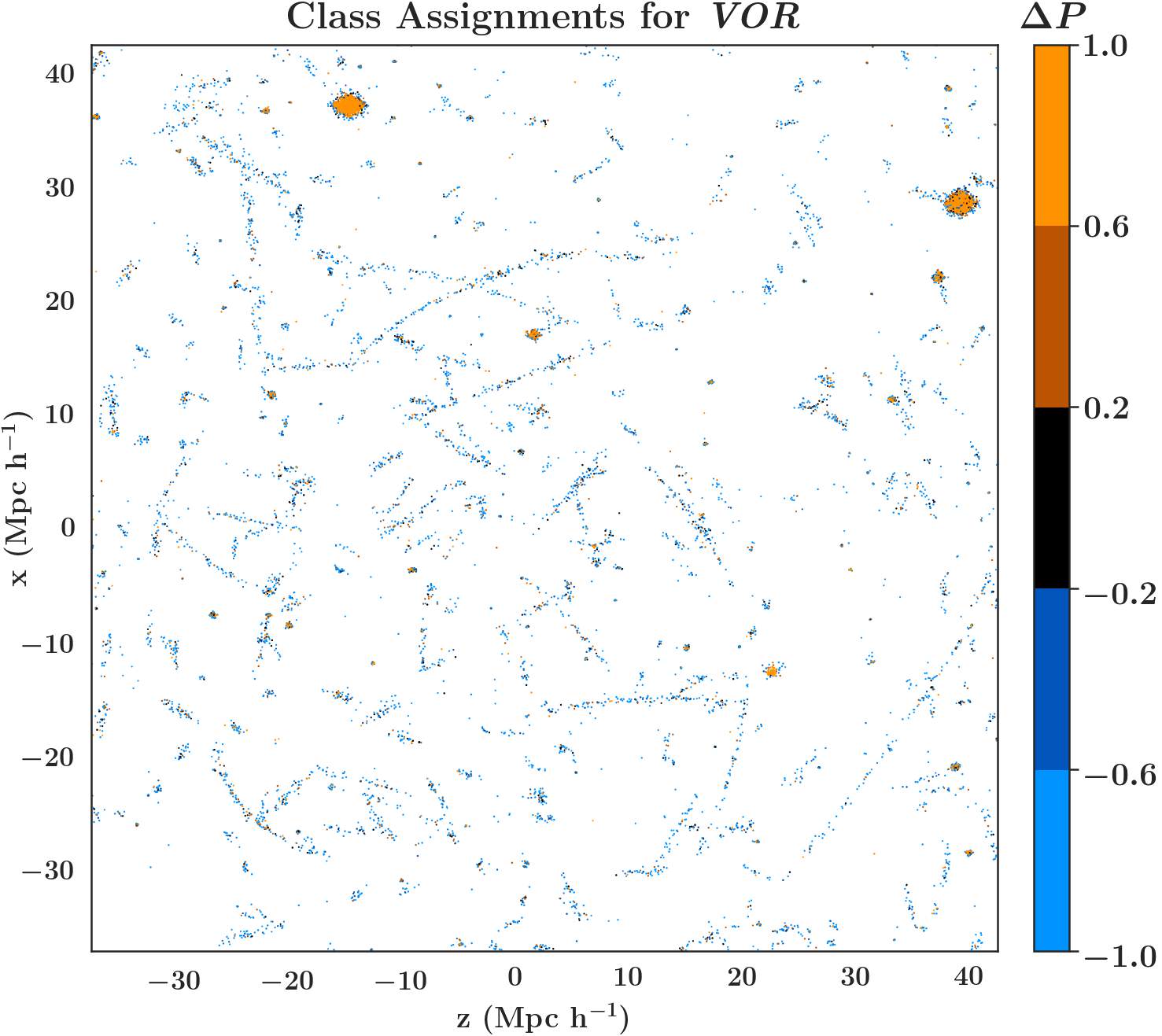}\label{fig:CRadCompTTa}
        } \\
        \subfloat[]{
            \includegraphics[width = \columnwidth]{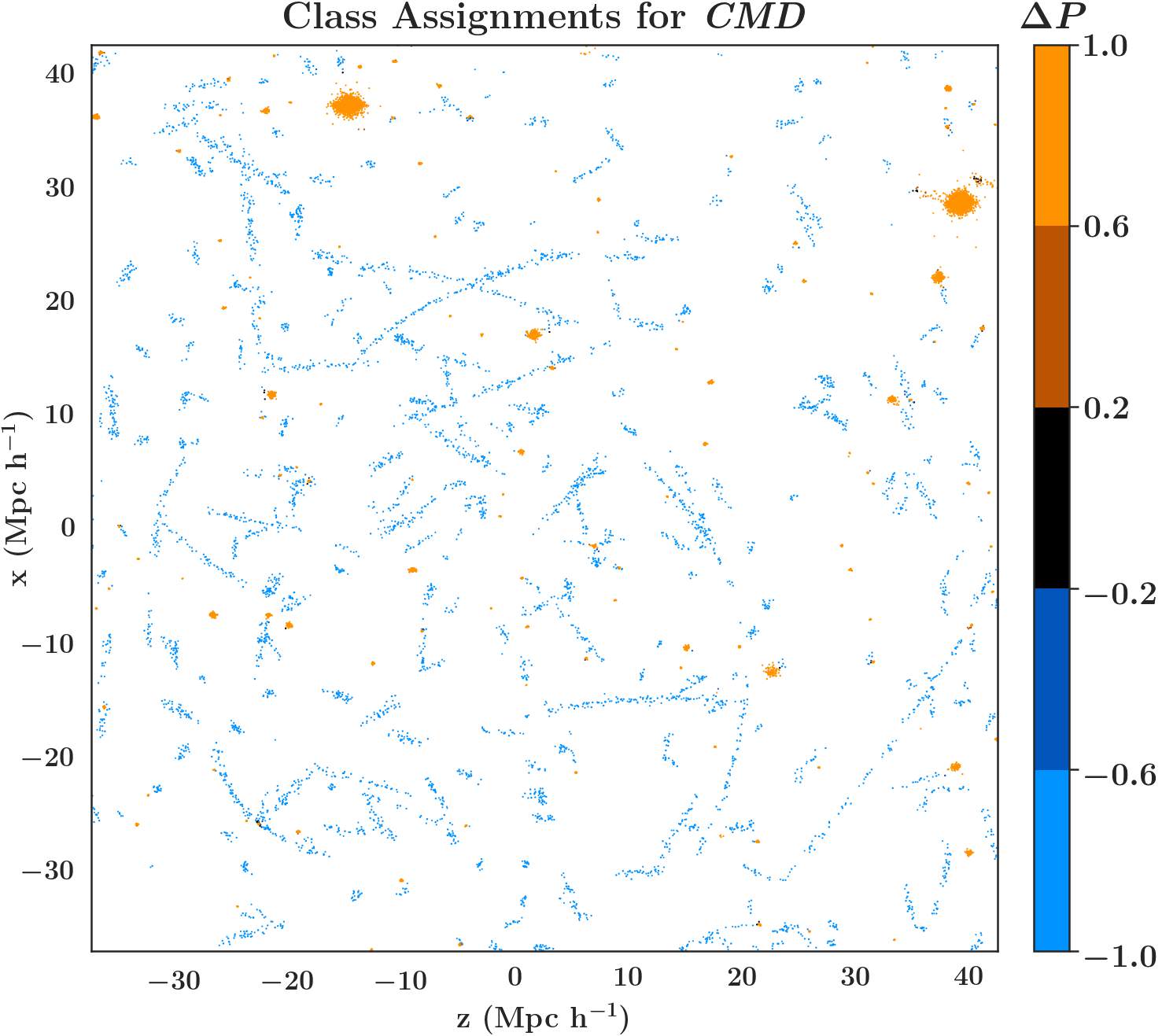}\label{fig:CRadCompTTb}
        }
        \end{tabularx}
    \caption{Class assignments for \texttt{TSIM}.  \textbf{(a)} and \textbf{(b)} show the halo and filament class assignments made using \EmphMeas{VOR} and \EmphMeas{CMD} measurements, respectively, colored based on each particle's probability contrast.  The particles seen are from a 7 Mpc thick slice of the 3D N-body simulation \texttt{TSIM}.  This slice includes the largest halo, which has $N = 5482$ particles and is centered at $(x,\,y,\,z) = (28.5,\,-7.3,\,39.2)$ Mpc $h^{-1}$.}
    \label{fig:CRadCompTT}
\end{figure}

\begin{figure*}
    \centering\captionsetup[subfloat]{labelfont=bf}
    \begin{minipage}{\textwidth}
        \centering
        \begin{tabularx}{\textwidth}{CC}
            \multicolumn{2}{c}{\bf{Toy-to-\texttt{TSIM} Probability Contrasts for \EmphMeas{KNN} (+ \EmphMeas{PCA})}} \\
            \hline
            \subfloat[]{
            \includegraphics[width = 0.5\textwidth]{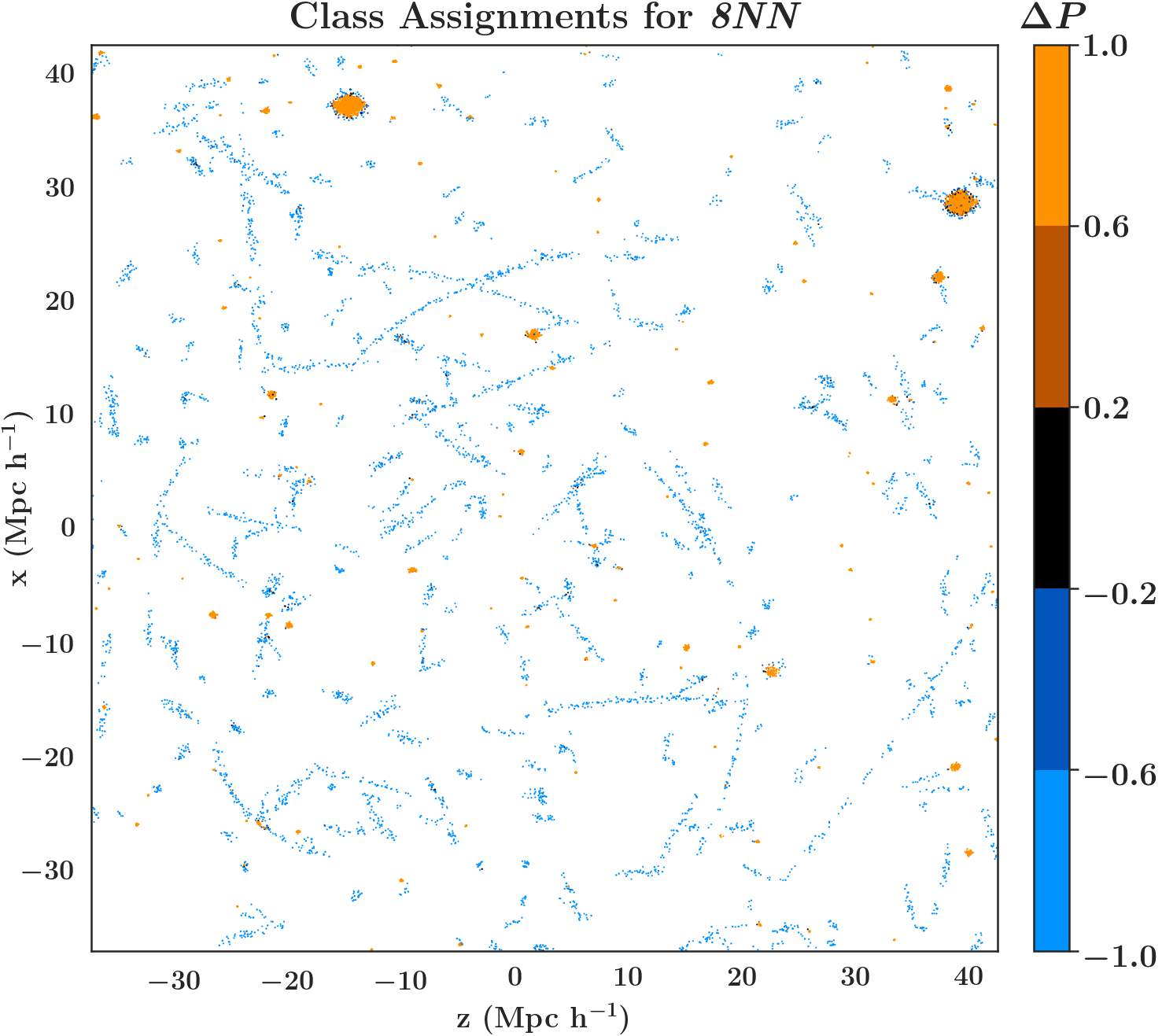}\label{fig:KNNCompTTa}
            } &
            \subfloat[]{
            \includegraphics[width = 0.5\textwidth]{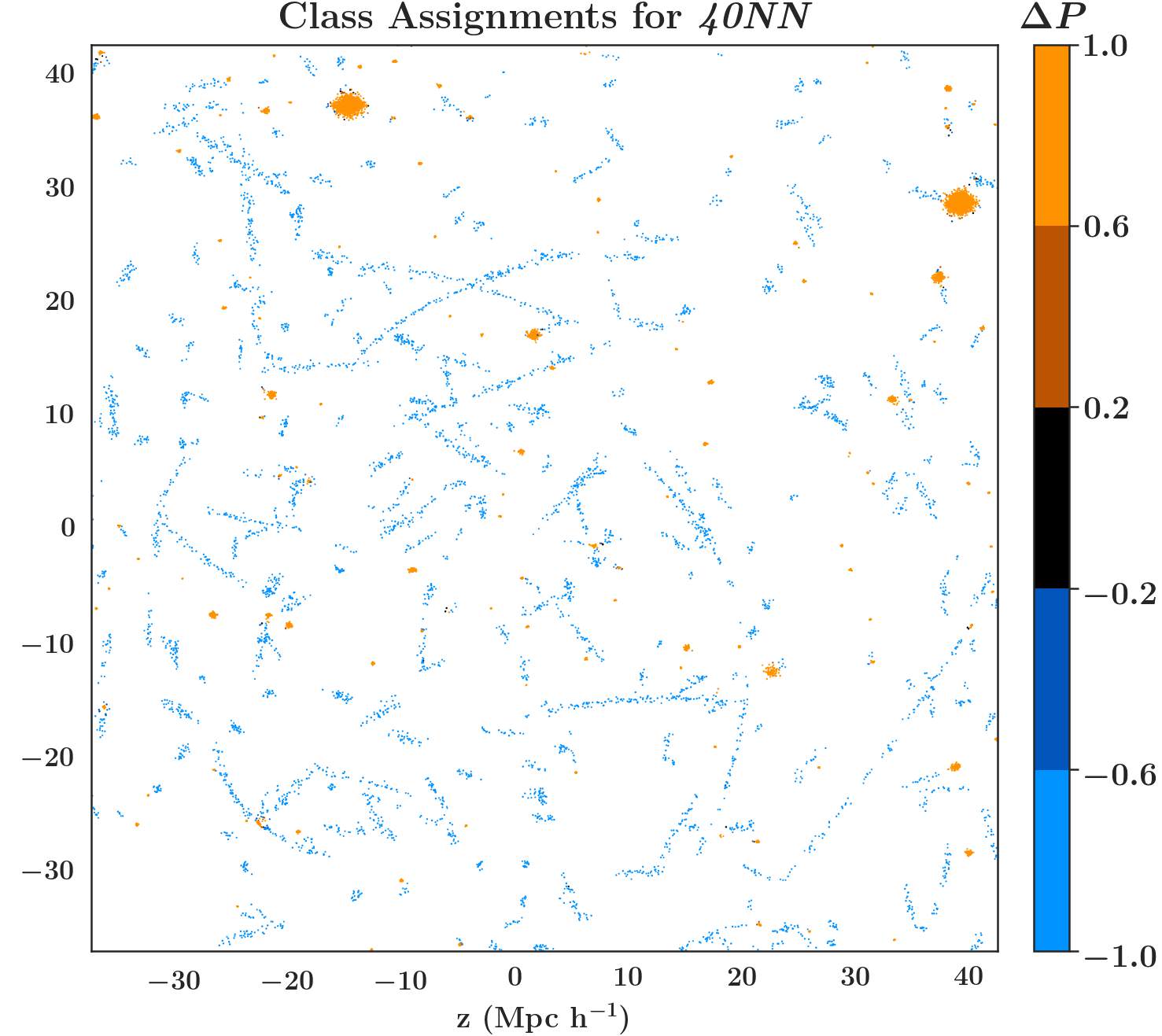}\label{fig:KNNCompTTb}
            } \\
            \subfloat[]{
            \includegraphics[width = 0.5\textwidth]{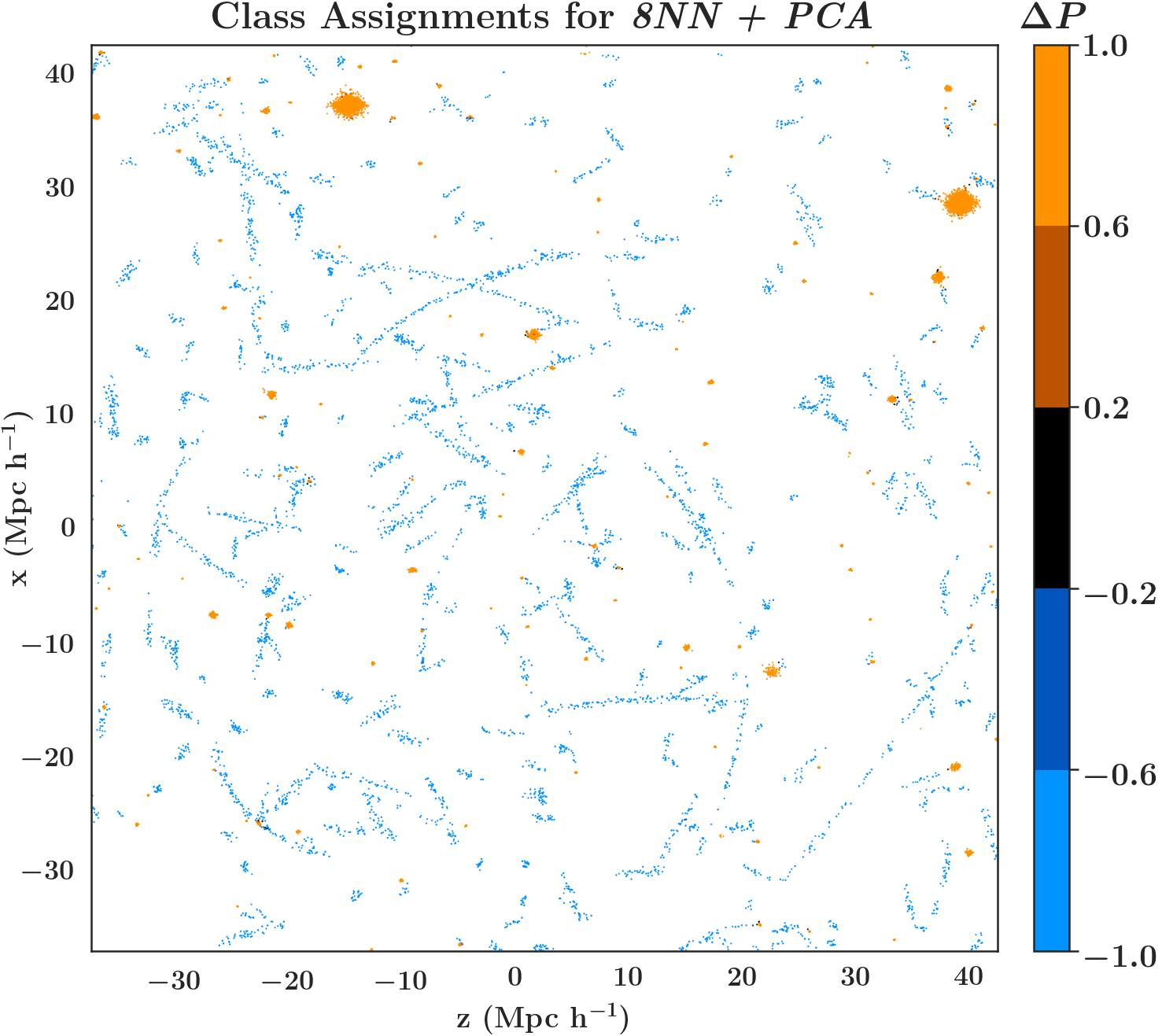}\label{fig:KNNCompTTc}
            } &
            \subfloat[]{
            \includegraphics[width = 0.5\textwidth]{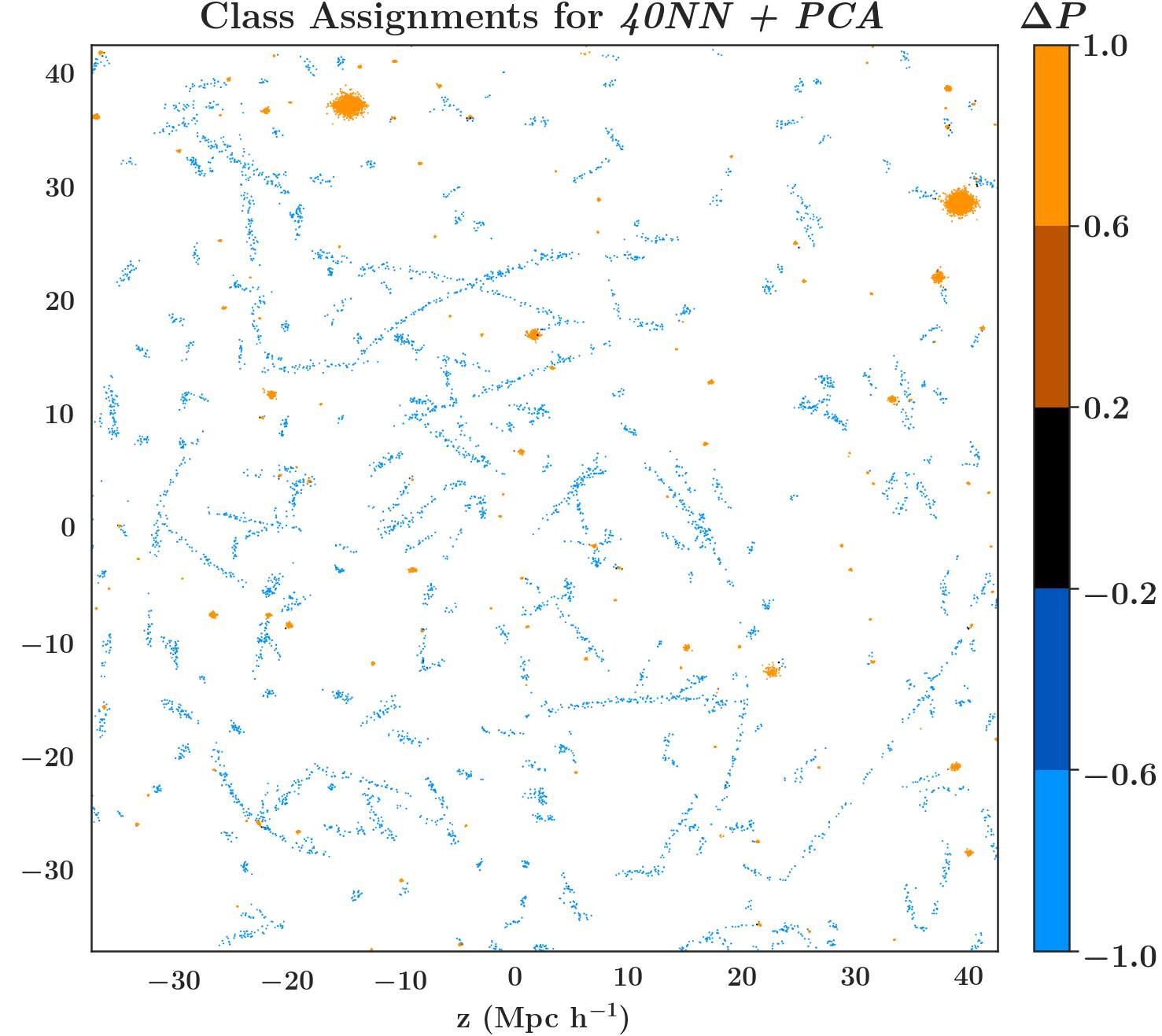}\label{fig:KNNCompTTd}
            }
        \end{tabularx}
    \end{minipage}
    \caption{Class assignments for \texttt{TSIM}.  The halo-filament assignments, colored based on the probability contrast, are shown in \textbf{(a)} (\EmphMeas{KNN}, $k \leq 8$), \textbf{(b)} (\EmphMeas{KNN}, $k \leq 40$), \textbf{(c)} (\EmphMeas{KNN} + \EmphMeas{PCA}, $k \leq 8$), and \textbf{(d)} (\EmphMeas{KNN} + \EmphMeas{PCA}, $k = 40)$.  The particles seen are from the same 7 Mpc thick slice of the 3D N-body simulation \texttt{TSIM} as \ref{fig:CRadCompTT}.}
    \label{fig:KNNCompTT}
\end{figure*}

The true class values for \texttt{SIM} are not known, and there is not a known verifiable method that can assign these true classes.  However, demonstrating that the statistics of our class assignments are similar for target datasets with very different structural properties demonstrates the robustness and lack of bias of our classifier.  We will do this by comparing the statistics of \texttt{SIM} particle assignments to classes assigned to a toy model \texttt{TSIM}.  A plot of \texttt{TSIM} (not shown) appears nearly identical to that of the training dataset seen in Figure \ref{fig:Train}.

Classes were assigned to \texttt{TSIM} particles using the same training dataset as \texttt{SIM}; these results may be seen in Figures \ref{fig:CRadCompTT} and \ref{fig:KNNCompTT}.  The large-scale properties of these class assignments were largely the same as those of in \texttt{SIM} seen in Figure \ref{fig:CRadComp} and \ref{fig:KNNComp}: \EmphMeas{VOR} and \EmphMeas{CMD} (\ref{fig:CRadCompTTa} and \ref{fig:CRadCompTTb}, respectively) generally overestimated halo abundance, with assignments made by \EmphMeas{VOR} having lower probability assignments than \EmphMeas{CMD}; \EmphMeas{KNN}-trained classifiers with only small $k$ (Figure \ref{fig:KNNCompTTa}) assigned lower probabilities to all particles and produced smaller halos than a classifier with large $k$ (Figure \ref{fig:KNNCompTTb}); and the addition of \EmphMeas{PCA} calculations (Figure \ref{fig:KNNCompTTc} and \ref{fig:KNNCompTTd}) helped remove the differences in class label assignments associated with different ranges in $k$-values.

One notable exception, however, is the classifier trained with only \EmphMeas{CMD} (\ref{fig:CRadCompTTb}).  Unlike in \texttt{SIM} (Figure \ref{fig:CRadCompTTb}), this classifier did not produce clumps of ambiguous particles near particularly dense halo-filament boundaries in \texttt{TSIM} (Figure \ref{fig:CRadCompTT}).  This is likely due to the fact that the training data used an identical generation procedure to \texttt{TSIM}.  This highlights the robustness of \EmphMeas{KNN} (and \EmphMeas{PCA}) calculations.  The assignments made by \EmphMeas{CMD} calculations are heavily tied to the generation algorithm, indicating that it is not suited for training using a simplified dataset such as the toy model we developed.  On the other hand, the class labels assigned by a classifier trained with \EmphMeas{KNN} calculations, especially when paired with \EmphMeas{PCA} calculations, are less affected by the exact structure of the toy model, enabling their use even when training was performed using a simplified toy model.  This indicates that \EmphMeas{KNN} calculations establish a natural length scale for halos and filaments together when performing classification, and the addition of \EmphMeas{PCA} calculations help distinguish between halos and filaments by establishing distinct length scales for these structures individually.  Note that \EmphMeas{PCA} calculations are most effective for structures with a length scale that is not much larger than $R_{\textrm{PCA}}$.  Additional discussion of these properties may be found in Section \ref{RobPCA}.

\begin{figure}
    \centering\captionsetup[subfloat]{labelfont=bf}
    \begin{tabularx}{\columnwidth}{C}
        \bf{Halo Mass Functions from \texttt{TSIM} Class Labels} \\
        \hline
        \subfloat[]{
            \includegraphics[width = \columnwidth]{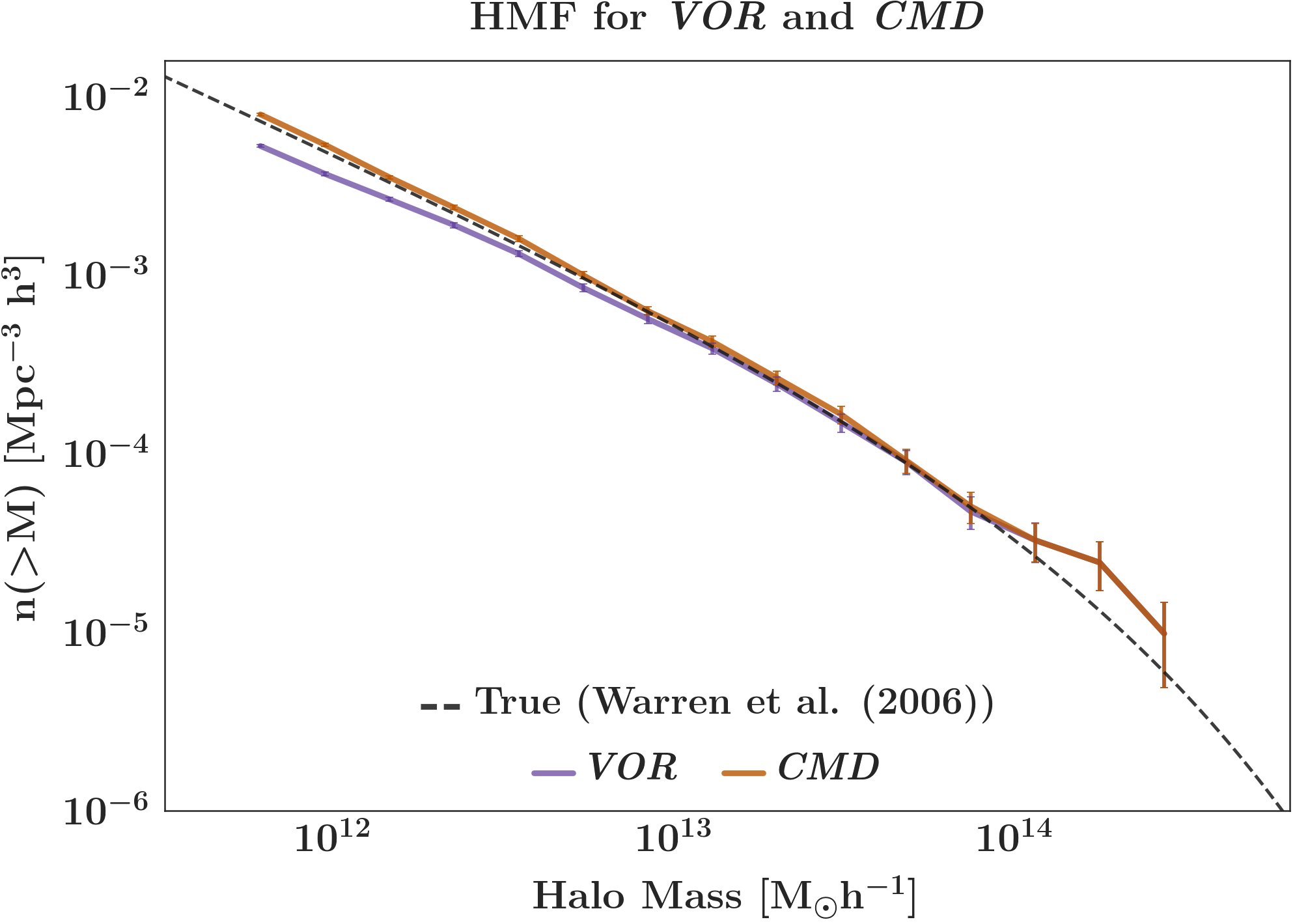}\label{fig:HMFTTa}
        } \\
        \subfloat[]{
            \includegraphics[width = \columnwidth]{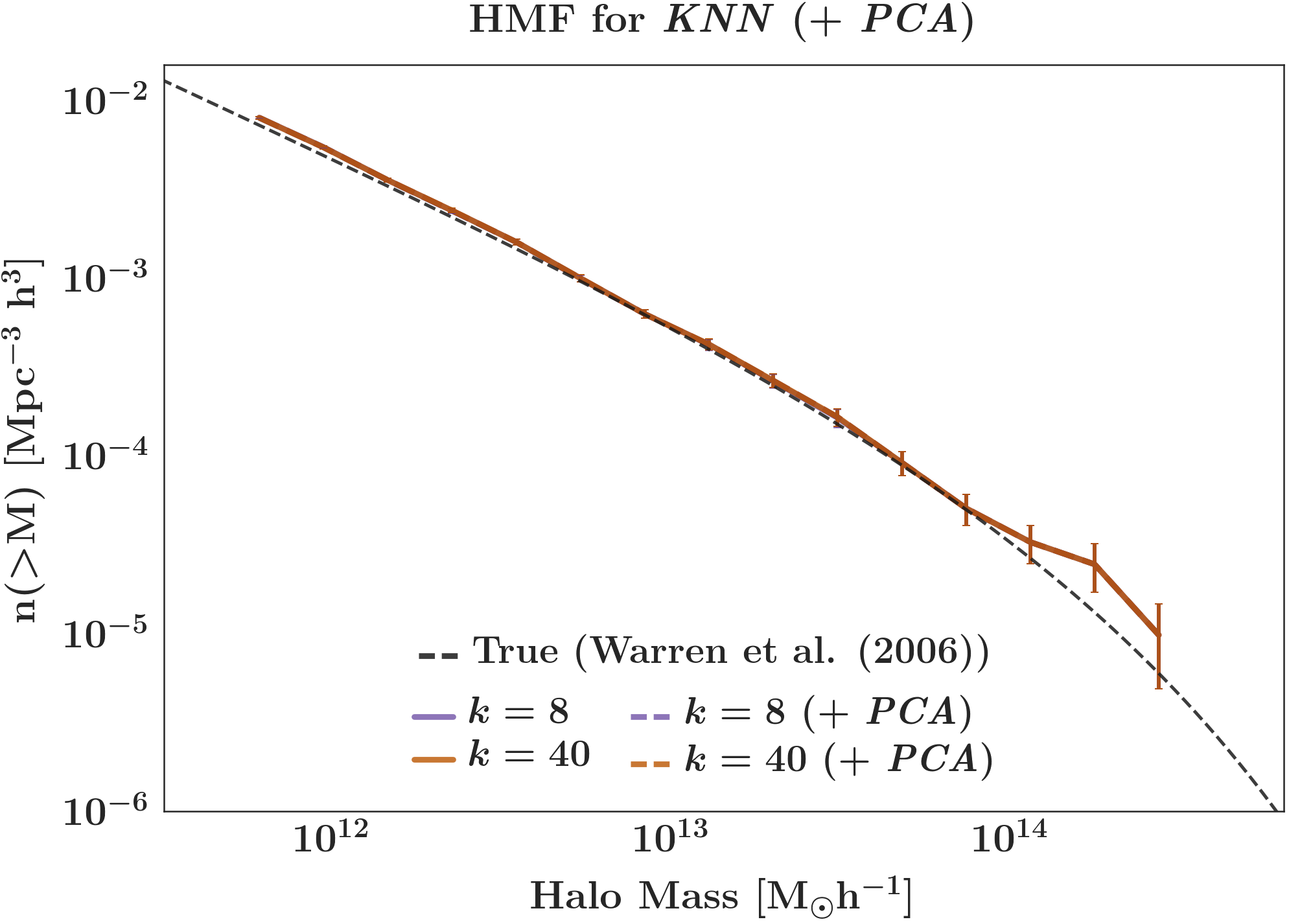}\label{fig:HMFTTb}
        }
    \end{tabularx}
    \caption{Halo mass functions for \texttt{TSIM} using \textbf{(a)} \EmphMeas{VOR} and \EmphMeas{CMD} and \textbf{(b)} \EmphMeas{KNN} (+ \EmphMeas{PCA}) for $k \leq 8,\,40$.}
    \label{fig:HMFTT}
\end{figure}

Figure \ref{fig:HMFTT} shows the HMFs derived from \texttt{TSIM}.  For all figures, it is clear that halo classification was least accurate at the extreme ends of the mass range.  The high-mass deviation was likely due to statistical fluctuations resulting from there being very few large-mass halos in \texttt{TSIM}.  The overprediction of low-mass halos by all classifiers other than \EmphMeas{VOR} may arise from the generation parameters.  In the toy model, the radii of low-mass halos was close to the radii of most filaments.  As a result, some filaments in \texttt{TSIM} were incorrectly classified as small halos, leading to the discrepancy.  The \EmphMeas{VOR} HMF is inaccurate for low-mass halos because \EmphMeas{VOR} is a poor differentiator between LSS classes (see Figures \ref{fig:CRadCompa} and \ref{fig:CRadCompTTa}).

\section{Analysis} \label{Analysis}

By correlating results from \texttt{SIM} with those of \texttt{TSIM}, we can demonstrate that the toy model is effective as a training data set: if the the properties of class assignments in \texttt{SIM} are statistically and/or visually similar to those in \texttt{TSIM}, which has markedly different properties, we have demonstrated that the methodology is robust enough to be applied to observed data.  The goal of this section is not only to demonstrate the validity of our classifier, but also to establish the importance of utilizing measurements of both local density magnitude and directionality to ensure that our class assignments are not strongly influenced by the somewhat arbitrary parameters used to generate structures, particularly filaments, in the toy model.

\subsection{Feature Importances} \label{FI}

\begin{figure}
  \centering\captionsetup[subfloat]{labelfont=bf}
  \begin{tabularx}{\columnwidth}{C}
    \bf{Feature Importances for Toy-to-\texttt{SIM}} \\ \hline
    \subfloat{
      \includegraphics[width = \columnwidth]{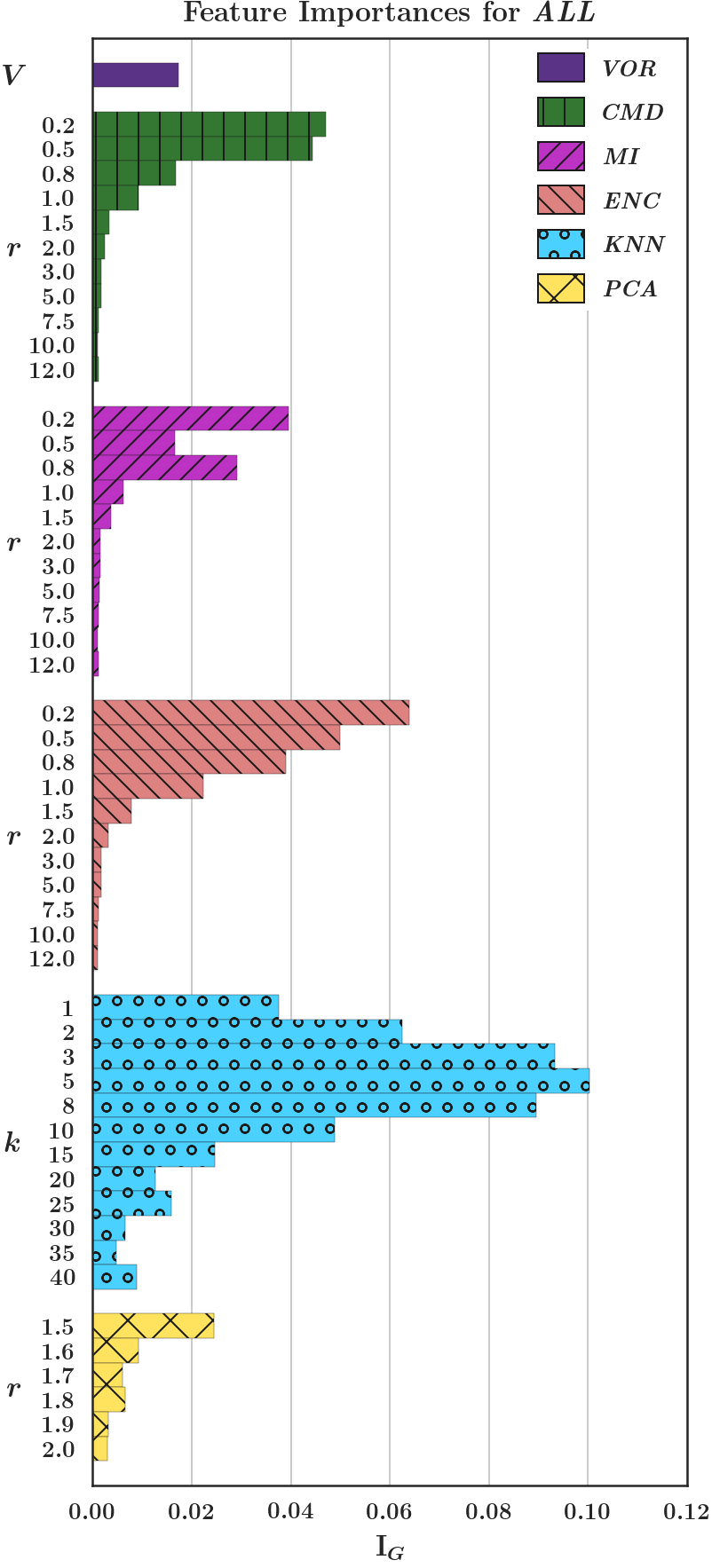}
    }
  \end{tabularx}
  \caption{The feature importances $\textrm{\textbf{I}}_G$ for the Toy-to-\texttt{SIM} calculations using all metrics.}
  \label{fig:FIAll}
\end{figure}

\begin{figure*}
  \centering\captionsetup[subfloat]{labelfont=bf}
  \begin{tabularx}{0.9\textwidth}{CC}
    \multicolumn{2}{c}{\bf{Feature Importances for Toy-to-\texttt{SIM}}} \\ \hline
    \subfloat[]{
      \includegraphics[width = 0.45\textwidth]{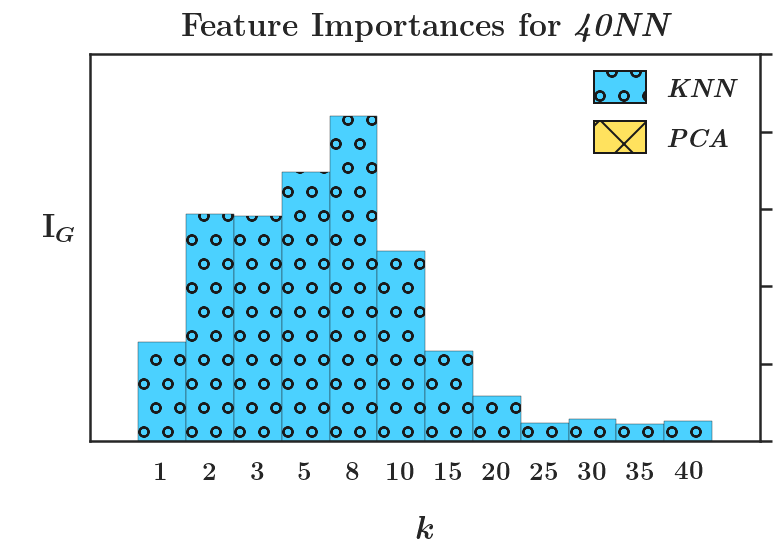}\label{fig:FIa}
    } &
    \subfloat[]{
      \includegraphics[width = 0.45\textwidth]{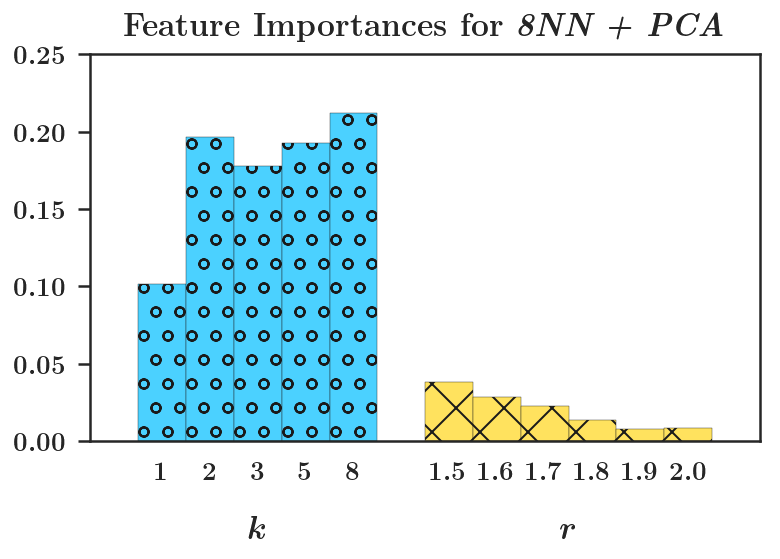}\label{fig:FIb}
    }
  \end{tabularx}
  \caption{The feature importances $\textrm{\textbf{I}}_G$ for the Toy-to-\texttt{SIM} calculations using \textbf{(a)} \EmphMeas{KNN} and \textbf{(b)} \EmphMeas{KNN} + \EmphMeas{PCA}.}
  \label{fig:FI}
\end{figure*}

First, we aimed to decrease the number of measurements required to minimize computation time and understand the role of each metric.  We utilized the classifier's feature importances, which describes how relevant each feature was when performing class assignment; this was determined by the frequency a metric was used to choose a branch as the classifier descended a tree.  Figure \ref{fig:FI} shows the feature importances for a variety of different metrics.  From here, we see that \EmphMeas{KNN} is weighted the most heavily, indicating that it may provide valuable information for the classifier.  This expectation correlates with the measurement histogram seen in Figure \ref{fig:MeasHistc}, where it may be seen that the distance to the eighth-nearest neighbor separates each of the classes distinctly from one another.

Figures \ref{fig:FIAll} and \ref{fig:FI} shows the feature importances for several different feature sets.  In general, small radii/$k$ were deemed most important.  From \ref{fig:FI}, the feature importances for all features, density magnitude calculations were weighted more heavily than \EmphMeas{PCA}, and of the density magnitude calculations, \EmphMeas{KNN} was weighted most heavily, reflecting the lack of differentiation between structures in the measurement histograms for the other density magnitude calculations.  In both Figures \ref{fig:FI} and \ref{fig:FIa}, the feature importances for \EmphMeas{KNN}, the most important measurements were those with $k \leq 8$, reaching a peak at $k = 8$.  This is likely due to the fact that the smallest halos had 8 particles, and as small halos dominated the halo mass function, they will be utilized most by the training algorithm to determine a feature's importance.  These phenomena were also reflected in \ref{fig:FIb} \EmphMeas{KNN} + \EmphMeas{PCA}.

\subsection{ROC AUC} \label{ROC}

\begin{figure*}
  \centering\captionsetup[subfloat]{labelfont=bf}
  \begin{minipage}{0.9\textwidth}
    \centering
    \begin{tabularx}{0.9\textwidth}{CC}
      \multicolumn{2}{c}{\bf{ROC Curves for \texttt{SIM} and \texttt{TSIM}}} \\
      \hline \\
      \textbf{\texttt{SIM}} & \textbf{\texttt{TSIM}} \\
      \subfloat[]{
        \includegraphics[width = 0.45\textwidth]{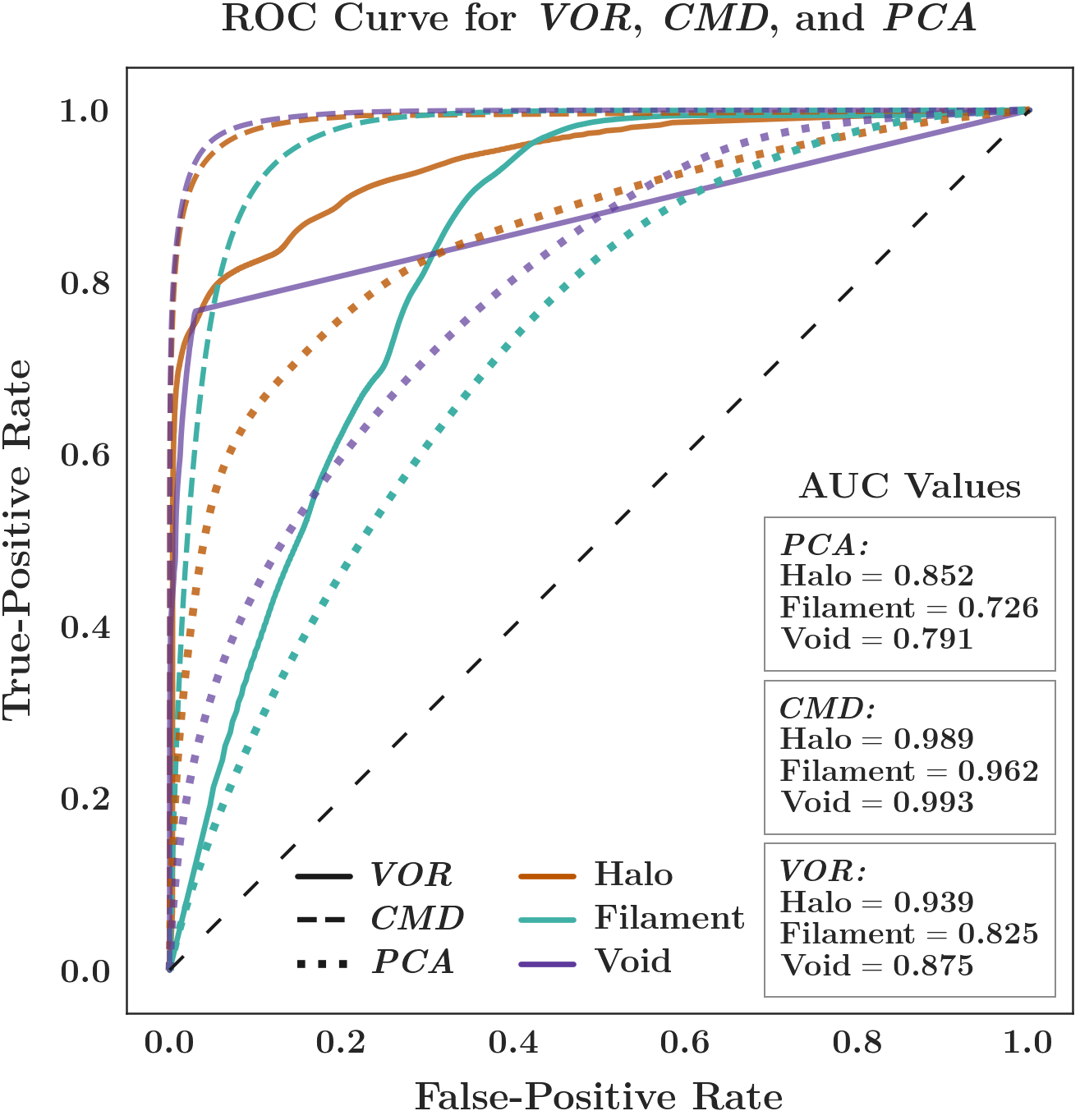}\label{fig:ROCa}
      } &
      \subfloat[]{
        \includegraphics[width = 0.45\textwidth]{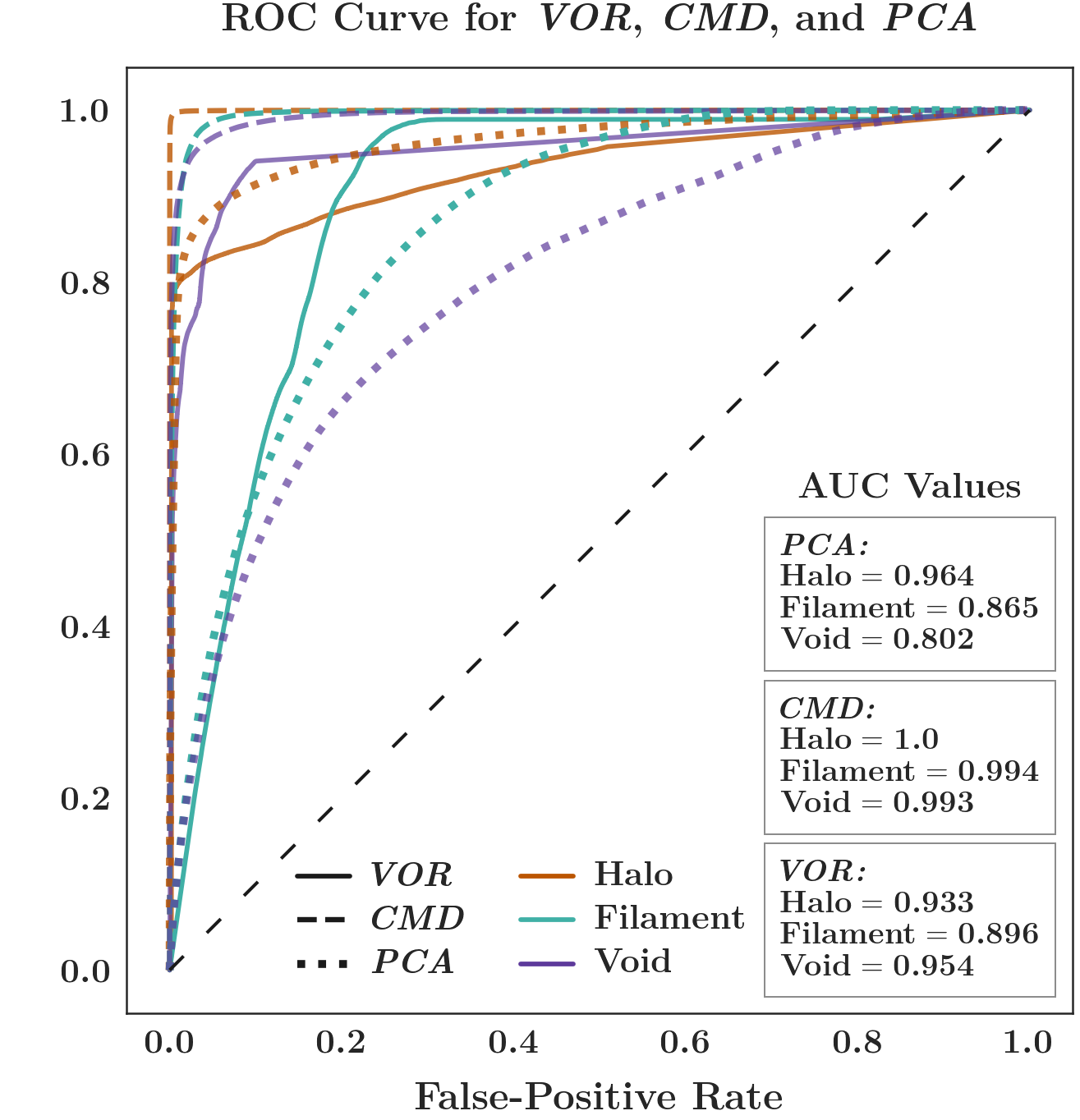}\label{fig:ROCb}
      } \\
      \subfloat[]{
        \includegraphics[width = 0.45\textwidth]{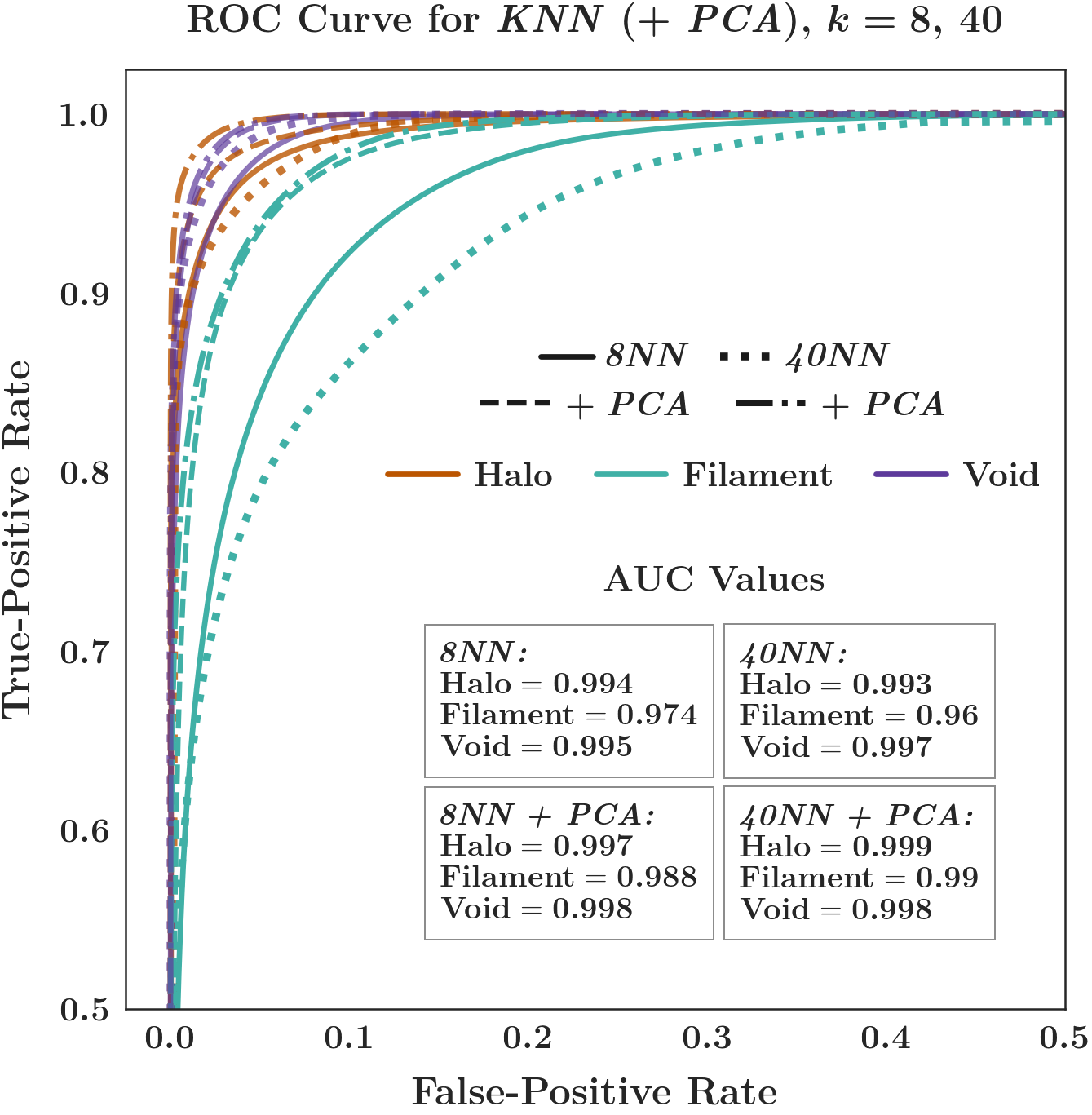}\label{fig:ROCc}
      } &
      \subfloat[]{
        \includegraphics[width = 0.45\textwidth]{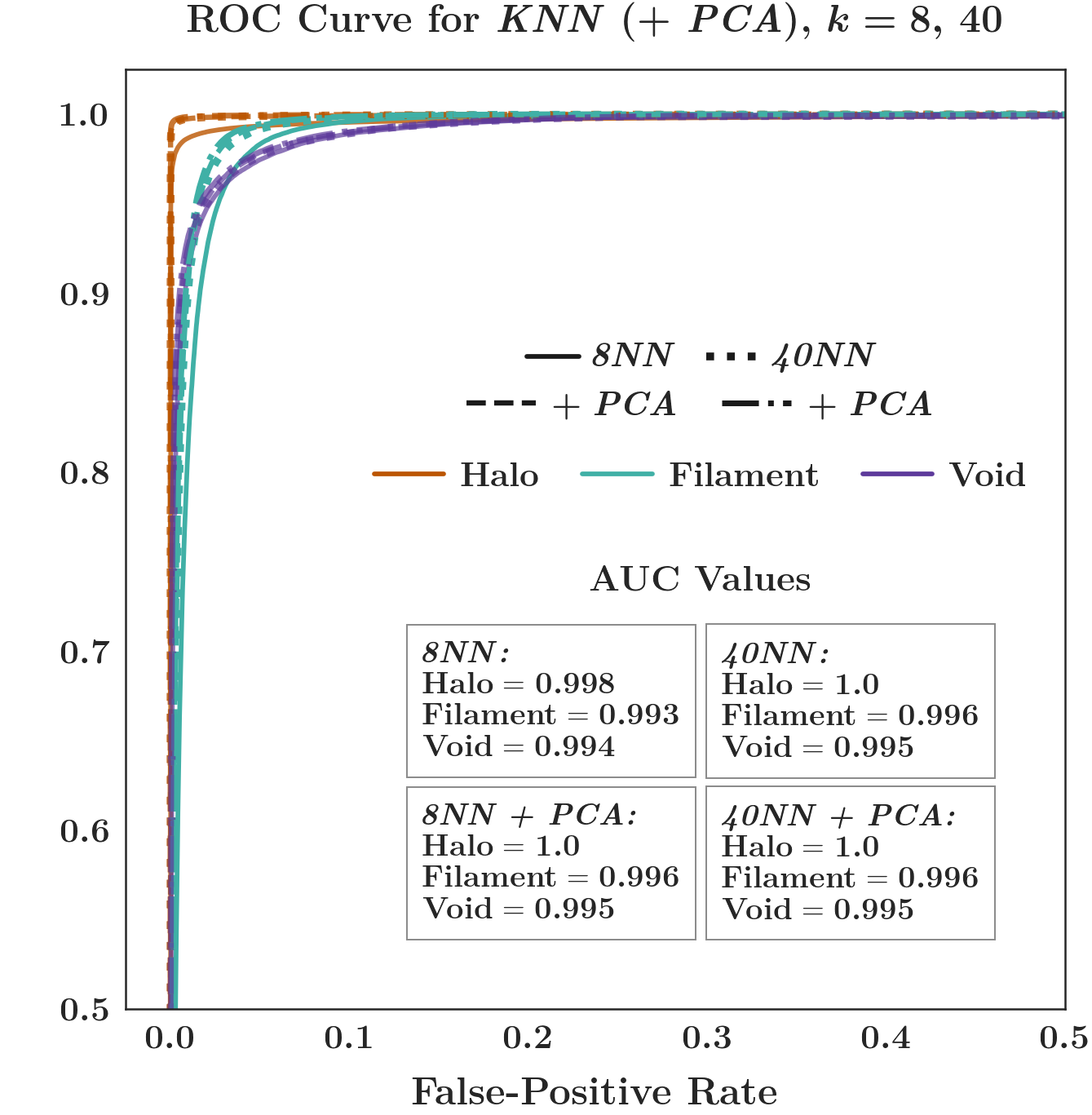}\label{fig:ROCd}
      }
    \end{tabularx}
  \end{minipage}
  \caption{ROC plots for \texttt{SIM} and \texttt{TSIM}.  \textbf{(a)} and \textbf{(b)} show the ROC curves for a classifier trained using \EmphMeas{VOR}, \EmphMeas{CMD}, an \EmphMeas{PCA}, and \textbf{(c)} and \textbf{(d)} show the same for \EmphMeas{KNN} (+ \EmphMeas{PCA}), $k \leq 8,\,40$.  The dashed diagonal line with unit slope in \textbf{(a)} and \textbf{(b)} corresponds with a classifier that assigns
labels by selecting a class at random; curves above this line indicate that the classifier using those features has predictive capabilities.  Note that this line is not visible in \textbf{(c)} and
\textbf{(d)}; this is because the bounds on these figures were altered for illustration purposes.}
  \label{fig:ROC}
\end{figure*}

The receiver operating characteristic (ROC) curve is a way of demonstrating a classifier's ability to accurately discriminate between classes.  It consists of a plot of the classifier's true positive rate as a function of its false positive rate.  A classifier that cannot discriminate between classes would have a $50\%$ probability of assigning the correct class to a given data point, and would have equal true positive and false positive rates; hence, its ROC curve would appear as a line with unit slope.  On the other hand, an effective classifier would have a much larger true positive rate than false positive rate.  The area under the curve (AUC) is a measure for a classifier's effectiveness: a classifier with a true positive rate much larger than its false positive rate would have an AUC value near unity, while one with poor discriminatory ability would have AUC = 0.5.

The first method we used to verify our results was through calculating the ROC AUC independently for each class; ROC plots \citep{sklearn,ROC1,ROC2} for some metric combinations can be found in Figure \ref{fig:ROC}.  For \texttt{TSIM}, the assigned class labels were compared with the true class values in \texttt{TSIM}; however, these values did not exist in \texttt{SIM}.  As a result, we chose the set of class labels assigned by a classifier trained using all features to use as a fiducial comparison dataset.  This provides the most generality, as it effectively allows the comparison of each classifier to all others simultaneously.  As we found that most metric sets produced similar results, and that the AUC for \texttt{TSIM} was maximized for all classes when using a classifier trained using all features, we believe that these class assignments will provide a sufficient approximation to the true class values to use as a fiducial comparison dataset.

In general, the ROC curves in Figure \ref{fig:ROC} demonstrate that filament classification was the most difficult.  In addition, classification on \texttt{TSIM} was more accurate than on \texttt{SIM}, as evidenced by the shape of the curves and the AUC values.

Figure \ref{fig:ROCa} and \ref{fig:ROCb} show the ROC curves for classifiers trained using only \EmphMeas{VOR}, \EmphMeas{CMD}, and \EmphMeas{PCA}.  From this, it can be seen that classifiers trained on \EmphMeas{VOR} or \EmphMeas{PCA} alone were not effective when classifying particles in \texttt{SIM} and on \texttt{TSIM}.  This further demonstrates that \EmphMeas{VOR} does not suffice as a proxy for local density magnitude.  The poor performance of \EmphMeas{PCA} alone may be attributed to the lack of information provided to the classifier about local density magnitude.

Figure \ref{fig:ROCc} and \ref{fig:ROCd} show the ROC curves for classifiers trained using \EmphMeas{KNN} (+ \EmphMeas{PCA}) measurements for $k \leq 8,\,40$.  Notably, the addition of \EmphMeas{PCA} calculations improved classifier performance for all LSS classes, particularly for filaments.  As seen in the \texttt{SIM} classes assigned by a classifier trained by \EmphMeas{KNN} + \EmphMeas{PCA} (see Figure \ref{fig:KNNComp}), the addition of \EmphMeas{PCA} calculations diminished the dependence of halo classification on the values of $k$ used.  Figure \ref{fig:ROCc} demonstrates that this stabilization applies to all classes, especially filaments.  The benefits of this stabilization are immense: using large $k$-values allows the classifier to take into account the global environment when performing classification, improving classification of large halos, but lessens its sensitivity to properties of the local environment.  Including \EmphMeas{PCA} calculations enables the global environment to be used in training without contaminating information about small-scale properties.  The combination of small-scale and large-scale information in training enables classification of halos isolated in large void volumes (a major issue discussed in \citet{LSSSup}, \citeyear{LSSSup} and \citeauthor{TracCWeb}, \citeyear{TracCWeb}).

\subsection{HMF Comparison: Mean Absolute Proportion} \label{MAP}

\begin{figure*}
  \centering\captionsetup[subfloat]{labelfont=bf}
  \includegraphics[width = 0.85\textwidth]{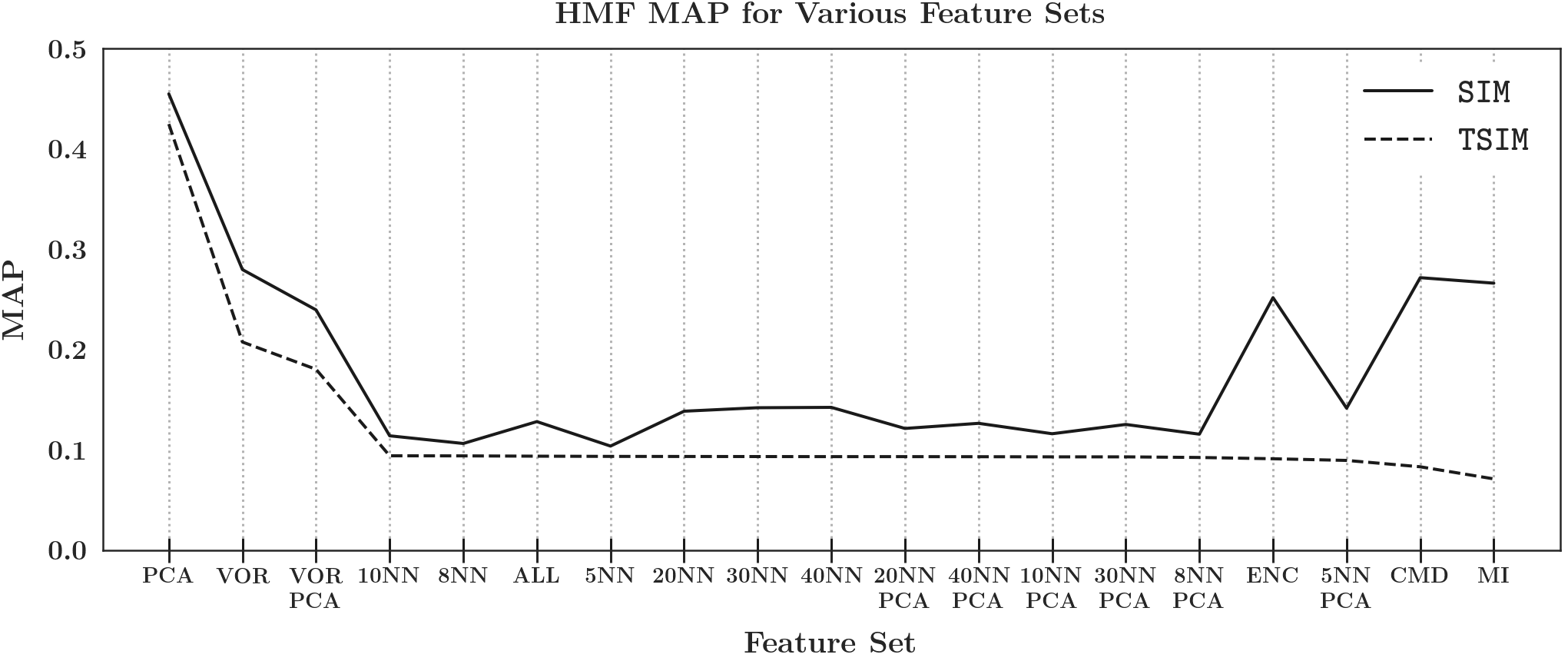}
  \caption{A plot of the HMF MAP; the metric sets are ordered such that the \texttt{TSIM} MAP decreased from left to right.}
  \label{fig:MAP}
\end{figure*}

We define the mean absolute proportion (MAP) as

\begin{align} \label{eq:MAP}
  \textrm{MAP} = \frac{\mu\left(\abs{n_\textrm{Pred} - n_\textrm{Warren}}\right)}{\mu\left(\abs{n_\textrm{Warren}}\right)},
\end{align}

where $\mu(n)$ is the mean of $n$ over all $M$ and $n_\textrm{Pred}$ and $n_\textrm{Warren}$ were the empirical and \citet{WarrenHMF} HMFs, respectively.  In Figure \ref{fig:MAP}, we show the MAP for each of the metrics for both \texttt{SIM} and \texttt{TSIM}.  The metrics are ordered based on the MAP value for \texttt{TSIM}.  From this plot, it can be seen that, as before, classifiers trained using only \EmphMeas{VOR} or \EmphMeas{PCA} performed substantially worse than all other feature combinations when classifying \texttt{TSIM} particles.  In addition, though not previously discussed, a classifier trained using \EmphMeas{VOR} + \EmphMeas{PCA} performed poorly when classifying \texttt{TSIM} particles, emphasizing the importance of a robust density magnitude metric.  For \texttt{SIM}, these three also performed very poorly; however, unlike in \texttt{TSIM} classification, classifiers trained using \EmphMeas{CMD}, \EmphMeas{MI}, and \EmphMeas{ENC} also exhibited a large MAP.  This further supports the conclusion that these methods are not effective due to their strong dependence on the training model generation algorithm.  Classifiers trained using \EmphMeas{KNN} (+ \EmphMeas{PCA}) generally performed better for \texttt{SIM}, possibly due to the similarity between \EmphMeas{KNN} measurements with FOF algorithms.

The analysis performed in Sections \ref{MeasHist} and \ref{FI} both indicated that \EmphMeas{KNN} calculations would likely be the most important, and the results from Sections \ref{ROC} and \ref{MAP} supported this by demonstrating that \EmphMeas{KNN} calculations were robust and generated the most accurate results for both \texttt{SIM} and \texttt{TSIM}.  As a result, in the following section, we will use \EmphMeas{KNN} alone as the proxy for local density magnitude.  In addition, Sections \ref{MeasHist}, \ref{ROC}, and \ref{MAP} suggest that the inclusion of \EmphMeas{PCA} calculations may also be of benefit, particularly for filament classification.  In the next section, we present additional measurements to emphasize the importance of \EmphMeas{PCA} calculations when creating a robust classifier.

\subsection{Robustness of \EmphMeas{PCA} Calculations} \label{RobPCA}

\begin{figure*}
  \centering\captionsetup[subfloat]{labelfont=bf}
  \begin{minipage}{\textwidth}
    \centering
    \begin{tabularx}{\textwidth}{CC}
      \multicolumn{2}{c}{\bf{Halo vs. Filament Probability Scatterplots for \EmphMeas{KNN}}} \\
      \hline
      \subfloat[]{
        \includegraphics[width=0.5\textwidth]{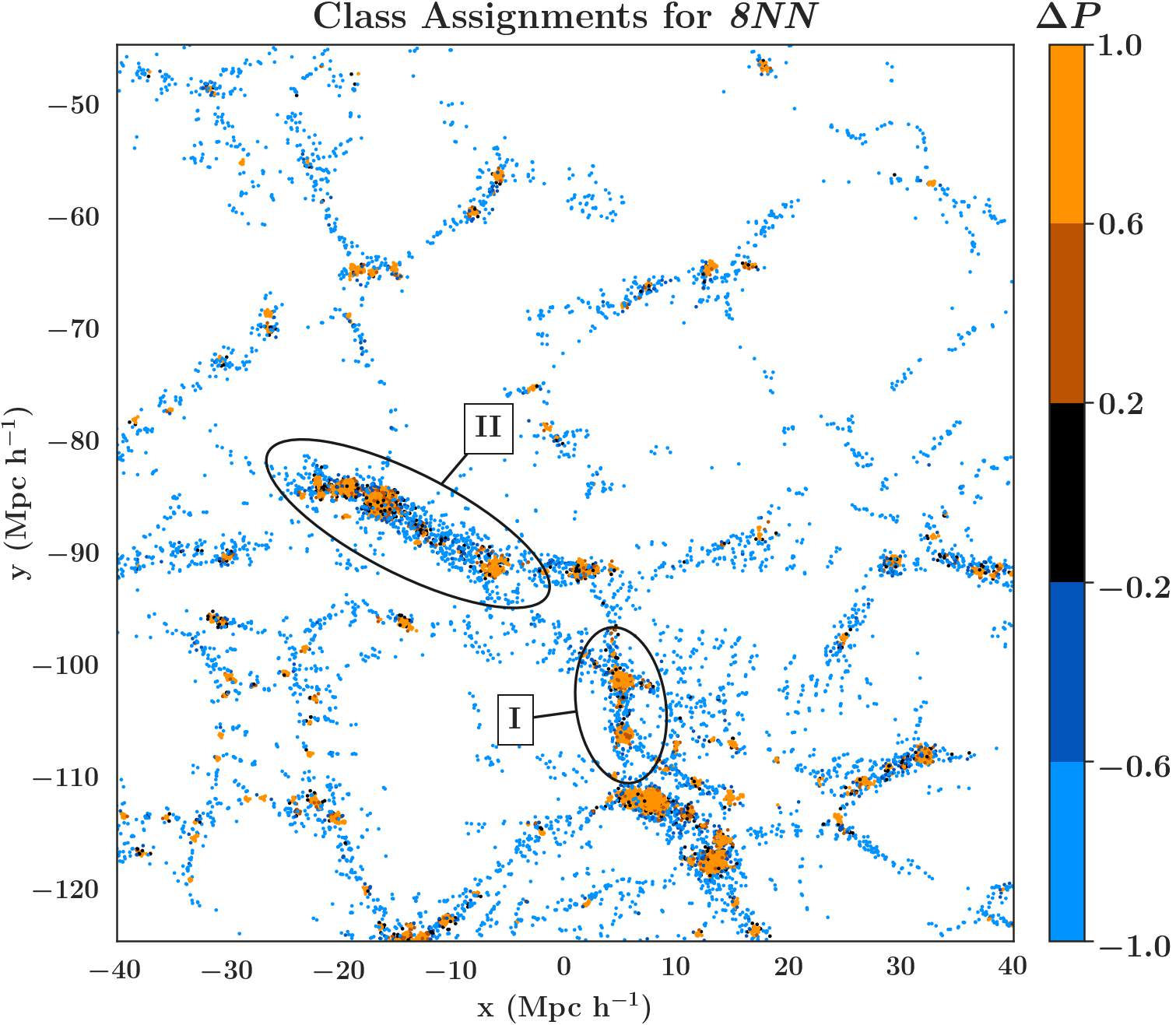}\label{fig:FilProba}
      } &
      \subfloat[]{
        \includegraphics[width=0.5\textwidth]{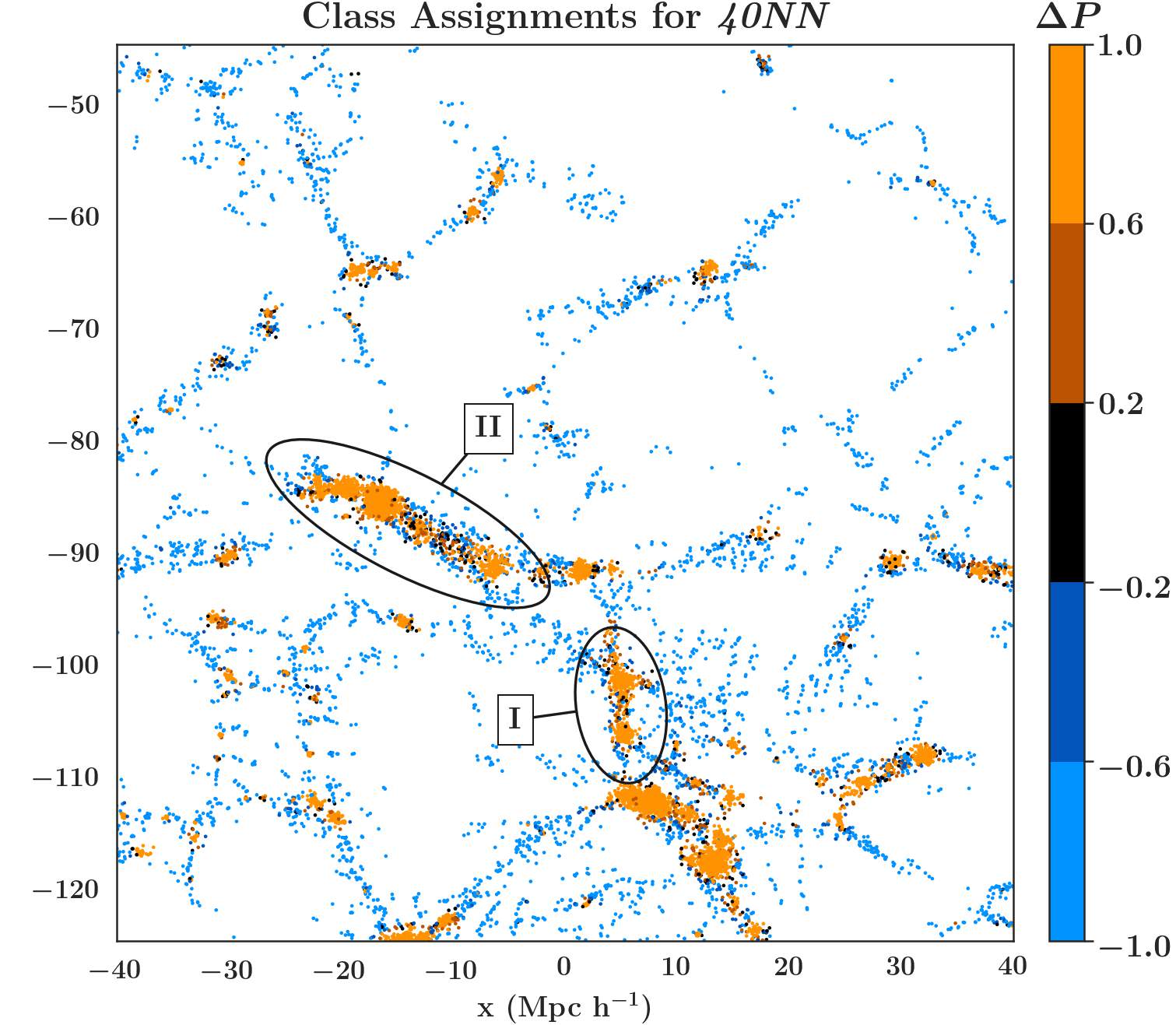}\label{fig:FilProbb}
      } \\
      \subfloat[]{
        \includegraphics[width=0.5\textwidth]{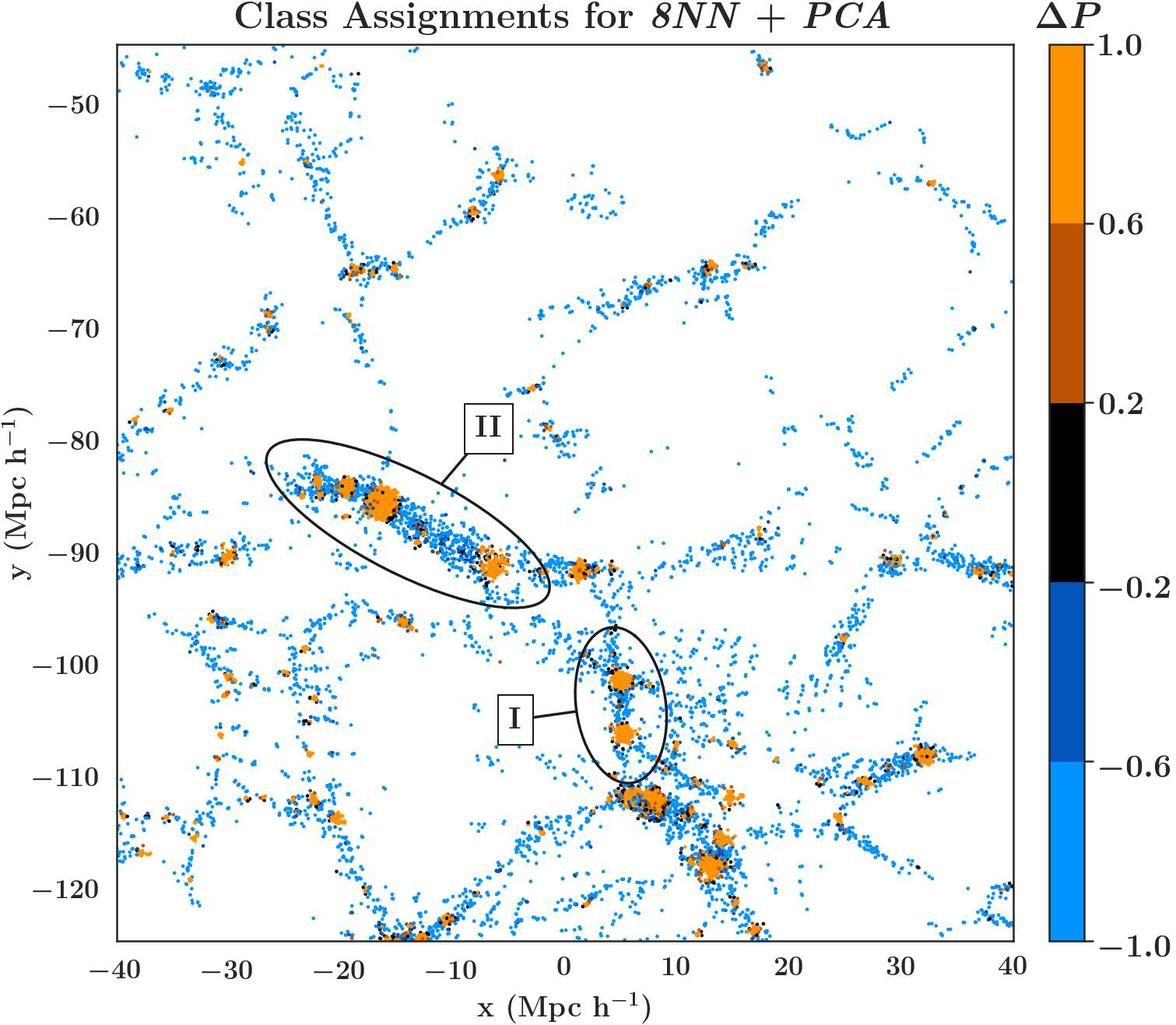}\label{fig:FilProbc}
      } &
      \subfloat[]{
        \includegraphics[width=0.5\textwidth]{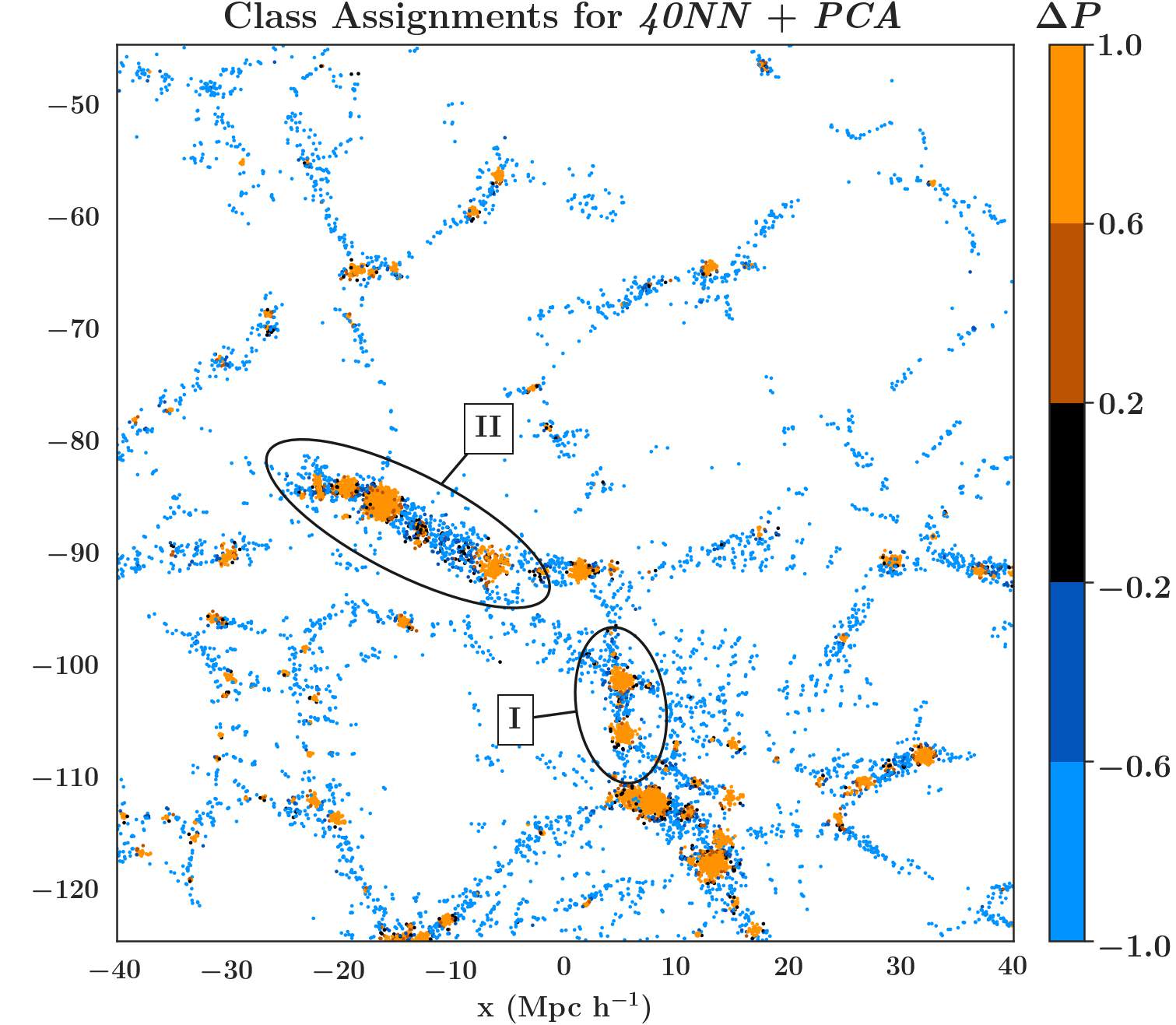}\label{fig:FilProbd}
      }
    \end{tabularx}
  \end{minipage}
  \caption{Halo vs. filament probability contrast scatterplots for \EmphMeas{KNN} (\textbf{(a)} and \textbf{(b)}) and \EmphMeas{KNN} + \EmphMeas{PCA} (\textbf{(c)} and \textbf{(d)}) for $k \leq 8,\,40$.  Particles are from a slice of width $2.0$ Mpc $h^{-1}$.  In this region, most halos had radii that were not substantially larger than the maximum radius used for \EmphMeas{PCA} calculations ($r = 2.0$ Mpc).}
  \label{fig:FilProb}
\end{figure*}

As discussed previously, \EmphMeas{PCA} calculations provide substantial benefit when paired with \EmphMeas{KNN} calculations; however, as noted in Section \ref{MAP}, these benefits are most apparent for structures with length scales no larger than the maximum radius used in \EmphMeas{PCA} calculations ($R_{\textrm{PCA}} = 2.0$ Mpc $h^{-1}$)  Halo vs. filament probability plots for a region where this is true can be seen in Figure \ref{fig:FilProb}.

Figure \ref{fig:FilProba} and \ref{fig:FilProbb} show the classification results for classifiers trained with \EmphMeas{KNN}, $k \leq 8,\,40$, respectively.  As discussed previously, introducing \EmphMeas{PCA} calculations (Figure \ref{fig:FilProbc} and \ref{fig:FilProbd}) curbed the classifier's dependence on the maximum value of $k$ and improved its ability to distinguish between halos and filaments.  Regions \textbf{I} and \textbf{II} in Figure \ref{fig:FilProb} demonstrate this clearly: in \ref{fig:FilProba} and \ref{fig:FilProbb}, the classes assigned to particles varied greatly, with the classifier with $k \leq 40$ substantially overestimating te number of halo particles.  This produced halos that are elongated along one axis, a hallmark trait of filaments.  However, \EmphMeas{PCA} calculations prevented this issue, as exemplified by the similarity between these regions in \ref{fig:FilProbc} and \ref{fig:FilProbd}.  These strong similarities also indicate that \EmphMeas{PCA} calculations help enable the use of global density measurements for classification.  While it is clear that the halo radii in regions \textbf{I} and \textbf{II} vary substantially between Figures \ref{fig:FilProba} and \ref{fig:FilProbb}, these values in \ref{fig:FilProbc} and \ref{fig:FilProbd} are much closer.  Measurements using large $k$ provide information about distant structures.  This information is very valuable; however, it can impede classification based on density magnitude by blurring the distinction between the characteristics of the morphological classes.  \EmphMeas{PCA} calculations help decontaminate the class properties, allowing the inclusion of information on the global density magnitude without losing information about small-scale structural properties.

In addition, due to the ambiguous class of particles on halo-filament boundaries, we expect the particles in these regions would have a probability contrast near zero (these particles are colored black in Figure \ref{fig:FilProb}).  However, as the density contrast between halos and voids is very large, we expect the class assignments on halo-void boundaries to have higher probabilities.  In Figure \ref{fig:FilProba}, there are an abundance of ambiguous particles on halo-filament and halo-void boundaries, as well as in the interior of halos; this is most visible in the halo in the upper-left of Region \textbf{II}.  In contrast, Figure \ref{fig:FilProbb} predominantly lacks ambiguous border particles.  The addition of \EmphMeas{PCA} calculations in \ref{fig:FilProbc} and \ref{fig:FilProbd} stabilize these border regions, removing most ambiguous particles inside halo interiors and on halo-void boundaries, and clarify the halo-filament boundary with a thin layer of ambiguous particles, consistently making particle classes more physically realistic.

Though most visible in Regions \textbf{I} and \textbf{II}, these conclusions apply to many halos shown in Figure \ref{fig:FilProb}.  As these halos lie in environments with widely varying densities, it is clear that directionality information helped relax the classifier's dependence on density magnitude, instead favoring the local density contrast.  Constructing a classifier that is consistent in varied environments has proven difficult \citep{LSSSup,TracCWeb}, but information on both the local density field magnitude and directionality provides a way to avoid this issue.

\subsubsection{Structural Mass Fractions} \label{MassFraction}

\begin{figure*}
  \centering
  \includegraphics[width = \textwidth]{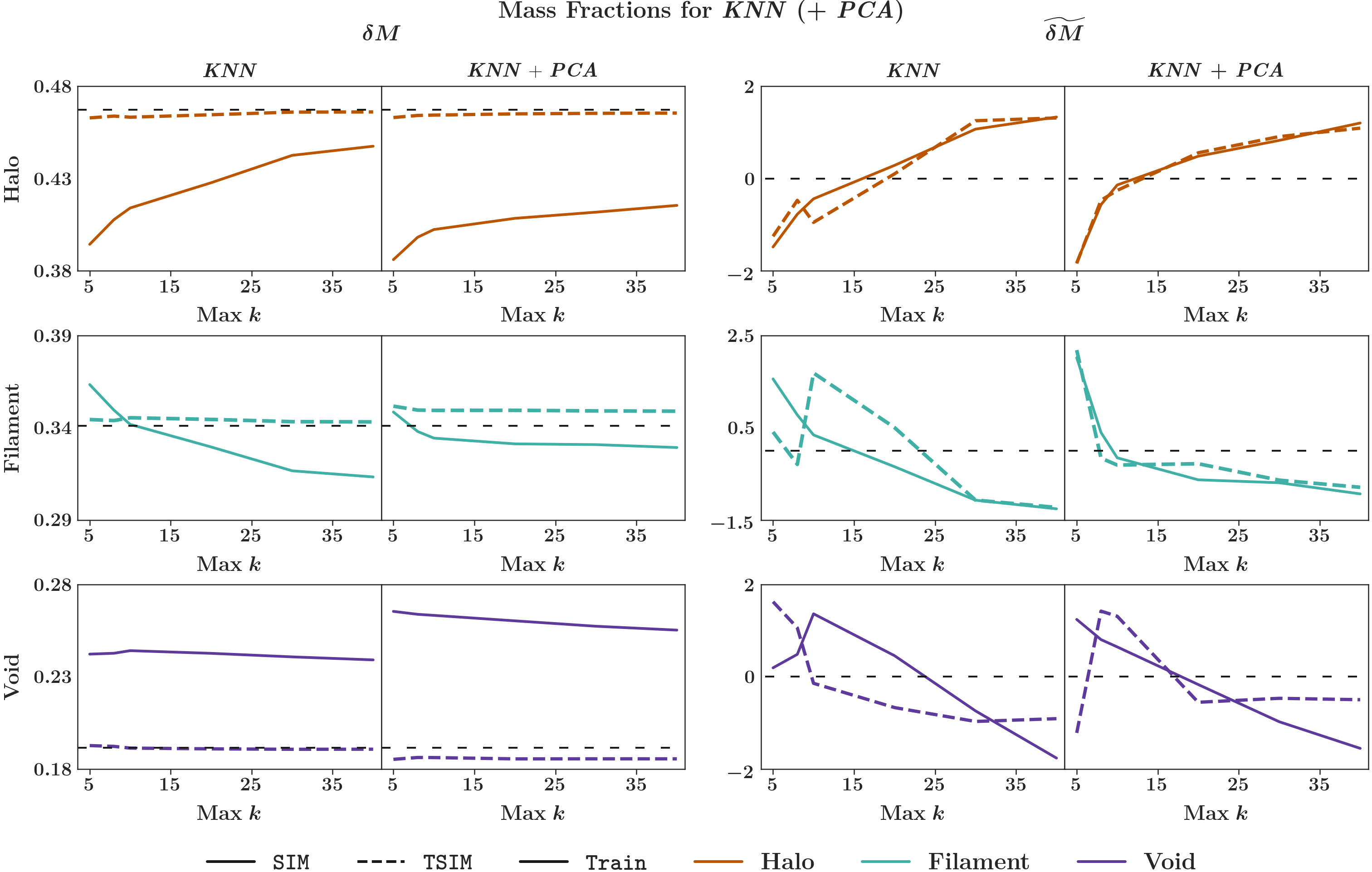}
  \begin{tabularx}{\textwidth}{CC}
     & \\
    \textbf{(a)} & \textbf{(b)}
  \end{tabularx}
  \caption{\textbf{(a)} The mass fractions $\delta M$ for halos, filaments, and voids for \texttt{SIM} and \texttt{TSIM} as a function of $k$. \textbf{(b)} The normalized mass fraction
$\widetilde{\delta M}$ (defined in Eqn. \ref{eq:CPropEqn}) of particles classified as halos, filaments, and voids as a function of $k$.  \EmphMeas{KNN} classifiers were trained using only \EmphMeas{KNN} measurements for $k\leq k_{\textrm{max}}$; each classifier trained using \EmphMeas{PCA} included measurements for all $r\in R_{\textrm{PCA}}$.}
  \label{fig:CProp}
\end{figure*}

In the previous sections, we demonstrate that \EmphMeas{KNN} and \EmphMeas{PCA} calculations together create a robust classifier.  Here, we aim to show that the exact construction of our toy model (particularly that of filaments, which was based on visual appearance) did not substantially bias our results.  Using this information, we may isolate the gain in classifier effectiveness provided by training using \EmphMeas{PCA}.

Figure \ref{fig:CProp} (a) shows mass fractions for halos, filaments, and voids for \texttt{SIM} and \texttt{TSIM}.  For all structures, with and without \EmphMeas{PCA}, it is clear that the mass fractions for \texttt{SIM} are different from those of \texttt{TSIM}.  However, while the mass fractions in \texttt{TSIM} vary little as $k$ changes, those of \texttt{SIM} do (for \EmphMeas{KNN} only).  With the addition of \EmphMeas{PCA} calculations, the \texttt{SIM} mass fractions tend to stabilize to a constant value as $k$ changes.  The addition of \EmphMeas{PCA} calculations affects minimally affects the mass fraction values of \texttt{TSIM}.  The black dashed line shows the true mass fractions for the training dataset (which corresponds very closely to the true mass fractions for \texttt{TSIM}).  The fact that the mass fractions for \texttt{SIM} were much further from the true mass fraction from the training dataset indicates that the exact parameters of the training dataset did not affect the results substantially.  This demonstrates that the filament chosen for the toy model did not bias the classes assigned to particles in \texttt{SIM}.

Next, we aim to demonstrate the robustness provided by the addition of \EmphMeas{PCA} calculations by determining the dependence of the mass fraction on $k$.  To determine whether the classifier's results were dependent on $k$, we assumed the null hypothesis: the assignments made by given classifier are independent of the set of features used for training, so the assignments made by a classifier trained using a particular feature set is taken to be a single trial in a set of predictions made by a single classifier.  Under this assumption, we may normalize the mass fractions using their mean and standard deviation, which were calculated using the mass fractions derived from the test data labelled by each classifier.

Consider a classifier $\mathcal{C}_f$ trained using a set of features $f\subseteq F$.  Let $\delta M_{f,\,\textrm{C}}$ be the fraction of particles assigned to a morphological class $\textrm{C}$ by $\mathcal{C}_f$.  Then define the normalized mass fraction $\widetilde{\delta M}_{f,\,\textrm{C}}$ by

\begin{align} \label{eq:CPropEqn}
    \widetilde{\delta M}_{f,\,\textrm{C}} = \frac{\delta M_{f,\,\textrm{C}} - \overline{\delta M}_{\textrm{C}}}{\sigma_{\textrm{C}}}.
\end{align}

Here, $\overline{\delta M}_{\textrm{C}}$ is the mean of the mass fraction over all $f\subseteq F$ for a class $\textrm{C}$ and $\sigma_{\textrm{C}}$ is the corresponding standard deviation.  For \texttt{SIM} and \texttt{TSIM}, we will compare $\widetilde{\delta M}_{f,\,\textrm{C}}$ for the set of classifiers trained using \EmphMeas{KNN} to those trained using \EmphMeas{KNN} and \EmphMeas{PCA}.

Plots of $\widetilde{\delta M}_{f,\,\textrm{C}}$ are shown in Figure \ref{fig:CProp} (b).  For a given classifier trained using all $k \leq k_{\textrm{max}}$, the value of $\widetilde{\delta M}_{k_{\textrm{max}}}$ corresponds with the number of standard deviations between $\delta M_{k_{\textrm{max}}}$ and the average of $\delta M_{k_{\textrm{max}}\textrm{C}}$ over all $k_{\textrm{max}}$.

While measured mass fractions for \texttt{SIM} and \texttt{TSIM} were different, the normalized mass fractions each have a mean of $0$ and a standard deviation of $1$, allowing \texttt{SIM} and \texttt{TSIM} to be compared to one another directly.

A plot of the normalized mass fractions can be seen in Figure \ref{fig:CProp} (b), where several trends may be seen.  For both \texttt{TSIM} and \texttt{SIM}, as $k$ increased, the normalized mass fraction for halos increased, while that of filaments decreased, regardless of whether or not \EmphMeas{PCA} calculations were used.  In addition, for both \texttt{SIM} and \texttt{TSIM}, \EmphMeas{PCA} calculations generally affected the values of the normalized mass fractions for halos and filaments minimally.  Without \EmphMeas{PCA} calculations, the \texttt{TSIM} normalized mass fraction for both halos and filaments varied widely for small values of $k$; for filaments in particular, this large difference was seen for $k < 30$.  In addition, for $k \leq 10$, the \texttt{TSIM} normalized mass fractions for halos and filaments did not exhibit a particular trend with increasing $k$.

However, the addition of \EmphMeas{PCA} calculations substantially improved the classifier's robustness by making the normalized mass fraction values and trends more consistent between \texttt{SIM} and \texttt{TSIM} for halos and filaments, particularly for $k \leq 10$.  This is likely because halos are only found as nodes on filaments in the toy model, so for very small halos, using only density calculations blurred the lines between halos and filaments through contamination of filament point measurements by nearby halos.  The incorporation of \EmphMeas{PCA} calculations helped differentiate filaments from halos, improving the robustness.  This claim is further supported by the very large filament normalized mass fraction for $k = 5$: as the smallest halos in the toy model had $8$ particles, measurements for $k \leq 5$ would not be able to include all particles in a halo.  As a result, many halos were classified as filaments due to the fact that directionality effects would dominate the local density magnitude for halos, causing overrepresentation of filaments.

Voids showed no recognizable trend for \texttt{TSIM}, regardless of whether or not \EmphMeas{PCA} calculations were included.  For \texttt{SIM}, the normalized mass fraction showed a general downward trend, though as before, the normalized mass fraction without \EmphMeas{PCA} calculations for $k \leq 10$ showed an inconsistent trend.

\section{Conclusions} \label{Conclusion}

We have presented a novel method for cosmic web classification, demonstrating that supervised machine learning using a simplified toy model as training data provides a potential avenue for robust and efficient cosmic web classification.  The simplicity of our toy model indicates that the amount of information required for cosmic web classification is relatively low: assuming appropriate metrics are used, accurate classification can be achieved using even minimally realistic training data.  While most known methods require the measurement or inference of the velocity field and/or knowledge of structural properties predicted by analytical methods, we demonstrate a methodology that requires only information about each galaxy's position.  The use of a random forest algorithm in particular provides a method for achieving probabilistic classification.  In addition, we found that the use of density field directionality measurements in tandem with local density magnitude measurements are crucial for distinguishing between halos, filaments, and voids.  In particular, we have provided a method that can classify isolated halos inside large voids, an outstanding problem discussed in \citet{LSSSup} and \citet{TracCWeb}.  Through calculating and comparing statistical data about our classifications, we found a method to verify our calculations, demonstrating that our algorithm is robust and is not biased by our training data creation algorithm.

The key advantage of our method is the speed and efficiency of toy model generation.  While N-body simulations require substantial computational expense and lack true class values, a new toy model can be generated much more efficiently, and this model provides enough information to accurately classify substantially more complicated N-body simulations and potentially observed data.  This makes the method especially suitable for large datasets: using a single training dataset, we were able to assign class values to an N-body simulation substantially larger than the toy model.  Due to the speed of generation, our method is extremely scalable, as generating additional training datasets would allow us to assign class values to very large N-body simulations at no cost.

In addition, the use of a toy model is particularly suited for cosmic web classification of observed data.  Observed datasets contain masked regions and areas with non-uniform depth.  The use of a toy model helps account for these issues: for each field, an individual toy model can be generated that matches the density, mask, and depth of that field.  By classifying each field using its corresponding training dataset, class assignments would be generated consistently for each field.  The use of periodic boundary conditions or padding (as we used here) could avoid issues associated with masked regions.

The ability to assign probabilistic classes to individual galaxies opens to door to a variety of novel data analysis techniques.  Observables such as density and physical composition are known to be linked to LSS class membership, so correlations with class probability would enable novel methods for understanding these relationships.  For example, spectral analysis may be used to understand chemical composition of galaxies in different environments.  Correlating R/G-band magnitudes with cluster-filament probability contrasts could help establish not only the differences in the chemical compositions of clusters and filaments, but also how that composition changes as the cluster-filament boundary is crossed.

As \EmphMeas{PCA} calculations clarify halo-filament boundaries, the application of a directionality metric can be used to differentiate halos and filaments in general, as well as study the fundamental properties of LSS.  Capturing snapshots of an N-body simulation over large time scales and tracking filament halos as they cross a halo-filament boundary could provide a deeper understanding of the matter inflow from filaments to halos, separating its role in halo formation and collapse from other processes.

While our algorithm is a robust classifier for halos, filaments, and voids, our feature selection is not ideal; in particular, we found that \EmphMeas{PCA} calculations did not perform well for filaments and halos with radii substantially larger than the radius used in the \EmphMeas{PCA} decomposition calculations.  As increasing this radius leads to substantial cross-contamination, future work should focus on identifying and implementing a directionality metric that can more effectively capture properties of large filaments and halos.  The information content \citep{InfoTheory,DecTheory} could be used to determine the utility of new metrics.

In addition, we chose not to differentiate between sheets/walls and filaments in our classifier to the complexity of creating a simple algorithm for this purpose.  Future work could be devoted to expanding this algorithm to allow sheet/wall classification.

Though untested, we expect our classifier to be just as effective in cosmology models other than $\Lambda$CDM.  The length scales for LSS are much larger than the scales at which deviations from $\Lambda$CDM are detectable.  As our training data only reproduces properties of LSS at these larger scales, we expect classification to be independent of the cosmology of the target data set, so the same training data sets may be used to classify fields with equivalent geometric parameters (e.g. average density) but different cosmologies.

\section{Softwares Used}

The Python packages \texttt{Bezier} \citep{Bezier}, \texttt{DBSCAN} \citep{DBSCAN}, \texttt{HMF} \citep{HMF}, \texttt{matplotlib} \citep{matplotlib}, \texttt{numpy} \citep{numpy}, \texttt{scikit-learn} \citep{sklearn}, \texttt{SciPy} \citep{SciPy}, \texttt{Shapely} \citep{Shapely}, and \texttt{yt} \citep{yt} were used extensively in this work.

\section{Acknowledgements}

This material is based upon work supported by the National Science Foundation Graduate Research Fellowship Program under Grant No. DGE --- 1746047

Matias Carrasco Kind has been supported by grant projects NSF AST 07-15036 and NSF AST 08-13543.

This research is part of the Blue Waters sustained-petascale computing project, which is supported by the National Science Foundation (awards OCI-0725070 and ACI-1238993) and the state of Illinois. Blue Waters is a joint effort of the University of Illinois at Urbana-Champaign and its National Center for Supercomputing Applications.

\section{Data Availability}

The data underlying this article will be shared on reasonable request to the corresponding author.  Additional figures and basic software demonstrations may be found at https://github.com/bmbuncher/Prob-CWeb.

\bibliographystyle{mnras}
\bibliography{main}

\bsp
\label{lastpage}

\end{document}